\documentclass[%
 reprint,
superscriptaddress,
nofootinbib,
nobibnotes,
 amsmath,amssymb,
 aps,
 prd,
 longbibliography,
 twocolumn,
]{revtex4-1}% 4-2 has problem with bibtex

\usepackage{siunitx}

\DeclareSIUnit \mm {\milli\meter}
\DeclareSIUnit \cm {\centi\meter}
\DeclareSIUnit \us {\micro\second}
\DeclareSIUnit \ms {\milli\second}
\DeclareSIUnit \pA {\pico\ampere}
\DeclareSIUnit \pC {\pico\coulomb}
\DeclareSIUnit \fC {\femto\coulomb}
\DeclareSIUnit \fF {\femto\farrad}
\DeclareSIUnit \pF {\pico\farrad}
\DeclareSIUnit \mV {\milli\volt}
\DeclareSIUnit \kV {\kilo\volt}
\DeclareSIUnit \V {\volt}
\DeclareSIUnit \GOhm {\giga\ohm}
\DeclareSIUnit \MOhm {\mega\ohm}
\DeclareSIUnit \ton {\tonne}
\DeclareSIUnit \kton {\kilo\tonne}
\DeclareSIUnit \kt {\kilo\tonne}
\DeclareSIUnit \Mt {\mega\tonne}
\DeclareSIUnit \eV {\electronvolt}
\DeclareSIUnit \keV {\kilo\electronvolt}
\DeclareSIUnit \MeV {\mega\electronvolt}
\DeclareSIUnit \GeV {\giga\electronvolt}
\DeclareSIUnit \km {\kilo\meter}
\DeclareSIUnit \kW {\kilo\watt}
\DeclareSIUnit \MW {\mega\watt}
\DeclareSIUnit \MHz {\mega\hertz}
\DeclareSIUnit \kHz {\kilo\hertz}
\DeclareSIUnit \mrad {\milli\radian}
\DeclareSIUnit \year {year}
\DeclareSIUnit \POT {POT}
\DeclareSIUnit \sig {$\sigma$}
\DeclareSIUnit\parsec{pc}
\DeclareSIUnit\lightyear{ly}
\DeclareSIUnit\foot{ft}
\DeclareSIUnit\ft{ft}
\usepackage[printwatermark]{xwatermark}
\usepackage{graphicx}% Include figure files
\usepackage{dcolumn}% Align table columns on decimal point
\usepackage{bm}% bold math
\usepackage{xcolor}
\usepackage[hidelinks]{hyperref}
\usepackage{lineno}
\usepackage{mathtools}
\usepackage{amsmath}
\usepackage{breqn}
\newlength{\figwidth}
\newlength{\fighalfwidth}
\newcommand{\dqdx}{$dQ/dx$}
\newcommand{\dedx}{$dE/dx$}
\newcommand\brabarb{\scalebox{.3}{(}\raisebox{-1.7pt}[0pt][0pt]{$-$}\scalebox{.3}{)}}

\begin{document}
\title{Cosmic Ray Background Rejection with Wire-Cell LArTPC Event Reconstruction in the MicroBooNE Detector}

% ****** Start of file apssamp.tex ******
% See the REVTeX 4 README file
%
%%\documentclass[aps,prd,twocolumn,showpacs,superscriptaddress]{revtex4}  
%%\usepackage{graphicx}% Include figure files
%%\usepackage{dcolumn}% Align table columns on decimal point
%%\usepackage{bm}% bold math

% Jan 2019 notes:
% invited author for CC pi0: H. Chen (as per IB vote, first PRL)
% invited author for CC inclusive and any papers using MCS: Polina Abratenko 

%%\begin{document}
%\title{MicroBooNE Author List - September 2020 (PRD Format)}
%%\title{MicroBooNE Author List - September 2020 - special author list for WC PRL (PRL Format)}

%%\title{MicroBooNE Author List - July 2020 -- special version for CCNp paper (PRD Format)}

% List of institutions in command form:
\newcommand{\Bern}{Universit{\"a}t Bern, Bern CH-3012, Switzerland}
\newcommand{\BNL}{Brookhaven National Laboratory (BNL), Upton, NY, 11973, USA}
\newcommand{\UCSB}{University of California, Santa Barbara, CA, 93106, USA}
\newcommand{\Cambridge}{University of Cambridge, Cambridge CB3 0HE, United Kingdom}
\newcommand{\StKates}{St. Catherine University, Saint Paul, MN 55105, USA}
\newcommand{\CIEMAT}{Centro de Investigaciones Energ\'{e}ticas, Medioambientales y Tecnol\'{o}gicas (CIEMAT), Madrid E-28040, Spain}
\newcommand{\Chicago}{University of Chicago, Chicago, IL, 60637, USA}
\newcommand{\Cincinnati}{University of Cincinnati, Cincinnati, OH, 45221, USA}
\newcommand{\CSU}{Colorado State University, Fort Collins, CO, 80523, USA}
\newcommand{\Columbia}{Columbia University, New York, NY, 10027, USA}
\newcommand{\FNAL}{Fermi National Accelerator Laboratory (FNAL), Batavia, IL 60510, USA}
\newcommand{\Granada}{Universidad de Granada, Granada E-18071, Spain}
\newcommand{\Harvard}{Harvard University, Cambridge, MA 02138, USA}
\newcommand{\IIT}{Illinois Institute of Technology (IIT), Chicago, IL 60616, USA}
\newcommand{\KSU}{Kansas State University (KSU), Manhattan, KS, 66506, USA}
\newcommand{\Lancaster}{Lancaster University, Lancaster LA1 4YW, United Kingdom}
\newcommand{\LANL}{Los Alamos National Laboratory (LANL), Los Alamos, NM, 87545, USA}
\newcommand{\Manchester}{The University of Manchester, Manchester M13 9PL, United Kingdom}
\newcommand{\MIT}{Massachusetts Institute of Technology (MIT), Cambridge, MA, 02139, USA}
\newcommand{\Michigan}{University of Michigan, Ann Arbor, MI, 48109, USA}
\newcommand{\Minnesota}{University of Minnesota, Minneapolis, MN, 55455, USA}
\newcommand{\NMSU}{New Mexico State University (NMSU), Las Cruces, NM, 88003, USA}
\newcommand{\Otterbein}{Otterbein University, Westerville, OH, 43081, USA}
\newcommand{\Oxford}{University of Oxford, Oxford OX1 3RH, United Kingdom}
\newcommand{\PNNL}{Pacific Northwest National Laboratory (PNNL), Richland, WA, 99352, USA}
\newcommand{\Pitt}{University of Pittsburgh, Pittsburgh, PA, 15260, USA}
\newcommand{\Rutgers}{Rutgers University, Piscataway, NJ, 08854, USA}
\newcommand{\StMarys}{Saint Mary's University of Minnesota, Winona, MN, 55987, USA}
\newcommand{\SLAC}{SLAC National Accelerator Laboratory, Menlo Park, CA, 94025, USA}
\newcommand{\SDSMT}{South Dakota School of Mines and Technology (SDSMT), Rapid City, SD, 57701, USA}
\newcommand{\Maine}{University of Southern Maine, Portland, ME, 04104, USA}
\newcommand{\Syracuse}{Syracuse University, Syracuse, NY, 13244, USA}
\newcommand{\TelAviv}{Tel Aviv University, Tel Aviv, Israel, 69978}
\newcommand{\Tennessee}{University of Tennessee, Knoxville, TN, 37996, USA}
\newcommand{\UTA}{University of Texas, Arlington, TX, 76019, USA}
\newcommand{\Tufts}{Tufts University, Medford, MA, 02155, USA}
\newcommand{\VTech}{Center for Neutrino Physics, Virginia Tech, Blacksburg, VA, 24061, USA}
\newcommand{\Warwick}{University of Warwick, Coventry CV4 7AL, United Kingdom}
\newcommand{\Yale}{Wright Laboratory, Department of Physics, Yale University, New Haven, CT, 06520, USA}
%%\newcommand{\listerThanks}{Now at University of Wisconsin, Madison}

% So that institutions appear in alphabetical order:
\affiliation{\Bern}
\affiliation{\BNL}
\affiliation{\UCSB}
\affiliation{\Cambridge}
\affiliation{\StKates}
\affiliation{\CIEMAT}
\affiliation{\Chicago}
\affiliation{\Cincinnati}
\affiliation{\CSU}
\affiliation{\Columbia}
\affiliation{\FNAL}
\affiliation{\Granada}
\affiliation{\Harvard}
\affiliation{\IIT}
\affiliation{\KSU}
\affiliation{\Lancaster}
\affiliation{\LANL}
\affiliation{\Manchester}
\affiliation{\MIT}
\affiliation{\Michigan}
\affiliation{\Minnesota}
\affiliation{\NMSU}
\affiliation{\Otterbein}
\affiliation{\Oxford}
\affiliation{\PNNL}
\affiliation{\Pitt}
\affiliation{\Rutgers}
\affiliation{\StMarys}
\affiliation{\SLAC}
\affiliation{\SDSMT}
\affiliation{\Maine}
\affiliation{\Syracuse}
\affiliation{\TelAviv}
\affiliation{\Tennessee}
\affiliation{\UTA}
\affiliation{\Tufts}
\affiliation{\VTech}
\affiliation{\Warwick}
\affiliation{\Yale}

% Authors in alphabetical order
%\author{P.~Abratenko} \affiliation{\Michigan} % for CC incl and any papers using MCS
\author{P.~Abratenko} \affiliation{\Tufts} 
\author{M.~Alrashed} \affiliation{\KSU}
\author{R.~An} \affiliation{\IIT}
\author{J.~Anthony} \affiliation{\Cambridge}
\author{J.~Asaadi} \affiliation{\UTA}
\author{A.~Ashkenazi} \affiliation{\MIT}
\author{S.~Balasubramanian} \affiliation{\Yale}
\author{B.~Baller} \affiliation{\FNAL}
\author{C.~Barnes} \affiliation{\Michigan}
\author{G.~Barr} \affiliation{\Oxford}
% ...

\collaboration{The MicroBooNE Collaboration}
\thanks{microboone\_info@fnal.gov}\noaffiliation

%%\maketitle

%%\end{document}

\date{\today}% It is always \today, today,
%\linenumbers
%\preprint{APS/123-QED}
\begin{abstract}
For a large liquid argon time projection chamber (LArTPC) operating on or near the Earth's surface to detect neutrino interactions, the rejection of cosmogenic background is a critical and challenging task because of the large cosmic ray flux and the long drift time of the TPC. We introduce a superior cosmic background rejection procedure based on the Wire-Cell three-dimensional (3D) event reconstruction for LArTPCs. From an initial 1:20,000 neutrino to cosmic-ray background ratio, we demonstrate these tools on data from the MicroBooNE experiment and create a high performance generic neutrino event selection with a cosmic contamination of 14.9\% (9.7\%) for a visible energy region greater than O(200)~MeV. The neutrino interaction selection efficiency is 80.4\% and 87.6\% for inclusive $\nu_\mu$ charged-current and $\nu_e$ charged-current interactions, respectively. This significantly improved performance compared to existing reconstruction algorithms, marks a major milestone toward reaching the scientific goals of LArTPC neutrino oscillation experiments operating near the Earth's surface.
\end{abstract}

\maketitle

%\tableofcontents

\section{Introduction}\label{sec:intro}
\begin{figure*}[htb]
	\includegraphics[width=\figwidth]{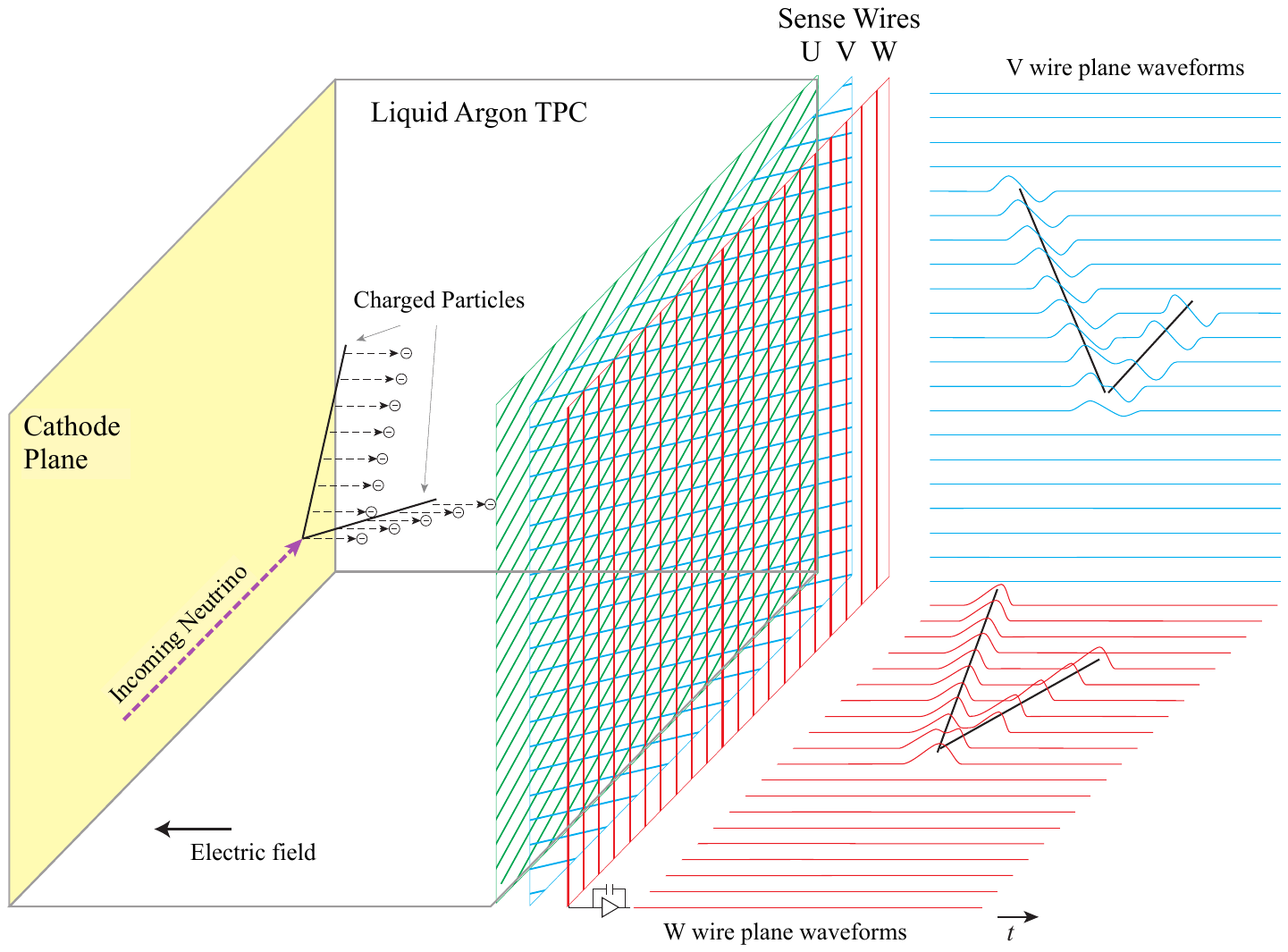}% Here is how to import EPS art
	\caption{\label{fig:tpc_principle} Illustration of a LArTPC detector. Taken from Ref.~\cite{Acciarri:2016smi}.
		A negative high voltage is applied on the cathode plane, producing a uniform drift electric field between the cathode and anode. Multiple parallel wire planes with different wire orientations are placed at the anode as a position-sensitive detector for the drifting ionization electrons. The readout time, $t$, indicates when ionization electrons arrive at the anode.}
\end{figure*}

% Explain the principle of LArTPC 
The liquid argon time projection chamber~\cite{rubbia77,Chen:1976pp,willis74,Nygren:1976fe}
(LArTPC) is a three-dimensional tracking calorimeter that is widely used in neutrino
physics~\cite{Amerio:2004ze,ArgoNeuT2012,Acciarri:2016smi,Badhrees:2012zz,Bhandari:2019rat,Hahn:2016tia,Cavanna:2014iqa,Abi:2020mwi}.
When charged particles traverse the liquid argon (LAr) detection medium, ionization 
electrons and scintillation photons are produced. The detection of the prompt scintillation photons
by a light detector (e.g. a photomultiplier) provides the time of the particle passage. 
Under the influence of an external electric field, the ionization electrons travel at a constant speed toward the anode plane. 
The transverse position of ionization electrons is determined using position-sensitive detectors (e.g. multiple parallel wire planes with different wire orientations as shown in Fig.~\ref{fig:tpc_principle}) at the anode. Given the electron drift velocity, the longitudinal position along the electric field is calculated from the time delay, or drift time, between the time of the particle passage seen by the light detectors and the arrival time of the ionization electrons at the anode. Together, a three-dimensional (3D) image of the particles with a millimeter-scale position resolution is achieved. In addition, the number of measured ionization electrons is proportional to the energy deposition of the charged particle, which can provide particle identification (PID) information. 

%Its capability 
Compared to the water Cherenkov or liquid-scintillator detector technology, the LArTPC technology is expected to have a higher efficiency in differentiating electrons from photons in neutrino interactions through gap identification and reconstructed energy per unit length (\dedx) measurement~\cite{Cavanna:2018yfk}. This capability improves the identification of $\nu_e$ charged-current interactions, which enables precision measurements 
of $\overset{\brabarb}{\nu_{\mu}}\rightarrow\overset{\brabarb}{\nu_{e}}$ oscillations. Utilizing the LArTPC
technology, the MicroBooNE experiment~\cite{Acciarri:2016smi} aims to understand
the nature of the low-energy excess of $\nu_e$-like events observed in the MiniBooNE
experiment~\cite{Aguilar-Arevalo:2012fmn} and to measure neutrino-argon
interaction cross sections in the $\approx$1~GeV scale~\cite{Abratenko:2019jqo,Adams:2018sgn, Abratenko:2020acr, Abratenko:2020sga}. The Short Baseline Neutrino (SBN) Program~\cite{Antonello:2015lea},
consisting of three large LArTPCs on the surface, is under construction to search for light sterile
neutrinos~\cite{Machado:2019oxb}. Moreover, the Deep Underground Neutrino Experiment
(DUNE)~\cite{dune-tdr-1}, which will employ a LArTPC with approximately 10,000~m$^3$ detector modules, plans to search 
for charge conjugation parity violation (CP violation) in the neutrino sector~\cite{Abi:2020evt} and to determine the neutrino mass
ordering~\cite{Qian:2015waa}. To ensure the success of these physics 
programs, the current-generation of large LArTPCs operating on the surface, such as MicroBooNE~\cite{Acciarri:2016smi} and ProtoDUNE~\cite{Abi:2020mwi}, are critical for developing and demonstrating the full capability of this detector technology.

%and its challenge in detecting neutrinos on surface ... 
For LArTPCs operating on the Earth's surface, the presence of cosmic ray muons occurring at a rate of approximately 0.2/m$^2$/ms is a major challenge to reconstructing neutrino interactions efficiently. This challenge is the result of the low rate of neutrino interactions, the slow timing of the TPC (the typical readout time is a few ms), and the decoupling of the ionization charge and scintillation light signals, which are measured by separate detectors. In this article, we present a high-performance cosmic muon background rejection procedure based on the Wire-Cell LArTPC event reconstruction techniques~\cite{Qian:2018qbv} to achieve a generic neutrino interaction selection
in the MicroBooNE experiment.

The MicroBooNE detector~\cite{Acciarri:2016smi} consists of a $\SI{2.56}{\meter} \times \SI{2.32}{\meter} \times \SI{10.36}{\meter}$ (approximately 85 metric tons of LAr) active TPC for ionization charge detection and an array of 32 photomultiplier tubes (PMTs)~\cite{Briese:2013wua} for scintillation light detection. It is located along the Booster Neutrino Beam (BNB)~\cite{AguilarArevalo:2008yp} of the Fermi National Accelerator Laboratory (FNAL) in Batavia, IL. The BNB starts with a proton beam divided in pulses, each pulse is called a spill and lasts 1.6~$\mu$s. Sitting on the beam axis, 463~m from the beam target, the MicroBooNE detector observes one neutrino interaction inside the TPC active volume per about 600 spills at the nominal beam intensity of approximately 4$\times$10$^{12}$ protons on target (POT) per pulse. When the BNB delivers a beam spill, a hardware trigger is initiated in the MicroBooNE DAQ that results in the recording of 4.8~ms of TPC data and 23.4~$\mu$s of PMT data that includes the beam-spill time window. This record is referred to as an event. In addition, self-triggered PMT readouts are taken during a period of 6.4~ms around the BNB trigger. Section~\ref{sec:uboone} provides more details about the MicroBooNE detector and its readout. 

To reduce the data size by selecting events consistent with a neutrino interaction, a software trigger that requires significant PMT signals to be coincident with the beam spill is applied in the data acquisition (DAQ). After rejecting the events with low light output, the recorded data rate is reduced by a factor of 22. After the software trigger, over 95\% of the remaining events have only cosmic rays within the trigger window. Furthermore, at a data rate of 5.5~kHz~\cite{Acciarri:2017rnj}, there are an average of 26 cosmic-ray muons in the full 4.8~ms readout window. Such a large number of cosmic rays creates significant challenges in selecting neutrino events~\cite{Adams:2018fud,Adams:2018sgn,Adams:2018lzd,Adams:2019iqc}. In this work, an offline light reconstruction procedure is applied to reject events triggered by cosmic rays arriving just before the beam spill, leading to a reduction of triggered events by a factor of four. Then, a novel TPC-charge to PMT-light matching algorithm, which requires digital signal processing of the TPC data followed by the reconstruction of 3D images and activity clustering, is applied to remove TPC activity from cosmic rays outside of the beam spill. Section~\ref{sec:foundation} briefly summarizes these techniques. The rejection of cosmic muons that stop in the detector requires a new set of tools to reconstruct the particle track trajectory and its reconstructed ionization charge per unit distance (\dqdx), which is described in detail in Sec.~\ref{sec:track_fitting}. The rejection of the remaining background, which is dominated by cosmic rays in time coincidence with the beam spill, is described in Sec.~\ref{sec:rejection_in_beam}. In particular, the rejection of through-going muons based on geometry information, the rejection of stopped muons based on the rise in \dqdx\ near the candidate stopping point, and the re-examination of mismatched charge-light pairs are described in Sec.~\ref{sec:TGM}, Sec.~\ref{sec:STM}, and Sec.~\ref{sec:incorrect_match}, respectively.  The final performance of this procedure on cosmic ray rejection and neutrino detection is found in Sec.~\ref{sec:performance} before the summary in Sec.~\ref{sec:summary}.

\section{MicroBooNE detector and readout}~\label{sec:uboone}
\begin{figure*}[!htb]
	\includegraphics[width=0.4\figwidth]{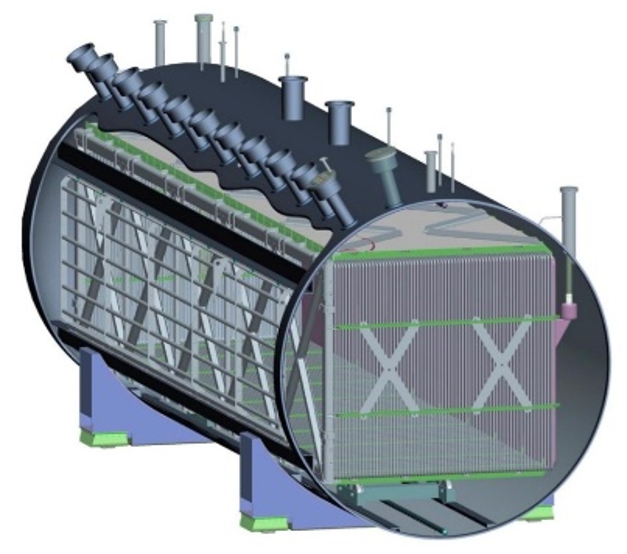}% Here is how to import EPS art
	\includegraphics[width=0.58\figwidth]{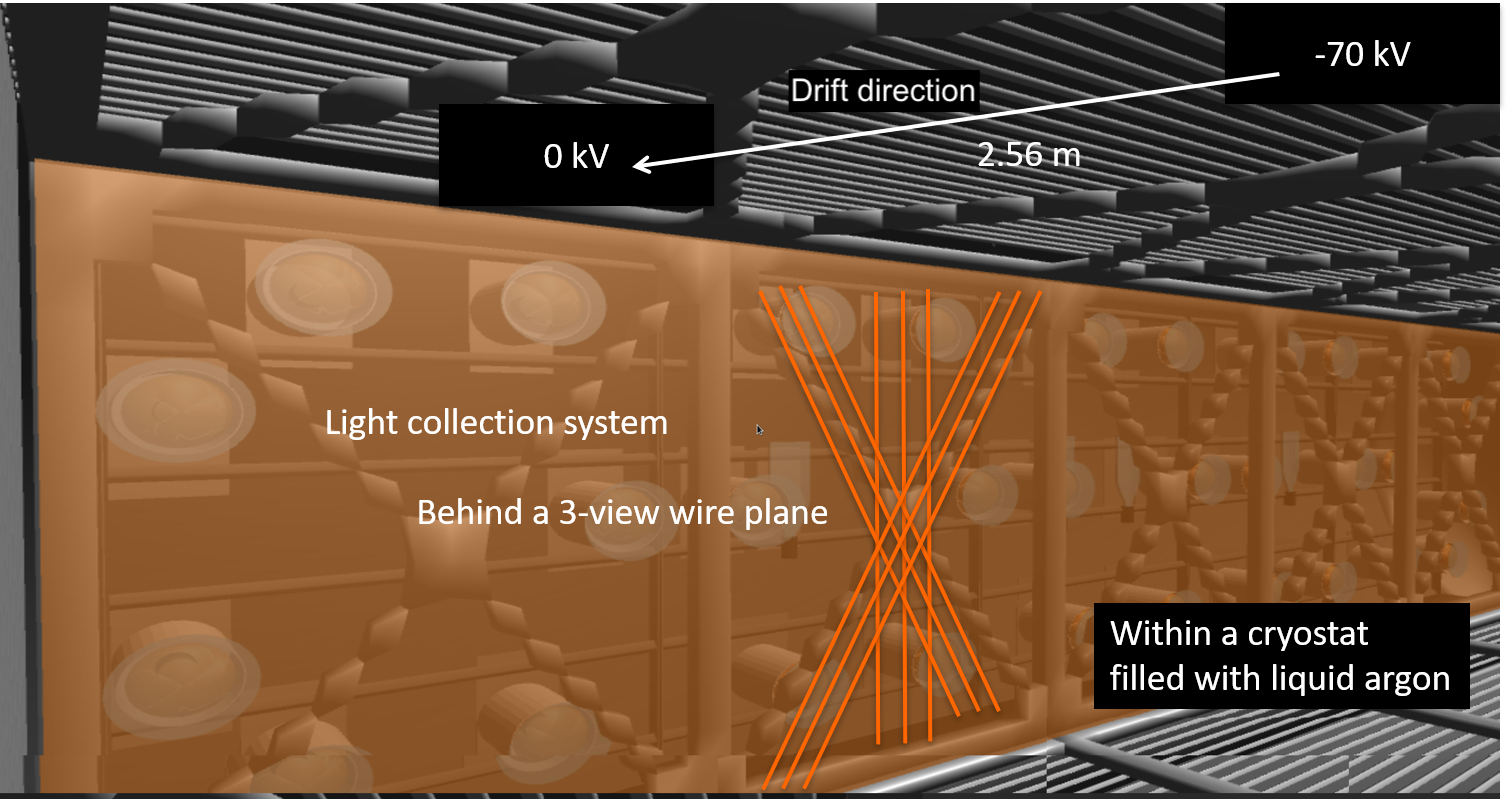}% Here is how to import EPS art
	\put(-490,5){(a)}
	\put(-270,5){(b)}
	\caption{\label{fig:uboone_det} (a) The MicroBooNE detector cryostat. The field cage is shown inside the cryostat. (b) Inside the cryostat of the MicroBooNE detector, visualized with the VENu software~\cite{DelTutto:2017vtk}. The maximum drift distance is 2.56~m with a drift electric field of 273~V/cm. The light-collection system, which consists of 32 PMTs, is located behind the three anode wire planes, which detect ionization charge. }
\end{figure*}

%Basic information regarding MicroBooNE detector
The MicroBooNE detector~\cite{Acciarri:2016smi} is a large LArTPC designed to observe neutrino interactions from the on-axis BNB~\cite{AguilarArevalo:2008yp} and the off-axis NuMI~\cite{Adamson:2015dkw} neutrino beam at FNAL. Figure~\ref{fig:uboone_det}a shows the MicroBooNE TPC, which is housed in a foam-insulated evacuable cryostat vessel. %The active TPC volume is embedded inside the field cage, which is about 10.36~m along the beam direction, 2.32~m in the vertical direction, and 2.56~m in the ionization electron drift direction. While the total mass of liquid argon (LAr) in the cryostat is 170 metric ton, the active mass of the TPC detector is 85 ton. 

As shown in Fig.~\ref{fig:uboone_det}b, the cathode-plane high voltage is set at -70~kV during normal operation, creating a drift field of 273~V/cm. In this field, the ionization electrons drift at a speed of 1.1~mm/$\mu$s~\cite{Li:2015rqa}, corresponding to a 2.3~ms drift time for the maximum 2.56~m drift distance. At the anode side, there are three parallel wire readout planes (see Fig.~\ref{fig:tpc_principle}). Along the drift direction, these planes are labeled as the ``U'', ``V'', and ``W'' planes, and the planes contain 2400, 2400, and 3456 wires, respectively. The wire spacing within a plane is 3~mm, and the planes are spaced 3~mm apart. The wires in the W plane are aligned vertically and the wires in the U and V planes are oriented at $\pm$60$^\circ$ with respect to the vertical direction. The vertical collection plane is labelled here as the W plane to avoid possible confusion with formulas in later sections (note that the collection plane is referred to as the Y plane in previous MicroBooNE publications). The different orientations of the wires allow for determination of the positions of the ionization electrons within the plane that is transverse to the drift direction. Bias voltages for the U, V, and W planes are -110~V, 0~V, and 230~V, respectively, which satisfies the transparency condition that all drifting electrons pass through the U and V (induction) wire planes and are fully collected on the W (collection) plane. The induced current on each wire is amplified, shaped, and digitized through a custom designed CMOS analog front-end ASIC~\cite{Radeka:2011zz} operating at 87~K in the liquid argon. The direct implementation of readout electronics in the cold liquid significantly reduces electronics noise, where the equivalent noise charge (ENC) on each wire is generally below 400 electrons, while a minimum ionizing particle usually produces in total 13000 electrons at a single wire if the particle trajectory is perpendicular to the wire orientation~\cite{Acciarri:2017sde}.

% PMT system description ...
Figure~\ref{fig:uboone_det}b also shows the light-collection system behind the anode wire planes. Thirty-two 8-inch Hamamatsu R5912-02MOD PMTs~\cite{Briese:2013wua}, providing approximately uniform coverage in the anode plane, are used to detect scintillation light from the LAr, which determines the timing of particle activity. A plate coated with tetraphenyl butadiene is installed in front of each PMT to shift the ultraviolet argon scintillation light to the visible part of the spectrum to which the PMT is sensitive. Each PMT is operated with a positive bias voltage, and the signal from the high-voltage line is split into two separated readouts with different gains (a low gain of $\times$1 and a high gain of $\times$10). The two readouts are merged offline and the overall dynamic range is enhanced. The magnitude of the detected light on each PMT provides position information for time-isolated particle activities, which is compared with the predicted light pattern from the ionization charge signals in the TPC. A successful match determines the association between individual TPC activity and light detection, and therefore the time of the corresponding TPC activity. An algorithm that performs this charge-light matching is described in Sec.~\ref{sec:flash_tpc_match}.

% readout and software trigger ... 
%In MicroBooNE experiment, 
Each event in MicroBooNE consists of data from both the TPC and the PMTs. The DAQ readout window for the TPC is 4.8~ms in duration, spanning from -1.6~ms to +3.2~ms relative to the BNB trigger time. This time duration is slightly more than twice the time needed for an ionization electron to drift across the full width of the detector (2.3~ms). At the digitization frequency of 2~MHz, 9600 samples (or time ticks) of the waveform from each wire channel is recorded. 

The PMT data contains two separated trigger streams. Within each event, 1500 samples (digitized at 64~MHz) covering the beam spill are recorded for every PMT channel, which is referred to as the {\it beam discriminator}. In addition, self-triggered PMT readouts called the {\it cosmic discriminator}, each with 40 samples (digitized at 64~MHz) only for the triggered PMT, are taken during a period of 6.4~ms around the trigger time. This is to record the cosmic activity that may result in particle activity recorded by the TPC, because of the relatively slow drift of ionization electrons.

\section{Review of fundamental reconstruction techniques}~\label{sec:foundation}
This section describes some fundamental reconstruction techniques for the TPC and PMT data implemented in the Wire-Cell LArTPC reconstruction. Since most of these techniques have been reported in detail elsewhere, they are briefly summarized for the completeness of this article. 

\subsection{PMT light reconstruction}~\label{sec:light_reco}
The PMT waveforms are processed offline to reconstruct a flash,  which is a cluster of PMT signals that occur within a short time interval. For the cosmic discriminator (40 samples), the photoelectrons (PEs) are calculated by integrating over within a specified time interval~\cite{wire-cell-uboone} after the baseline (estimated from the first sample) is subtracted.%\footnote{For one particular PMT where the electronics response is distinct with a much shorter RC time constant, the PE calculation relies on the peak height after the baseline subtraction instead of the integral.}. 

A flash is then formed by requiring a 100~ns coincidence window among all PMTs, which takes into account the intrinsic light flight time and the timing difference of the PMTs. For the beam discriminator (1500 samples), a deconvolution using a fast Fourier transformation (FFT) is performed to remove the electronics responses from the signal shaper and the splitter, respectively. A flash is then formed if it satisfies the requirements of multiplicity ($>$2 with a threshold of 1.5~PE) and
total PE ($>$6) in a 100~ns window. Unless another flash with larger PE and significantly different PMT hit pattern is found after 1.6~$\mu$s, a flash lasts 7.3~$\mu$s in order to properly include the contribution from the late scintillation light and to exclude the effect from excess noise. The time bin with the maximum PE marks the time of the flash. When the beam discriminator data is present, the data for the same flash from the cosmic discriminator which is not as accurate as that from the beam discriminator is ignored. 
The number of photons in a flash can reach O($10^4$), if the activity is close to the PMT system. 
With this offline light reconstruction, 32\% of the BNB events from the software trigger remain after requiring a time coincidence between the flash and the beam spill. More details of the PMT light data processing are found in Ref.~\cite{wire-cell-uboone}.

\subsection{TPC charge reconstruction}~\label{sec:tpc_reco}
In this section, we describe a series of techniques to reconstruct TPC charge in the three-dimensional space from the original digitized wire waveforms.

\subsubsection{TPC digital signal processing}~\label{sec:tpc_sp}
%% TPC Signal Processing
The first stage of the TPC charge data reconstruction includes noise filtering~\cite{Acciarri:2017sde} and signal processing~\cite{Adams:2018dra,Adams:2018gbi}. The noise filtering step removes the excess noise on the wire channels, including noise from the high-voltage power supply for the cathode plane, and noise from the low-voltage regulator for the cold electronics through a coherent noise subtraction in the time domain. In addition, about 10\% of the nonfunctional channels are identified on an event-by-event basis. More details are found in Ref.~\cite{Acciarri:2017sde}. After noise filtering, the TPC signal-processing step reconstructs the ionization charge distribution from the digitized wire waveforms. The impulse response function includes the field response, which describes the induced current from a moving charge in the TPC, and the electronics response, which characterizes the amplification and shaping of the induced current. Since this function does not depend on the absolute time and position of the ionization electron cloud, a deconvolution technique using a FFT is used. Compared to the one-dimensional deconvolution~\cite{Baller:2017ugz} used in previous work, the signal processing in this work adopts the two-dimensional (2D) deconvolution technique~\cite{Adams:2018dra} which takes into account also nearby wires, significantly improving the performance of the induction wire planes. As a result, the deconvoluted waveforms from three wire planes are demonstrated to be matched both in their magnitudes and in their shapes. The TPC signal also shows good agreement between data and the improved TPC simulation, which takes into account the long-range and fine-grained position-dependent field response functions~\cite{Adams:2018gbi}. 

\subsubsection{Tomographic 3D image reconstruction}\label{sec:wire-cell-imaging}
The signals on the three wire planes provide three co-axial projected views of particle activities in the TPC. The three reconstructed 2D (time vs.~wire) ionization charge distributions are then fed into an advanced tomographic 3D-image reconstruction algorithm: Wire-Cell~\cite{Qian:2018qbv}, which consists of the following steps:

{\it Geometric tiling}: Along the drift direction, a 2D cross-sectional image is reconstructed within every 2~$\mu$s time slice in a tiling procedure. In each cross-sectional image, the consecutive triggered wires are merged to form a wire bundle. Regions called blobs, which represent the overlapping area of these wire bundles from each of the three views within the time slice, are created. A blob is therefore the geometric unit in the Wire-Cell reconstructed 3D image. The resulting image represents the most constrained possibility that is geometrically compatible with the measurements.

{\it Charge solving}: Under the assumption that the same amount of ionization charge is seen by each wire plane, linear equations are constructed which connect the unknown true charges of the blobs and the measured charges on wires. The linear equations are under-determined in many cases, which is the result of loss of information from $O(n^2)$ pixels to $O(n)$ wire measurements in each 2D image. In those cases, constraints from the sparsity, non-negativity, and connectivity information are used to solve the equations using the compressed sensing technique~\cite{cs}. 

{\it 3D imaging}: The 3D image of the event is then reconstructed by simply concatenating all the 2D cross-sectional images in the time dimension. A natural by-product of the Wire-Cell 3D image reconstruction is the 3D charge of each blob, which plays a crucial role in the charge-light matching as will be described in Sec.~\ref{sec:clustering}.
\begin{figure}[!htb]
	\centering
	\includegraphics[width=0.48\textwidth]{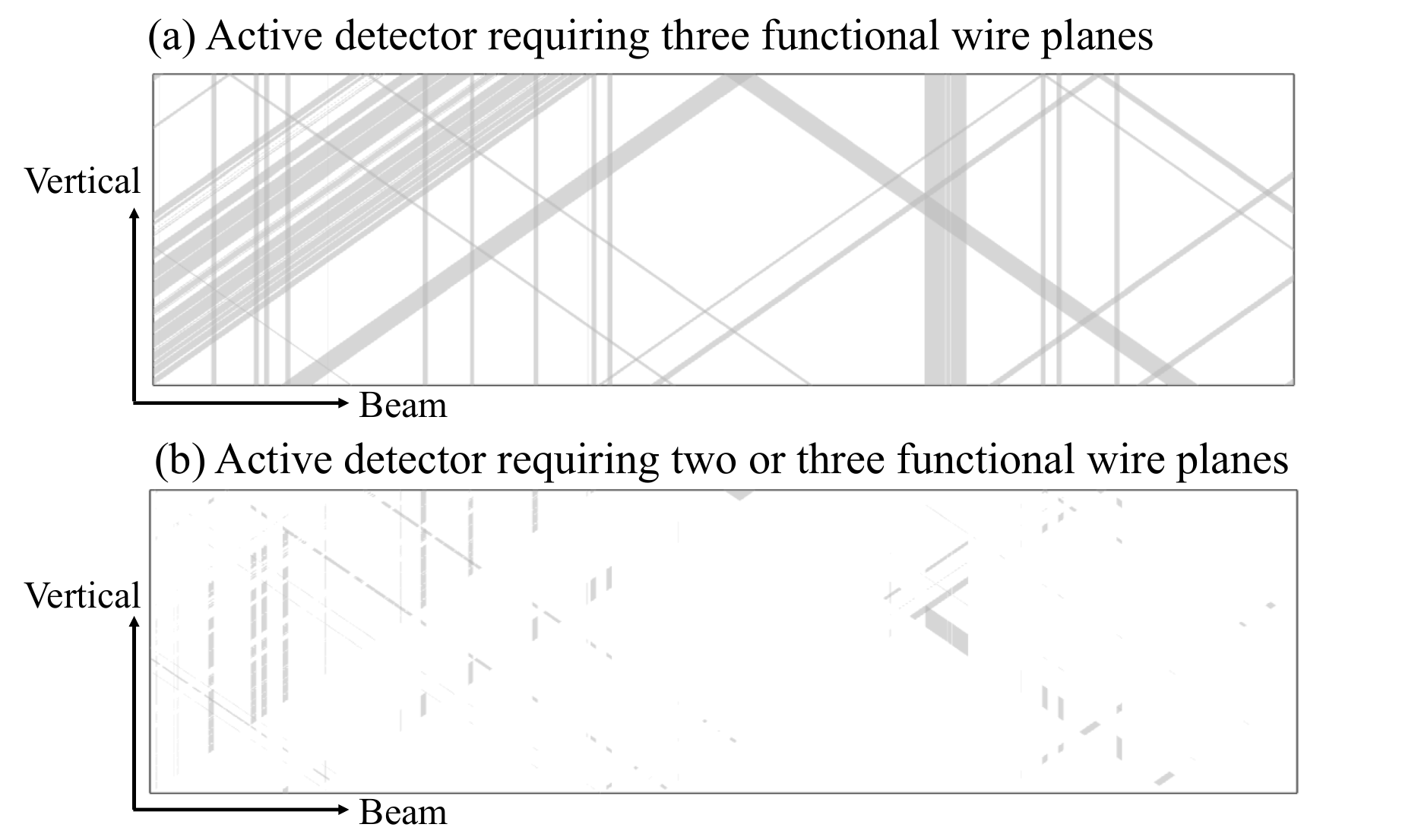}
	\caption{(a) The active detector regions in the Y-Z plane when requiring all three planes to be functional. (b) The active detector regions in the Y-Z plane in white when requiring two out of three planes to be functional. The Wire-Cell 3D image reconstruction allows tiling with only two functional planes, which reduces the dead region area percentage from $\approx$30\% to $\approx$3\%.
		% \url{https://www.phy.bnl.gov/wire-cell/bee/set/342b1227-74fa-4ea6-88c6-b4f69cf42e15/event/0/}.
		% \url{https://www.phy.bnl.gov/wire-cell/bee/set/71a3e15b-eed0-47f3-a70e-1d068f96cdc3/event/0/}.
		%\url{https://www.phy.bnl.gov/wire-cell/bee/set/uboone/imaging/dead-region/event/0/}.
		%\url{https://www.phy.bnl.gov/wire-cell/bee/set/uboone/imaging/dead-region-2/event/0/}.    
	}
	\label{fig:dead_region1}
\end{figure}

In MicroBooNE, due to the existence of $\approx$10\% nonfunctional channels, a 3D image reconstruction requiring all three wire planes to be live would yield about 30\% nonfunctional volume, as shown in Fig.~\ref{fig:dead_region1}(a). Instead, an alternative procedure requiring only two live planes is developed (with the wires in the third plane allowed to be either always on or nonfunctional). This reduces the nonfunctional volume to $\approx$3\%, as shown in Fig.~\ref{fig:dead_region1}(b), at the cost of increasing the number of spurious blobs. Additional algorithms such as iterative reconstruction and deghosting are implemented to improve the quality of the 3D event images. Since the Wire-Cell 3D-image reconstruction only uses very general constraints, the reconstruction of the event is independent of its topology (e.g.~tracks or electromagnetic showers). More details are found in Ref.~\cite{wire-cell-uboone}.

% \begin{figure*}[!htb]
% 	\includegraphics[width=\figwidth]{imaging.pdf}% Here is how to import EPS art
% 	\caption{\label{fig:imaging} %Reconstructed 3D image of run 3493, event 41075 and its associated 2D projections. 
% 		Reconstructed 3D image of an example event and its associated 2D projections.
% 		The vertical (horizontal) axis represents the drift time (wire number). The gray boxes represents the TPC readout time window.
% 		The 3D view is shown with the reconstructed 3D charge information. 
% 		The lighter blue color represents a higher reconstructed 3D charge. See text for 
% 		more discussions. }
% \end{figure*}

%Figure~\ref{fig:imaging} shows an example reconstructed 3D image of a BNB event with the 3D charge displayed in color. A neutrino interaction can be seen. 
%The three 2D projection views, mimicking the input data, are shown as well. 

\subsubsection{3D clustering}~\label{sec:clustering}
% \begin{figure*}[!htb]
% \includegraphics[width=\figwidth]{match_1.pdf}% Here is how to import EPS art
% \caption{\label{fig:clustering} %A $\nu_e$ charge-current (CC) interaction candidate from run 5906, event 3710. 
% A $\nu_e$ charge-current (CC) interaction candidate event. The image after 3D clustering is shown. Different clusters are labeled with different colors,and the black cluster is the neutrino interaction candidate event. }
% \end{figure*}
%% 3D clustering
The reconstructed 3D image consists of thousands of blobs for a typical BNB event. It is important to form group blobs into {\it clusters}, which represent individual physics signals from cosmic rays or neutrino interactions. Since a LArTPC is a fully active detector, tracks from a charged particle are expected to leave continuous energy depositions, which leads to connected blobs in the 3D image. Therefore, a set of 3D clustering algorithms based on 3D proximity and directionality is used. The use of 3D directionality is necessary to cluster electromagnetic showers together.

Special algorithms are implemented to deal with gaps in the 3D image. A gap could result from the $\approx$3\% nonfunctional volume due to the $\approx$10\% nonfunctional channels, the inefficiency introduced by the coherent noise removal step in the noise filtering~\cite{Acciarri:2017sde}, or the inefficiency introduced by the signal processing step for the prolonged track topology~\cite{Adams:2018dra} with tracks nearly orthogonal to the wire planes, in which the TPC signals are typically longer than tens of $\mu$s. In addition, coincidental overlap could happen when ionization charge produced at different times and different TPC locations (e.g.~from two muons) arrives at the anode plane at the same time and position. This leads to two separated clusters being identified as a single one. A special algorithm is created to separate such merged clusters assuming they follow a long-track-like topology. More details of the 3D clustering algorithms are found in Ref.~\cite{wire-cell-uboone}.

% \begin{figure*}[!htb]
% \includegraphics[width=0.8\textwidth]{match_2.pdf}% Here is how to import EPS art
% \caption{\label{fig:matching} %A $\nu_e$CC interaction candidate from run 5906,  event 3710.  
% A $\nu_e$CC interaction candidate event. Compared to Figure~\ref{fig:clustering}, the image after applying 
% the TPC-charge/PMT-light matching is shown. Only the $\nu_e$ CC cluster candidate is left. 
% The red (green) circles represent the observed (predicted) PEs at each PMT. The area of 
% the circle is proportional to the PE. The consistency between the measured and predicted 
% light pattern indicates a good match. %The effective detector boundary as a result of the space charge is shown as red dashed lines. 
% }
% \end{figure*}

\subsection{Matching between charge and light}\label{sec:flash_tpc_match}
Compared to other types of tracking calorimeters, such as NO$\nu$A~\cite{Adamson:2016xxw}, MINER$\nu$A~\cite{Aliaga:2013uqz}, or MINOS~\cite{Michael:2008bc}, the event topology information (from ionization charge) and the timing information (from scintillation light) in a LArTPC are decoupled. In the MicroBooNE detector, within the readout window of 4.8 ms, the typical number of TPC clusters in the active volume is 20--30. However, the typical number of PMT flashes, which is sensitive to activity in the LAr both inside and outside the active volume, is 40--50. There is no direct association between a TPC cluster and a PMT flash.

A new MicroBooNE charge-light matching algorithm is used to properly find the corresponding PMT flash for every TPC cluster~\cite{wire-cell-uboone}. Instead of matching the reconstructed positions of the light and charge directly, hypothetical pairs of TPC clusters and PMT flashes are created and tested. For each hypothesis, the observed PMT light pattern is compared to the predicted light pattern by assuming that the scintillation light yield is proportional to the reconstructed ionization
charge~\cite{wire-cell-uboone}. Since the electron drift start time is assumed to be the PMT flash time, the position of ionization charge along the drift direction is corrected for each cluster-flash pair, which determines the light production positions. This allows the light prediction at each PMT to take into account the light propagation and acceptance as parameterized by a photon library generated by GEANT4~\cite{Geant4}. If the prediction matches the measured flash light pattern, the hypothesis is accepted.

All possible hypotheses of pairs of TPC cluster and PMT flash are constructed after taking into account geometry constraints. For example, if a TPC cluster is not fully contained in the maximum in-time drift window given by PMT flash time, the hypothesis of this pair is not considered. To select the best hypotheses, the compressed sensing technique that was used in the 3D image reconstruction (Sec.~\ref{sec:wire-cell-imaging}) is again adopted, which not only greatly reduces the computational cost, but also naturally takes into account the following situations:

\begin{enumerate}
	\item One TPC cluster can match one PMT flash, which is the majority of cases.
	\item One TPC cluster can match zero PMT flashes due to the inefficiency in the light detection system, especially for low energy activities near the cathode plane (since the PMTs are behind the anode planes).
	\item One PMT flash can match zero TPC clusters because the light system is sensitive to activities outside the TPC active volume.
	\item One PMT flash can match multiple TPC clusters if the clustering step fails to group the same interaction activities together.
\end{enumerate}

The average accuracy of the charge-light matching algorithm is roughly 95\% with a few percent uncertainty, evaluated both with Monte Carlo simulation (by truth information) and with data (by performing hand scans).  More details of the matching algorithms and performance are found in Ref.~\cite{wire-cell-uboone}. After the charge-light matching, the TPC cluster bundle (one or more clusters) that is matched to an in-beam flash becomes a neutrino interaction candidate. 

\section{Track trajectory and \dqdx~determination}~\label{sec:track_fitting}
In the previous section, the existing foundational event reconstruction techniques that lead to the selection of neutrino interaction candidates are summarizied. In this section a new set of MicroBooNE tools used to determine the track trajectory and associated reconstructed charge per unit length (\dqdx) are described. For generic neutrino detection, these tools are essential in rejecting one of the main remaining background events: stopped muons (STMs). An STM is a muon that enters the active TPC volume in coincidence with the beam spill and stops inside the active volume. While tracks from a neutrino interaction originate inside the active volume and travel outward, an STM enters from outside and travels into the TPC volume; therefore the primary difference is the direction of the track. The direction of a stopped track is best determined by searching for a rise in \dqdx~near the candidate stopping point (often referred to as the Bragg peak). Multiple Coulomb Scattering (MCS) is another tool to determine a track's direction~\cite{Antonello:2016niy,Abratenko:2017nki}, although it is not used in this work because of the good performance of \dqdx~alone.

To fully realize the tracking and calorimetry capability of the LArTPC, 3D tracking and \dqdx~measurements are performed through a fit comparing a track hypothesis (a set of  ordered 3D points with their associated ionization charge) with the three sets of 2D wire plane measurements of the reconstructed ionization charges as a function of the drift time and the wire number. 
This approach solves several existing challenges including i) non-uniform signal response 
from the projected wire readout (e.g. a vertical cosmic muon will be completely parallel to the 
collection wire, where the collection-plane signal becomes useless in measuring dQ/dx. In this
case the induction plane signal is crucial in measuring dQ/dx), ii) non-uniform signal 
response from prolonged signal in the induction wire channels, iii) non-uniform signal response
from the isochronous signals and excess noise filtering, and iv) existence of non-functional
channels. This approach of simultaneously fitting of all three wire planes ensures the 
proper reconstruction of track trajectory and dQ/dx for tracks in all angles, which is important
in rejecting STM backgrounds.

In principle, a simultaneous fit to track trajectory and \dqdx~is performed. In practice, the execution of such a fit is computationally challenging because of its non-linear nature. Instead, a two-step fit is adopted, first focusing on the determination of the track trajectory then on the \dqdx~extraction. After decoupling these two problems, the stability of each step is ensured by applying several advanced linear algebra techniques. 

%The performance of the fits relies on a good construction of the initial seed of the track trajectory, which is set up in Sec.~\ref{sec:prep} and Sec.~\ref{sec:graph} and discussed in Sec.~\ref{sec:seed}. In Sec.~\ref{sec:track} and Sec.~\ref{sec:dQ_dx}, the basic principle of the track trajectory and dQ/dx determination is described, respectively. The overall performance of these techniques is shown in Appendix.~\ref{app:fitting_performance}.

\subsection{Track trajectory fit}~\label{sec:track}
The goal of the track trajectory fit is to determine a fine-grained 3D trajectory that is consistent with the intrinsic position resolution of the detector. The final result of the fit is a set of ordered 3D points $S\{x_j, y_j, z_j\}$ for each TPC cluster, which when projected onto the three wire planes best matches the measured 2D trajectories: $U\{u_i, t_i\}, V\{v_i, t_i\}$ and $W\{w_i, t_i\}$. Each cluster is assumed to be a single track-like object. This simplified assumption is sufficient for most of the cosmic background removal tasks described in this work. 

To perform the track trajectory fit, both the 2D images from wire plane measurements after signal processing (Sec.~\ref{sec:tpc_sp}) and the Wire-Cell 3D imaging results (Sec.~\ref{sec:wire-cell-imaging}) are used. The Wire-Cell reconstructed 3D image has coarse resolution because of: 1) diffusion of ionization electrons during their transportation, 2) application of the software filter in signal processing, and 3) geometric degeneracy of isochronous tracks leading to large blobs. 
However, the 3D image is important as it provides a base data structure from which graph theory algorithms are extensively used to find the initial seed of the 3D trajectory. The 3D seed is then utilized to associate nearby 2D pixels to the trajectory fit. This preparatory work is crucial for the fitting procedures described below, but mathematically is rather complex, and exhaustive care is taken to deal with the approximately 10\% nonfunctional channels. The details of the preparatory work are presented in Appendix~\ref{sec:trajectory_appendix}.

Since a TPC cluster typically has a limited number of 3D points, rather than using localized Kalman-filter approaches~\cite{kalman_filter}, a global track fitting strategy inspired by the projection matching algorithm~\cite{Antonello:2012hu} is adopted. For a given 3D trajectory $S\{x_j,y_j,z_j\}$, an empirical test statistic $T$ based on a charge-weighted distance is constructed to compare the projected 2D trajectory with the 2D measurement from each wire plane:
\begin{equation}\label{eq:chi2}
	T\left(S\{x_j, y_j, z_j\}\right) = \sum_{k=u,v,w} T_k, 
\end{equation}
where the index $k$ sums over the U, V, and W wire planes. For instance, the test statistic for the U plane is: 
\begin{eqnarray}\label{eq:dis}
	T_u &=& \sum_j \sum_i \frac{q_i^2}{\delta q_i^2} \cdot (\Delta{L}_u)^2_{ij}, \\
	(\Delta L_u)^2_{ij} &=& \Delta u^2 \cdot \left(u_i - u_j\left(y_j, z_j\right) \right)^2 \nonumber \\
	& & + \Delta x^2 \cdot \left(t_i - t_j\left(x_j\right) \right)^2, 
\end{eqnarray}
where $q$ and $\delta q$ represent the measured charge and its uncertainty within a 2D pixel (time vs. wire) from a wire plane readout. 
$\Delta x$, $\Delta u$ are the width of the time slice and wire pitch of the U plane.
Each time slice corresponds to a 2~$\mu$s readout window, equivalent to 2.2~mm drift distance along $x$ direction.
Here, $j$ represents the index of each 3D point in the track trajectory to be determined, and $i$ represents the index of a nearby 2D pixel in the data measurement from a wire plane. The association between a 3D point and its nearby 2D pixels is precalculated using the initial seed of the track trajectory, so that only a limited number of 2D pixels are included in the fit (Appendix~\ref{sec:trajectory_appendix}). $\Delta L$ represents the distance between the pixel $i$ and the 2D projection of the associated 3D point $j$ on the trajectory. The 2D coordinates ($u_i$, $t_i$) are the wire number and time slice number of pixel $i$, while ($u_j$, $t_j$) are the projected 2D coordinates from the associated 3D point $j$ on the trajectory. The projection from $(x_j, y_j, z_j)$ to $(t_j, u_j, v_j, w_j)$ is calculated as follows:
\begin{equation}\label{eq:trans}
	\begin{split}
		t = \frac{1}{\Delta x} \cdot x + t_0,  \\
		u = \frac{1}{\Delta u} \cdot \left(-\sin(\theta_u) y + \cos(\theta_u) z \right) + u_0,  \\
		v = \frac{1}{\Delta v} \cdot \left(-\sin(\theta_v) y + \cos(\theta_v) z \right) + v_0,  \\
		w = \frac{1}{\Delta w} \cdot \left(-\sin(\theta_w) y + \cos(\theta_w) z \right) + w_0,
	\end{split}
\end{equation}
where $\Delta x$, $\Delta u$, $\Delta v$, $\Delta w$ are the width of the time slice and wire pitches of the U, V, W plane, respectively.
$\theta_u$, $\theta_v$ and $\theta_w$ are the wire orientations with respect to the vertical direction for each wire plane, and $(t_0, u_0, v_0, w_0)$ are the coordinates of the origin. Finally, $q_i$ and $\delta q_i$ in Eq.~\eqref{eq:dis} are the deconvolved charge and its associated uncertainty at pixel $i$. The ratio of the two provides a weight to the distance $\Delta L$, which enhances the contribution of high charge pixels and suppresses the contribution from the pixels with large charge uncertainties. This weighting strategy is necessary given the presence of nonfunctional channels.

By substituting Eq.~\eqref{eq:trans} and Eq.~\eqref{eq:dis} into Eq.~\eqref{eq:chi2}, the test statistic $T$ can be rewritten in a compact matrix form:
\begin{equation}\label{eq:chi2_matrix}
	T = \sum_{k=u,v,w} \left(M_k - R_k \cdot S \right)^2,
\end{equation}
where $S\{x_j, y_j, z_j\}$ is a vector representing the 3D trajectory to be determined, $M_k$ is the charge-weighted 2D pixel coordinates and $R_k$ is the charge-weighted projection matrix derived from Eq~\eqref{eq:trans} for each wire plane. The best-fit 3D trajectory $S$ after minimizing $T$ is the solution to the following equation:
\begin{equation}\label{eq:solv}
	\left(\sum_{k=u,v,w} R_k^T R_k \right) \cdot S = 
	\left( \sum_{k=u,v,w} R_k^T \cdot M_k \right).
\end{equation}
In practice, the dimension of the matrix $\left(\sum_k R_k^T R_k \right)$ is very large, and its direct inversion is challenging computationally. Instead, the biconjugate gradient stabilized method (BiCGSTAB)~\cite{BiCGSTAB} is used, and it is an iterative method to numerically solve a linear system with fast convergence.

%We should note that the choice of the test statistics in Eq.~\eqref{eq:dis} is modified 
%from the following expression:
%\begin{equation}\label{eq:approx}
%\chi_{U/V/W}^2 = \sum_i q_i \cdot dis(U/V/W)_i^2,
%\end{equation}
%which is equivalent to an unbinned likelihood fit assuming a point charge cloud at the 
%generation. The advantages of using Eq.~\eqref{eq:approx} over the standard practice of fitting a Gaussian distribution to a histogram are twofold. First, Eq.~\eqref{eq:approx} does not depend on the width of the Gaussian distribution, which leads to less dependence on the accuracy of the model. Second, the minimization of Eq.~\eqref{eq:approx} leads to linear equations, while the fit of a binned histogram with a Gaussian distribution has a non-linear nature. The weight in Eq.~\eqref{eq:approx} is modified by the ratio of $\frac{q_i}{\delta q_i^2}$, which leads to suppression of 2Dpixels with large charge uncertainties (i.e. large $\delta q_i$). This modification is particularly crucial given the presence of nonfunctional channels. 

This track trajectory fitting process is iterated twice, first with a coarse spacing (1.2 cm) of the 3D trajectory points, and then again with a fine spacing (0.6 cm). For each iteration, the following procedures are applied in order: 
\begin{enumerate}
	\item The initial 3D trajectory seed is determined with a coarse or fine spacing. The 
	selection of seed is equivalent to determine the initial values of a fit. See Appendix~\ref{sec:trajectory_appendix} for more details in how to determine the seed. 
	\item The association between the initial 3D seed and a limited number of nearby 2D pixels is formed. The 3D position of the seed is projected to each view. The nearby 2D pixels within
	a certain range are selected. 
	\item The test statistic $T$ as described in Eq.~\eqref{eq:chi2_matrix} is constructed and the best-fit trajectory points to minimize $T$ are calculated based on Eq.~\eqref{eq:solv}.
	\item The best-fit trajectory points are evaluated to ensure proper ordering and consistent charge distribution. For example, it is possible that the two adjacent best-fit trajectory
	points are too close. In this case, one of them can be removed.
\end{enumerate}
The final result is a fine-grained 3D track trajectory that best describes the wire plane measurements, allowing for accurate \dqdx~determination in the next step.

\subsection{\dqdx~fit}~\label{sec:dQ_dx}
%\subsubsection{Principle}~\label{sec:dQ_dx_prin}
With the track trajectory $S\{x_j, y_j, z_j\}$ determined previously, the goal of the \dqdx~fit is to assign a charge $Q_j$, which is proportional to the number of ionized electrons, to each 3D trajectory point $(x_j, y_j, z_j)$. The \dqdx~along the track trajectory can then be calculated easily. 
Given the predicted charges associated with 3D trajectory points, an empirical test statistic $T'$ is constructed to compare the charges projected onto 2D pixels with the measured charge (time vs. wire) for each wire plane:
\begin{equation}\label{eq:dQ_dxchi2}
	T'\left(S\{Q_j\}; S\{x_j, y_j, z_j\}\right) = \sum_{k=u,v,w} T'_k + T'_{\rm reg},
\end{equation}
where the index $k$ sums over the U, V, and W wire planes, and $T'_{\rm reg}$ is a regularization term. For instance, the test statistic for the U plane is:
\begin{equation}\label{eq:q_compare}
	T'_u = \sum_i \frac{1}{\delta q_i^2} \cdot \left( q_i - \sum_j R^u_{ij} Q_j\right)^2.
\end{equation}
where $j$, $i$, $q_i$ and $\delta q_i$ have the same meaning as in Eq.~\ref{eq:dis}.

%As for the track trajectory fit, $j$ represents the index of each 3D point in the track trajectory, and $i$ represents the index of a pre-calculated nearby 2D pixel in the data measurement from a wire plane. $Q_j$ is the charge to be determined for the trajectory point $j$, and $q_i$ and $\delta q_i$ are the deconvolved charge and its uncertainty of the associated 2D pixel $i$, respectively. Again, the weighting of $\delta q_i$ is particularly necessary due to presence of nonfunctional channels.

$R^k_{ij}$ in Eq.~\eqref{eq:q_compare} is a conversion factor to enable the comparison between the charge $Q_j$ at generation and the measured charge $q_i$ at a wire. In theory, such a conversion involves the entire TPC signal formation and processing chain, which includes: 1) diffusion of the charge cloud $Q_j$ as it travels toward the anode plane, 2) induced current on the sensing wires due to the TPC field response, 3) amplification and shaping of the current due to the electronics response, and 4) digital signal processing to remove noise and deconvolve the induced signal back to the number of ionization electrons. In practice, this chain of processes requires significant computation that precludes direct inclusion in the fit. Instead, an effective signal formation model based on a Gaussian approximation is used. In this model, the diffusion coefficients $D_L$ and $D_T$ for the longitudinal (along the electric field) and transverse (perpendicular to the electric field) directions are assumed to be 6.4 and 9.8 cm$^2$/s, respectively~\cite{Li:2015rqa}. Since the interaction time $t_0$ of the TPC cluster has been determined during charge-light matching step~\cite{wire-cell-uboone}, the broadening of the charge cloud due to diffusion is predicted to be:
\begin{equation}
	\begin{split}
		\sigma_{D_L} =  \sqrt{2 D_L \cdot t_{\rm drift}},\\
		\sigma_{D_T} =  \sqrt{2 D_T \cdot t_{\rm drift}},
	\end{split}
\end{equation}
with $t_{\rm drift}$ being the overall drift time. Additional broadening of the reconstructed charge comes from the software filters during signal processing. This broadening is approximated as $\sigma_{F_t}=1.57$~mm, $\sigma_{F_u}=0.36$~mm, $\sigma_{F_v}=0.60$~mm,
and $\sigma_{F_w}=0.11$~mm, for the drift direction, U, V, and W planes, respectively. These broadening widths are added in quadrature for each wire plane to produce the final width of the Gaussian smearing in the effective model, from which $R^k_{ij}$ in Eq.~\ref{eq:q_compare} is calculated. 

%\begin{table}[!htbp]
%	\caption{Summary of charge broadening parameters in the effective signal formation model.}
%	\label{tab:charge_smear}
%	\begin{center}
%		\begin{tabular*}{0.45\textwidth}[c]{c @{\extracolsep{\fill}}c cccc}\hline
%			$D_L$  & $D_T$ & $\sigma_{F_t}$ & $\sigma_{F_u}$ & $\sigma_{F_v}$ & $\sigma_{F_w}$ \\
%			(cm$^2$/s) & (cm$^2$/s) & (mm) & (mm) & (mm) & (mm) \\\hline
%			6.4 & 9.8 & 1.57 & 0.36 & 0.60 & 0.11 \\\hline
%		\end{tabular*}
%	\end{center}
%\end{table}

%Figure~\ref{fig:simple_model_performance} shows the comparison of the results from the 
%simple model with that of the full simulation for a point-like charge cloud. To further 
%improve the precision of $R_{ij}$, short line segments (each approximated by multiple
%point charge clouds) are used to model the initial charge distribution. 

%\begin{figure}[!thb]
%	\centering
%	\includegraphics[width=0.48\figwidth]{simple_model_performance.pdf}
%	\caption{Performance comparison of the simple model with the full detector
%		simulation and signal processing for a point-like charge cloud at generation. 
%		Different wire planes are separated by vertical lines. }
%	\label{fig:simple_model_performance}
%\end{figure}

Finally, $T'_{\rm reg}$ in Eq.~\eqref{eq:dQ_dxchi2} is a regularization term that incorporates the smoothness of the \dqdx~curve along the track trajectory into the fit. It is defined as:
\begin{equation}\label{eq:dqdx_reg}
	T'_{\rm reg} =  \sum_i \left( \sum_j F_{ij} \cdot \frac{Q_j}{s_j} \right)^2,
\end{equation}
where $s_j$ represents the length of the $j$th segment, which is taken as the average distance between point $j$ and its previous and next points, i.e.~$s_j = (|\vec r_j-\vec r_{j-1}| + |\vec r_j-\vec r_{j+1}|)/2$. Effectively, $Q_j/s_j$ represents the \dqdx~for the 3D trajectory point $j$. $F$ is the regularization matrix with the following format:
\begin{eqnarray}
	F = \eta \cdot \begin{bmatrix}
		-1 & 1 & 0 & \dots & 0 & 0 & 0\\
		1 & -2 & 1 & \dots & 0 & 0 & 0\\
		\vdots & \vdots & \vdots& \ddots & \vdots & \vdots & \vdots\\    
		0 & 0 &0  & \dots&1 &-2 & 1 \\
		0  & 0 & 0 & \dots&0 & 1 & -1
	\end{bmatrix},
\end{eqnarray}
where $\eta$ is the regularization strength. The regularization term effectively calculates the overall second-order derivative of the $Q_j/s_j$ curve, and penalizes those points with large local curvatures. This term is important in the \dqdx~fit to mitigate the impact of ill-defined points, especially when the 2D pixels are inside or close to the nonfunctional channels. The regularization strength $\eta$ is set to be 0.3 or 0.9 if the nonfunctional channels belong to induction or collection wire planes, respectively. Further adjustment to $\eta$ is made for each trajectory point $j$ if the adjacent points share a large number of nearby 2D pixels. 

With the test statistic $T'$ defined in Eq.~\eqref{eq:dQ_dxchi2}, the best-fit set of charge depositions $S\{Q_j\}$ for all 3D trajectory points is obtained by minimizing $T'$ with respect to $Q_j$. Since the trajectory itself is fixed in the previous step (Sec.~\ref{sec:track}), the minimization of $T'$ leads to a system of linear equations similar to those in Eq.~\eqref{eq:solv} and is solved numerically using the BiCGSTAB method when the dimension is high. Reducing the problem to a linear system significantly improves the stability and speed of the fit. Finally, as defined in Eq.~\eqref{eq:dqdx_reg},  \dqdx~for each point $j$ is calculated as the ratio between $Q_j$ and its corresponding segment length $s_j$.

\begin{figure}[!thb]
	\centering
	\includegraphics[width=0.48\figwidth]{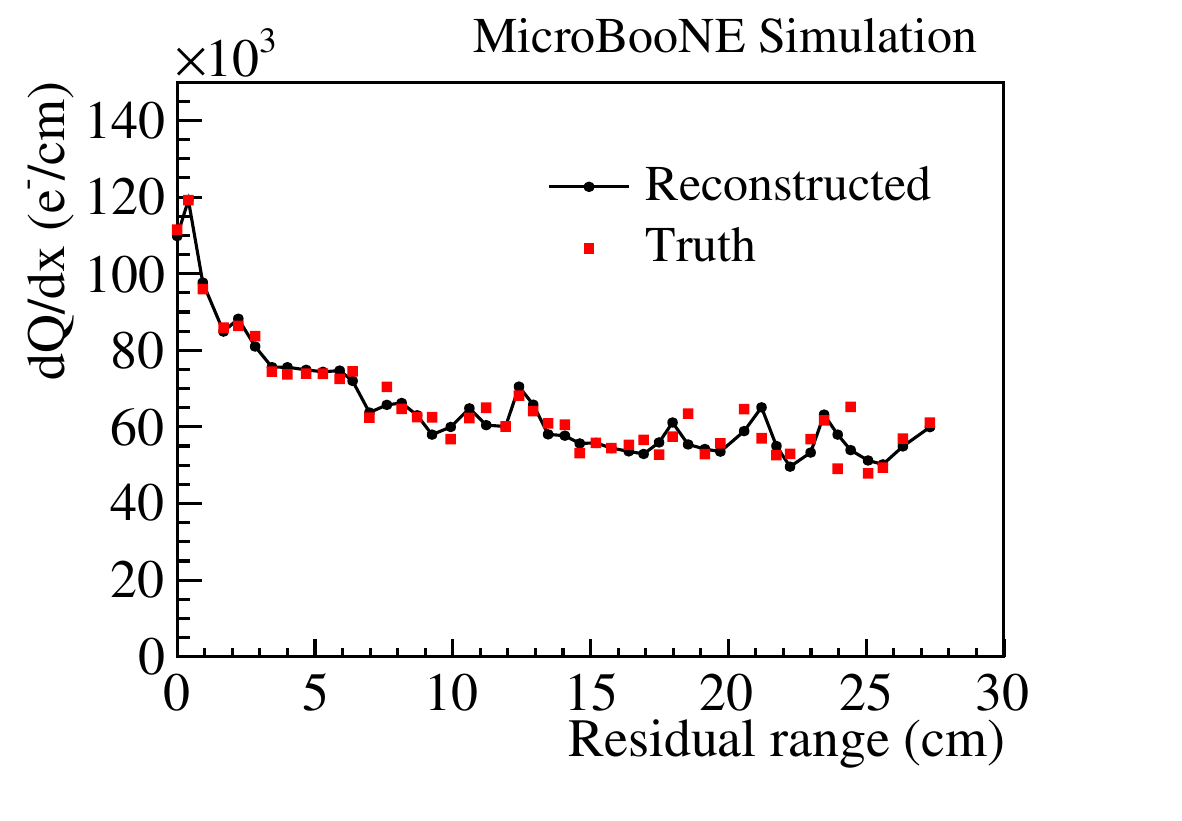}
	\caption{The best-fit \dqdx~(in black) along a simulated muon track trajectory is compared with the true \dqdx~(in red) as a function of the residual range (distance along the track with respect to the stopping location).
	}
	\label{fig:dQ_dx_fitting_example}
\end{figure}

The accurate determination of track trajectory and \dqdx~is vital to rejecting many of the cosmic ray backgrounds described in this work and plays a central role in subsequent steps such as performing particle identification. Figure~\ref{fig:dQ_dx_fitting_example} shows the performance of the \dqdx~determination for a simulated muon track. The reconstructed \dqdx~is consistent with the true \dqdx~along the trajectory, which is sufficient in identifying the increase of \dqdx at
the Bragg peak. In the next subsection, the performance of the full track trajectory and \dqdx~fitting procedure with a few representative data events from MicroBooNE are shown for illustration. 

\subsection{Performance}~\label{app:fitting_performance}
\begin{figure*}[!htbp]
	\centering
	\includegraphics[width=\figwidth]{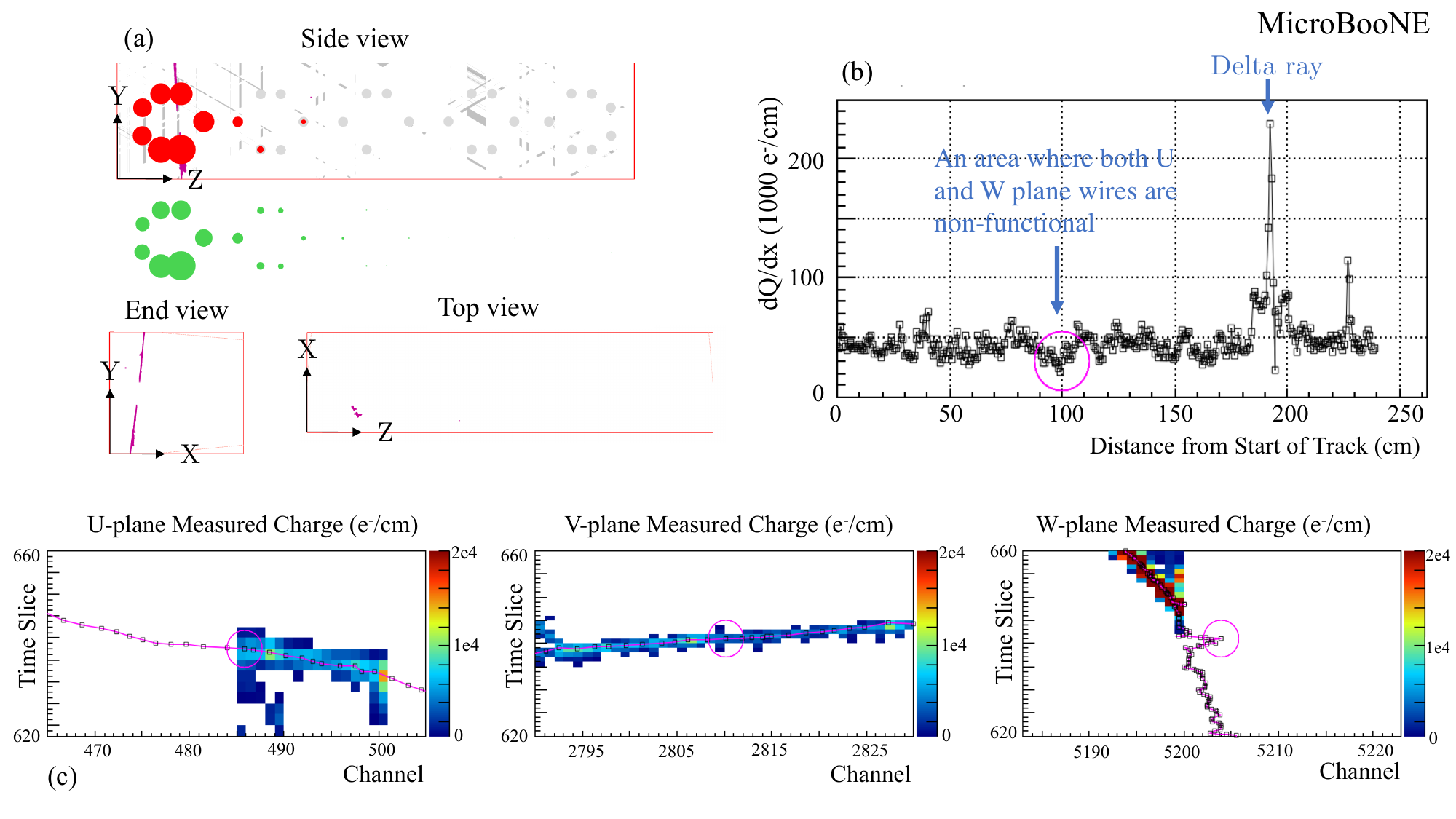}
	\caption{Example of a typical ``bad topology" track from MicroBooNE data. It is isochronous and compact in the W plane (see top view), where the ``cross" shape comes from the cosmic-ray muon and its delta-rays. It has gaps due to nonfunctional channels in both U and W planes.}
	\label{fig:perform1}
\end{figure*}
For most tracks, the trajectory and \dqdx~fitting procedures are robust and accurate due to the excellent tracking and calorimetry performance of the MicroBooNE detector. However, there are several difficult cases where extra care is taken to ensure a high quality fit:
\begin{itemize}
	\item An isochronous track: A track that is parallel to the wire planes, such that all TPC activity is recorded at approximately the same time. This leads to large ambiguities in determining the trajectory.    
	\item A track that is compact in the collection (W) plane view. This leads to difficulty in assigning correct charges to the trajectory points.
	\item A track with segments in the nonfunctional channels, which leads to gaps in the track. This creates difficulty in both trajectory and \dqdx~determination, as they have to be inferred from the other wire plane views in the same time slice.
\end{itemize}
The key to dealing with these difficult cases is in the initial trajectory seed determination, which is described in detail in Appendix~\ref{sec:trajectory_appendix}. Figure~\ref{fig:perform1} shows a typical ``bad topology" track from MicroBooNE data. It poses difficulties in all three categories. It is isochronous, compact in the W plane, and has gaps in the measurement due to nonfunctional channels in both the U and W planes. Figure~\ref{fig:perform1}(a) shows the track topology from the
side, end, and top views. The predicted light pattern (in green) is consistent with the measured light pattern (in red). Unresponsive areas due to nonfunctional channels are shown in dark gray. Figure~\ref{fig:perform1}(b) shows the best-fit \dqdx~curve as a function of the track length. Since this track is a through-going muon (TGM), the \dqdx~fit is consistent with one minimum ionizing particle (MIP), observed to be about 45k e$^-$/cm in data, for most of the track segments. The high \dqdx~region corresponds to the segments with a delta-ray electron, and the dip near 100~cm is the result of an incorrect track trajectory fit near nonfunctional channels. Figure~\ref{fig:perform1}(c) shows the three projection views. The channels that have no measurement are nonfunctional. The magenta lines are the projections of the best-fit 3D trajectory in each wire plane view. The magenta circles correspond to the bad fit in \dqdx~around 100~cm. Despite this imperfection, the majority of the trajectory is successfully determined, including a bridging of the gap corresponding to 100-170~cm in the best-fit \dqdx~curve.

\begin{figure*}[!htbp]
	\centering
	\includegraphics[width=\figwidth]{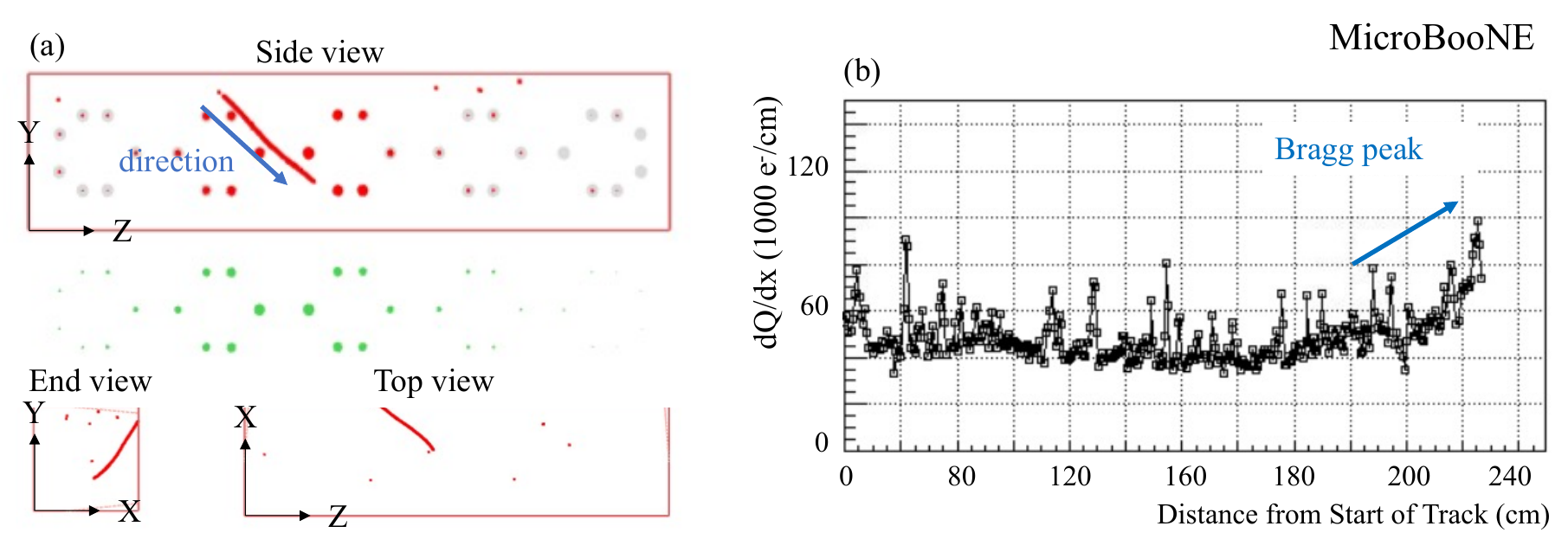}
	\caption{Example of a stopped muon from MicroBooNE data.}
	\label{fig:perform2}
\end{figure*}

As mentioned previously, accurate determination of \dqdx~is crucial in rejecting one of the main backgrounds to neutrino detection: the STM background. Figure~\ref{fig:perform2} shows such an example from a MicroBooNE data event. The side, end, and top views are shown in Fig.~\ref{fig:perform2}(a). The STM entered on the cathode side and stopped inside the detector. Figure~\ref{fig:perform2}(b) shows the best-fit \dqdx~curve as a function of the track length. The track is consistent with one MIP for most of the segments, with a rise (Bragg peak) in \dqdx~at the end, which is a clear evidence of the muon stopping inside the detector. Further details on STM background rejection are described in Sec.~\ref{sec:STM}.  

\begin{figure*}[!htbp]
	\centering
	\includegraphics[width=0.50\figwidth]{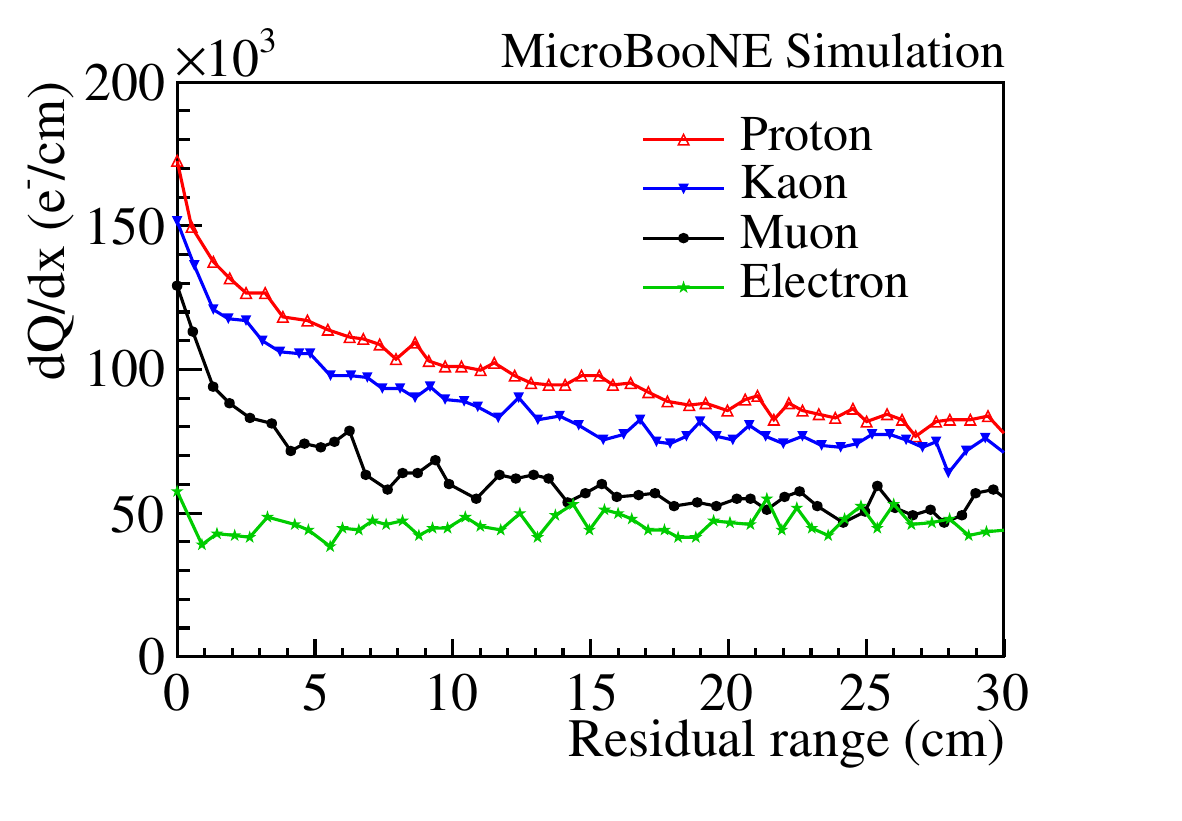}
	\includegraphics[width=0.50\figwidth]{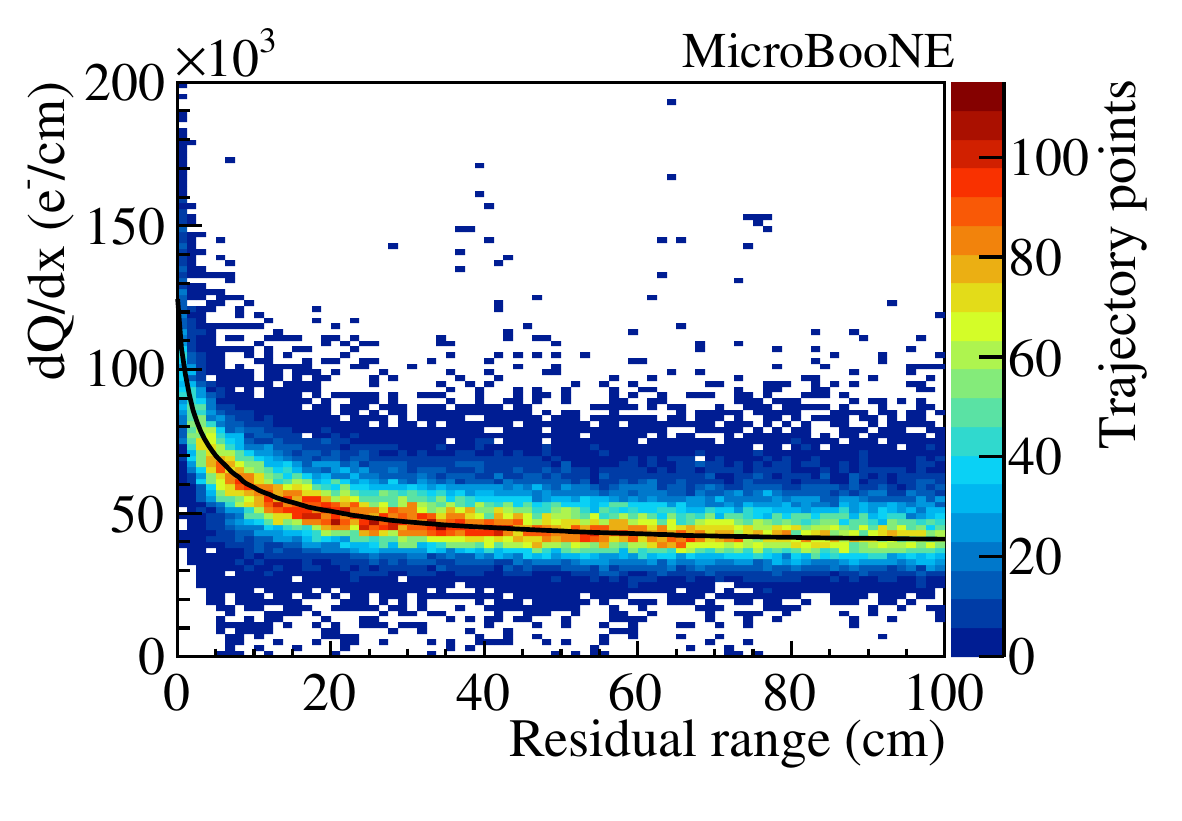}	
	%	\put(-400,-10){a)}
	\put(-445,140){(a)}
	\put(-195,140){(b)}
	\caption{
		%		(Left) Expected dE/dx for various simulated charged particles. The rise of dE/dx near the stopping point (i.e. the Bragg peak) is an important piece of information in determining the direction of the track. The separation in dE/dx for different particles are crucial in performing the particle identification (PID). 
		(a) Examples of best-fit \dqdx~curves for different simulated stopped charged particles  as a function of residual range (distance along the track with respect to the stopping location) using the fitting procedures described in this section. (b) The distribution of best-fit \dqdx~vs. residual range from a sample of $\approx$2000 stopped muon tracks in MicroBooNE data. The color indicates number of trajectory points. The shape of the \dqdx~distribution is consistent with the model-predicted \dqdx~curve (black curve) of the  muon. More details of this analytical model are described in Sec.~\ref{sec:STM}.%The shape of the \dqdx~distributions are consistent with those from simulated muons. %\red{right plot can be improved in the axis label, title and aspect ratio.} 
	}
	\label{fig:pid_example}
\end{figure*}

The track trajectory and \dqdx~determination is important for achieving good particle identification. Although not directly used in this work, such particle identification capabilities are shown in Fig.~\ref{fig:pid_example} with different simulated stopped charged-particle tracks (Fig.~\ref{fig:pid_example}(a)) and a sample of $\approx$2000 stopped muon tracks from MicroBooNE data (Fig.~\ref{fig:pid_example}(b)). The shape of the \dqdx~distributions from the STM data sample is consistent with those from simulated muons. 

\begin{figure*}[!htbp]
	\centering
	\includegraphics[width=\figwidth]{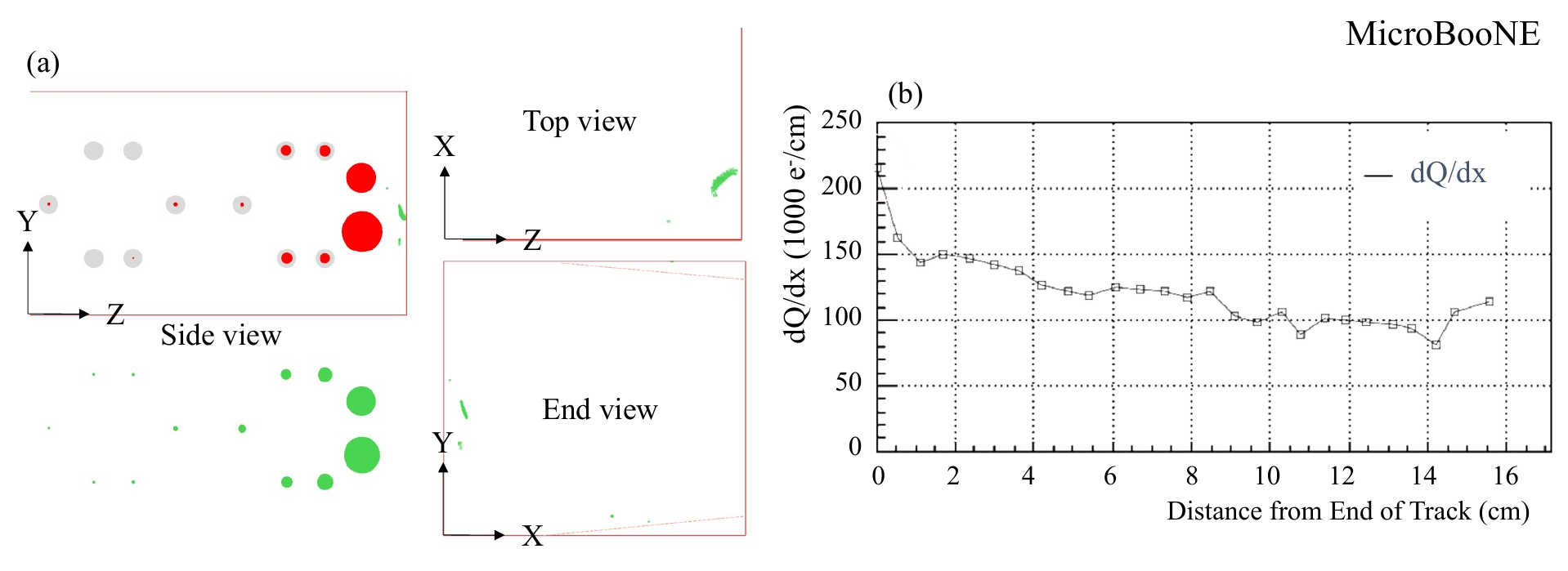}
	\caption{Example of a stopped proton from MicroBooNE data.}
	\label{fig:perform3}
\end{figure*}

Figure~\ref{fig:perform3} shows an example of a stopped proton from MicroBooNE data. The zoomed-in side, top, and end views are shown. Figure~\ref{fig:perform3}(b) shows the best-fit \dqdx~as a function of track length. While a typical MIP gives 45k e$^-$/cm, this track gives about 100k e$^-$/cm. A rise in \dqdx~(Bragg peak) is clearly seen. The shape of the \dqdx~distribution is consistent with that of a simulated proton.

\begin{figure*}[!htbp]
	\centering
	\includegraphics[width=\figwidth]{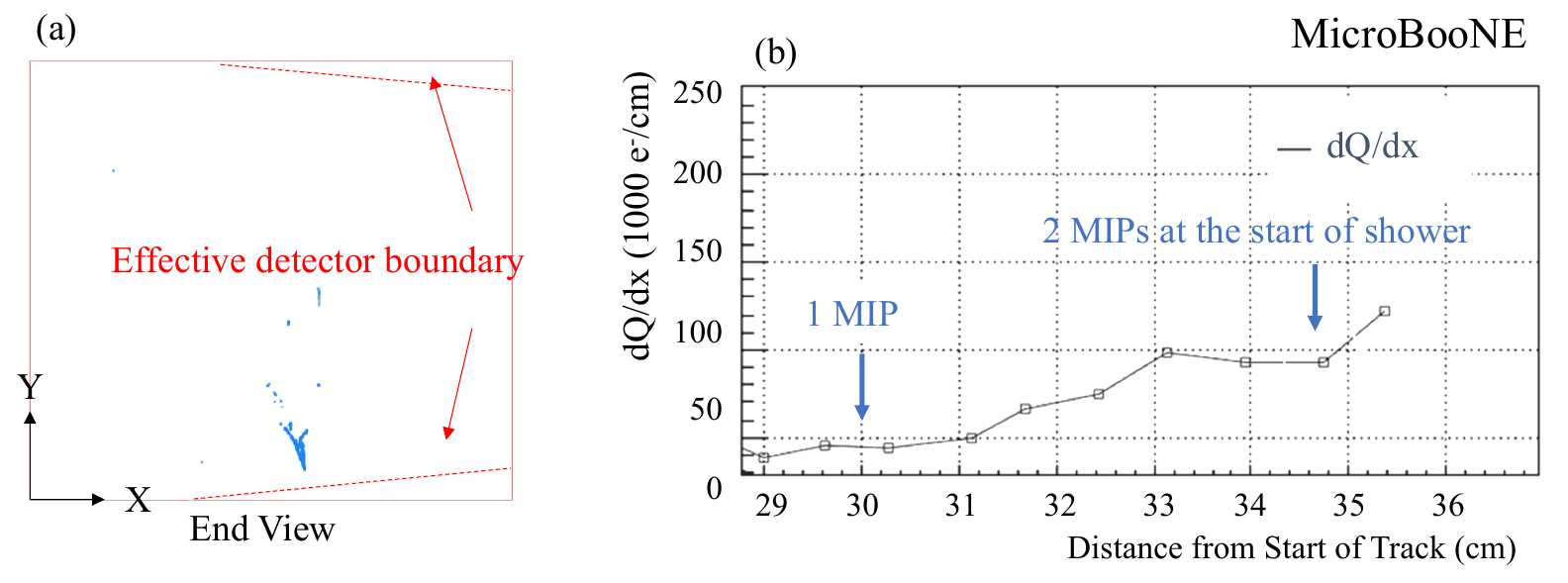}
	\caption{Example of a gamma pair production from MicroBooNE data.}
	\label{fig:perform4}
\end{figure*}

Figure~\ref{fig:perform4} compares one and two MIPs in a photon pair production. Figure~\ref{fig:perform4}(a) shows the top and end views. One energetic delta ray (second MIP) is split from one MIP. At the beginning of the split, two MIPs are overlapping. Figure~\ref{fig:perform4}(b) shows the best-fit dQ/dx. When two MIPs are overlapped, the \dqdx~is approximately 100k~e$^-$/cm, then reduces to $\approx$45k~e$^-$/cm (1 MIP) after the two tracks separate. This separation is crucial in achieving $e/\gamma$ separation with a LArTPC. 

\section{Rejecting in-beam cosmic-ray backgrounds}\label{sec:rejection_in_beam}
As described in Sec.~\ref{sec:flash_tpc_match}, after the charge-light matching, the TPC cluster bundle that is matched to an in-beam flash is a neutrino interaction candidate. However, many of these candidates are actually cosmic-ray backgrounds that are in random time-coincidence with the in-beam flash. This is one of the major challenges for an on-surface LArTPC detector, such as MicroBooNE. The track trajectory and \dqdx~tools described in the previous section allow for further identification
and rejection of these backgrounds, which will be illustrated in this section. Two main cosmic-ray backgrounds are TGMs and STMs: The identification of these muons requires accurate knowledge of the effective detector boundary, which is introduced in Sec.~\ref{sec:boundary}. TGMs and STMs are described in Sec.~\ref{sec:TGM} and Sec.~\ref{sec:STM}, respectively. Finally, the charge-light matching results are re-examined in Sec.~\ref{sec:incorrect_match} to remove certain incorrectly matched candidates, and such events are defined as light-mismatch events.

\subsection{Effective boundary and fiducial volume}~\label{sec:boundary}
%\subsubsection{Effective Detector Boundary}\label{sec:TGM_space_charge}
%
%\begin{figure*}[!thb]
%  \centering
%  \includegraphics[width=0.55\figwidth]{tpc-filter/spacecharge_1.png}
%  \includegraphics[width=0.43\figwidth]{tpc-filter/spacecharge_2.png}
%  \caption{ Physical positions of cosmic ray muons after performing TPC-charge/PMT-light 
%    matching. Left and right panels show different view angles. In particular, the right 
%    panel shows the projection in the Y-X plane with Y and X being vertical and drift 
%    direction, respectively. At large drift distance (right side along the X axis), gaps 
%    can be observed at both top and bottom corners. These gaps are expected because of 
%    the presence of the space charge. }
%  \label{fig:spacecharge}
%\end{figure*}
%There are two effects that impact the effective boundary of the detector active volume. One is the space charge effect, and the other is the existence of nonfunctional channels. 
While the active TPC volume is a rectangular cuboid defined by the rectangular wire-planes at the anode and the corresponding cathode at the opposite end, the reconstructed TPC boundary from trajectories of charged particles deviate from the physical boundary because of the space charge effect. The space charge effect~\cite{Adams:2019qrr, Abratenko:2020bbx} is caused by the drift of positively charged argon ions toward the cathode plane. Since the mass of the argon ion is much larger than the mass of the electron, the drift velocity of the ion is about five orders of magnitude slower. As a result, ions could take several minutes to travel the entire drift distance. For on-surface LArTPC detectors such as MicroBooNE, cosmic-ray muons provide a constant source of positively charged ions, leading to a large accumulation of positive charge inside the active volume and the distortion of the local electric field. As ionization electrons drift toward the anode plane, they are attracted by the positively charged ions toward the detector center. Consequently, the reconstructed position along the wire plane appears to be closer to the detector center compared to its true position, making the effective detector boundary smaller than the actual active TPC boundary. The more time the ionization electrons spend inside the active volume, the larger the position distortion is, which means that the deviation of the effective boundary from the physical boundary is larger for longer drift distances. The detector boundary is mapped out from the observed distribution of entry and exit points of cosmic muons. To enhance the accuracy of this mapping, small, non-muon like clusters, and clusters at the beginning or end of the TPC readout window, which are likely incompletely recorded, are removed. Understanding this effective TPC boundary is the key to identifying if a particle track is contained, or if it enters or exits the detector.

%\begin{figure*}[htpb!]
%  \centering
%  \includegraphics[width=0.4\figwidth]{SCB/canv_opt_scb_YX_2d_01.pdf}
%  \includegraphics[width=0.4\figwidth]{SCB/canv_opt_scb_YX_2d_02.pdf}
%  \includegraphics[width=0.4\figwidth]{SCB/canv_opt_scb_YX_2d_03.pdf}
%  \includegraphics[width=0.4\figwidth]{SCB/canv_opt_scb_YX_2d_04.pdf}
%  \includegraphics[width=0.4\figwidth]{SCB/canv_opt_scb_YX_2d_05.pdf}
%  \includegraphics[width=0.4\figwidth]{SCB/canv_opt_scb_YX_2d_06.pdf}
%  \includegraphics[width=0.4\figwidth]{SCB/canv_opt_scb_YX_2d_07.pdf}
%  \includegraphics[width=0.4\figwidth]{SCB/canv_opt_scb_YX_2d_08.pdf}
%  \includegraphics[width=0.4\figwidth]{SCB/canv_opt_scb_YX_2d_09.pdf}
%  \includegraphics[width=0.4\figwidth]{SCB/canv_opt_scb_YX_2d_10.pdf}
%  \caption{Illustration of locations of the effective detector boundary (black lines)
%    in the Y-X plane (high Y and large X) as a function of different Z locations. 
% The proposed universal effective detector boundary (red line) is also shown.}
%  \label{fig:SCB_YX_2d}
%\end{figure*}

\begin{figure*}[!htb]
	\centering
	\includegraphics[width=0.4\figwidth]{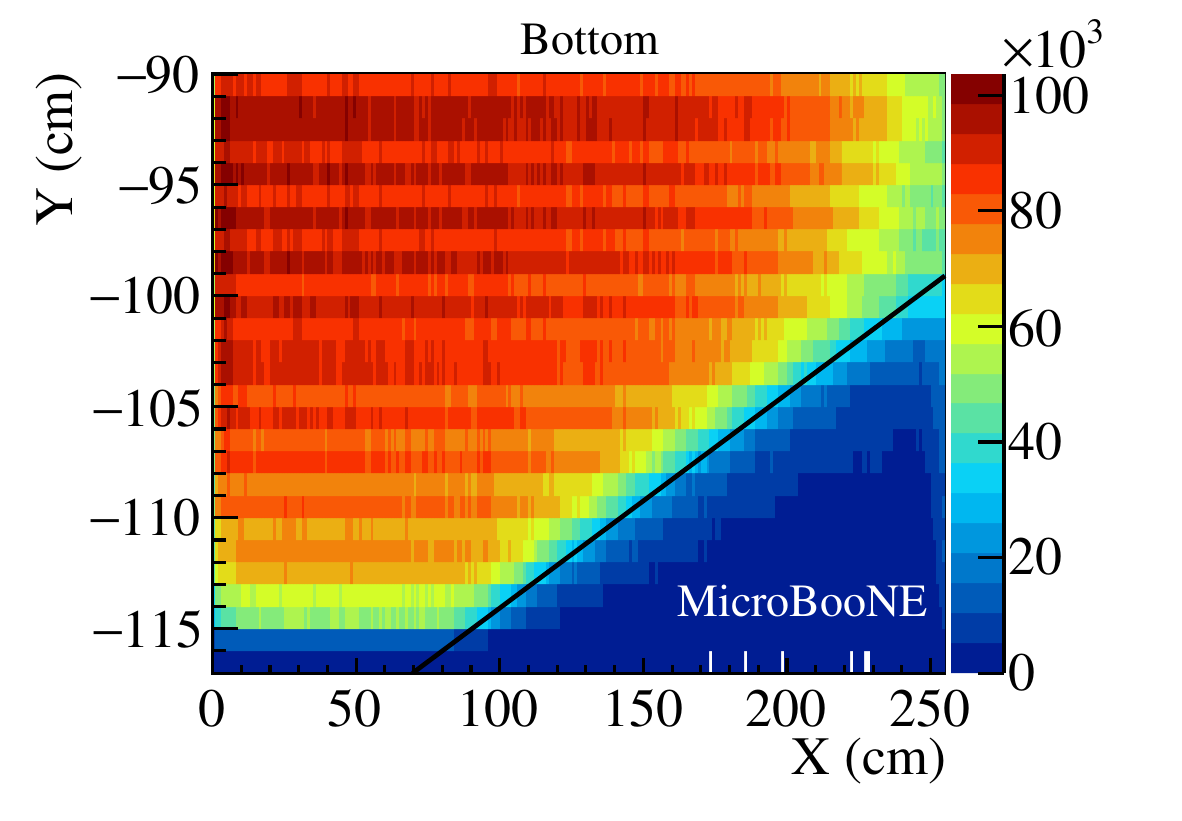}
	\includegraphics[width=0.4\figwidth]{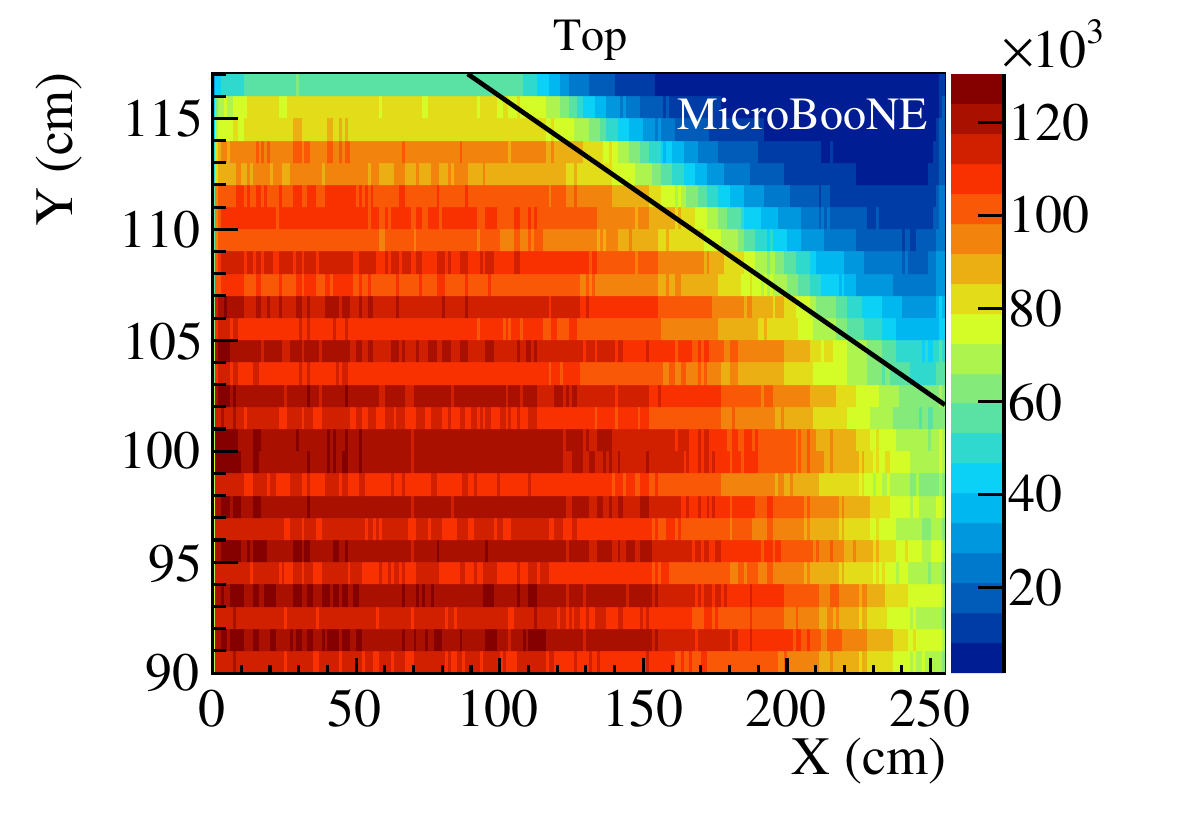}
	\includegraphics[width=0.4\figwidth]{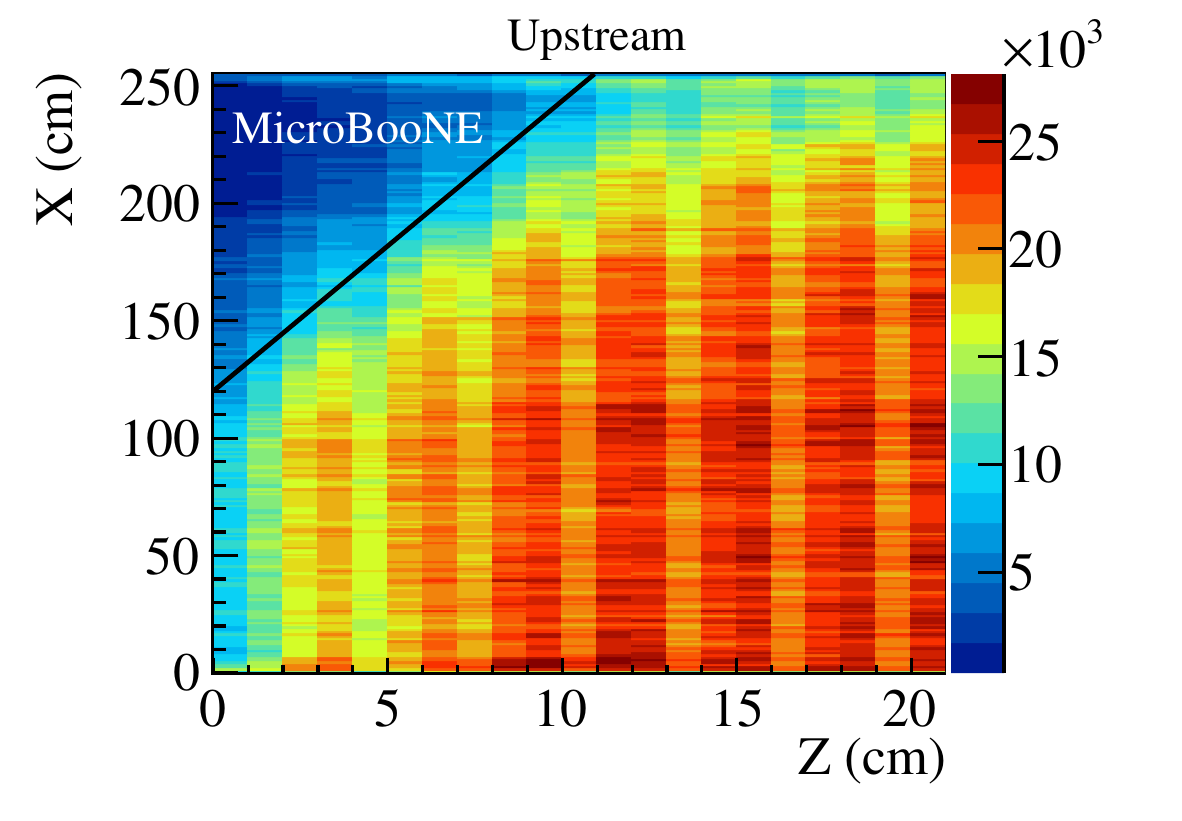}
	\includegraphics[width=0.4\figwidth]{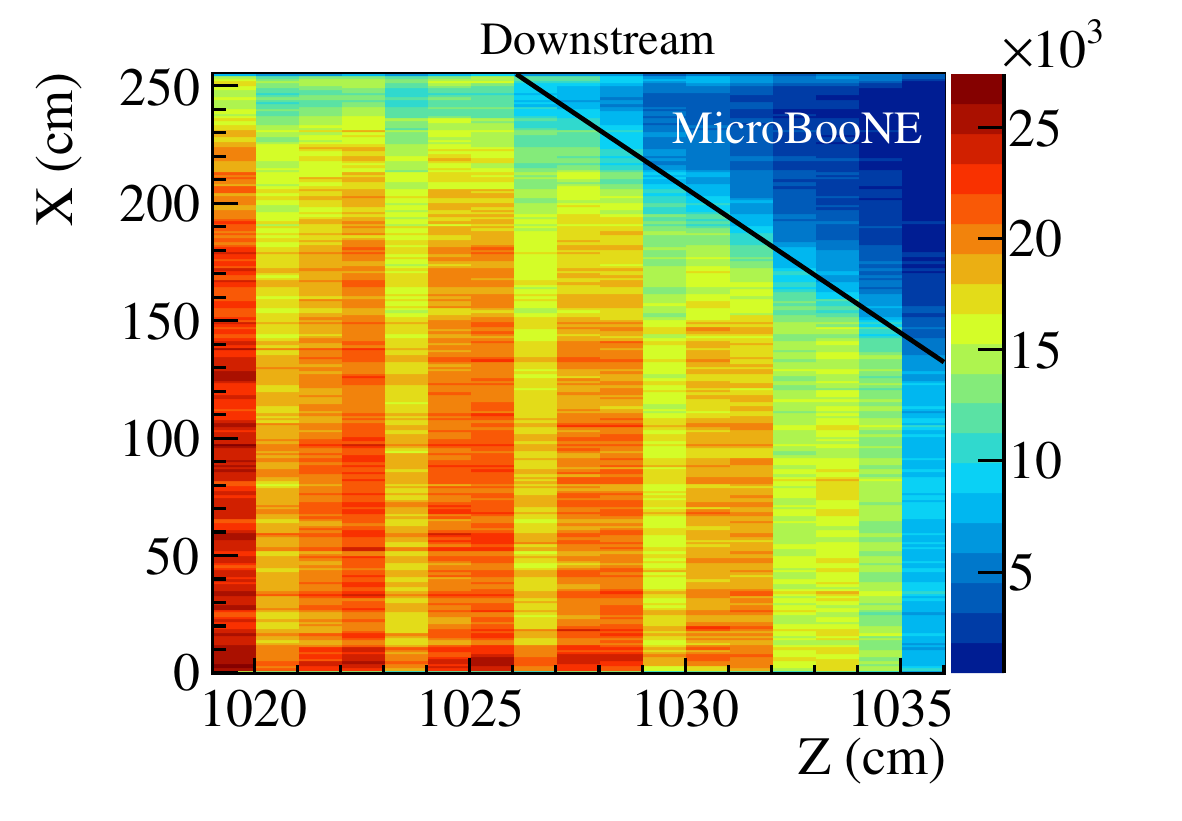}
	\caption{The effective detector boundary (black lines) at the four corners on the cathode side of the detector. The color scale shows the map of cosmic-muon charge clusters in the detector.}
	\label{fig:sc_boundary_1}
\end{figure*}

A sample of approximately 1700 events from MicroBooNE data, each containing 20-30 cosmic muons, are used to map the effective detector volume. The reconstructed 3D image points of the drift-time--corrected clusters are projected onto the X-Y (end view) and the X-Z (top view) planes. Figure~\ref{fig:sc_boundary_1} shows the zoomed-in views of the four corners at large drift distances, i.e.~near the cathode plane at $\approx$256 cm in the X-direction, where the space charge effect is largest. 

\begin{figure}[!htb]
	\centering
	\includegraphics[width=0.45\figwidth]{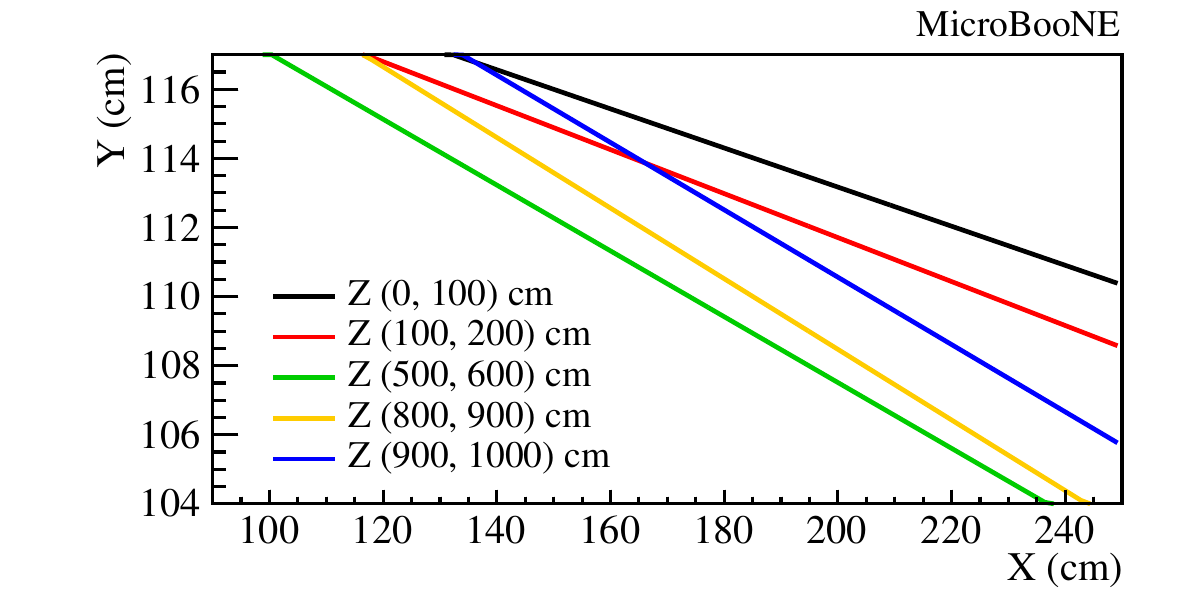}
	\caption{Effective detector boundaries in the Y-X plane for the sub-volume at different Z-slices. A clear position dependence is shown. The color represents the number of reconstructed points; regions with little activity (blue) are outside the active TPC volume. The black lines represent the proposed effective detector boundary. %\red{cz: remove the space after ``('', consider change half of the lines to dashed lines so you can pick less distinctive colors.} 
	}
	\label{fig:SCB_YX_2d}
\end{figure}

The projection of the effective detector boundary on the X-Y plane (end view) has a slight Z-dependence. This is studied by dividing the active TPC volume into 10 sub-volumes along the Z-axis. The effective X-Y boundary of each sub-volume is calculated. The results for large X and Y (top corner) are shown in Fig.~\ref{fig:SCB_YX_2d}, and the Z-dependence is clear. Note that the proposed detector boundary in Fig.~\ref{fig:sc_boundary_1} is conservatively estimated to be the inner boundary of all Z-slices. 
%The same conclusion applies to the rest of corners in the X-Y and X-Z boundary. 
The effective boundary is also checked for different time periods during MicroBooNE data taking, and no clear time-dependence is observed.

%%%% say something about this ...
The cosmic ray rejection and neutrino selection analysis uses a stricter fiducial volume to mitigate the uncertainty in determining the effective TPC boundary. The fiducial volume is defined as the inner volume at 3~cm away from all sides of the effective detector boundary. The total fiducial mass of liquid argon is 80 tons, which is 94.2\% of the full TPC active volume.

%The second modification to the effective detector boundary originates from the existence of nonfunctional channels. As shown in Fig.~\ref{fig:dead_region1}, the known nonfunctional channels leads to nonfunctional regions in the Y-Z plane, which extend into the active volume along the X (drift) direction. When a track enters into the active volume from one of these nonfunctional regions, the initial part of the track is not visible. With the nonfunctional regions precisely known~\cite{Acciarri:2017sde}, this effect can be taken into account in defining the effective detector boundary.

%To examine if a space point is 
%inside nonfunctional regions or not, the U, V, and W channel numbers as well as the time slice 
%are calculated from the 3D position. If two or more channel numbers are inside the known 
%nonfunctional channel list given the corresponding time slice, this point is regarded as being 
%inside the nonfunctional region. 

%Beside these two effects, gaps as the result of the inefficient signal processing and the 
%coherent noise removal can also lead to gaps between TGMs with the effective detector 
%boundary. The algorithm to identify TGM also need to take these into accounts. 

\subsection{Through-going muons (TGM)}~\label{sec:TGM}
The relatively slow drift velocity of the ionization electrons in the LArTPC results in a milisecond-level delayed TPC electronics readout. In general, there are 20--30 cosmic-ray backgrounds within each 4.8~ms TPC readout window in MicroBooNE. After charge-light matching (Sec.~\ref{sec:flash_tpc_match}), the selected in-beam candidates are still dominated by cosmic-ray muons, with a neutrino signal to cosmic-ray background ratio of 1:6.4 (Table~\ref{tab:cosmicrejection}). Most of the cosmic-ray muons traverse the active TPC volume; therefore, they are named through-going muons. 

It is straightforward to identify a TGM with the effective boundary and fiducial volume defined previously. First, a set of extreme points of the corresponding TPC cluster are found, including: 
\begin{itemize}
	\item the highest and lowest points in all three directions: vertical (Y) direction, drift direction (X), and beam direction (Z).
	\item the highest and lowest points in the vertical direction along the principle axis, determined by the principle component analysis (PCA), of the cluster.
\end{itemize}
If two of the extreme points are outside the fiducial volume boundary, this cluster is identified as a TGM, and these two points are defined as the two end points of the TGM. As a by-product, an event is tagged as fully contained if all extreme points are inside the fiducial volume.

Two cases need special care to improve the TGM tagging accuracy:
\begin{itemize}
	\item Gaps in the cluster caused by either nonfunctional channels or inefficient signal processing, which could lead to misplacement of the extreme points. This issue is mitigated by re-examining test points along the principle axis of the cluster against the known locations of the nonfunctional channels, and against the deconvolved signals from the original wire plane measurements. 
	\item A neutrino interaction cluster where there are two separate particle tracks exiting the fiducial volume boundary, mimicking a TGM. This issue is caused by the simplified assumption that each cluster is a single track-like object. Although a full multiple-track fitting algorithm is not developed for this work, a simplified algorithm to detect any large angle deflection along the track trajectory of the cluster is applied to protect against this case.
\end{itemize}

\begin{figure*}[t!hb]
	\centering
	\includegraphics[width=0.98\figwidth]{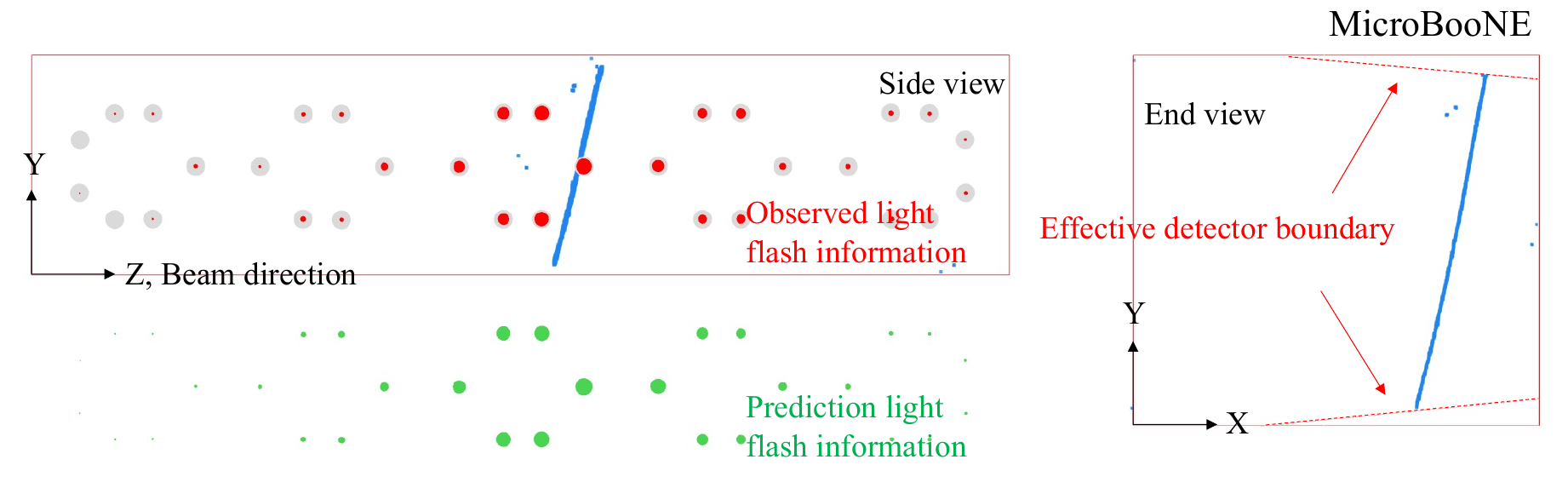}
	\caption{Example of a TGM entering and exiting through the TPC effective boundary due to the space charge effect. 
		%    The beam dump view represents the Y-X projection viewing from the beam dump (vertical-drift direction). The top view, which is also the collection W plane projection, represents the X-Z view (drift direction-beam direction). The side view represents the Y-Z view (vertical-beam direction). 
		The red (green) circles represent the observed (predicted) light flash information. The consistency between them indicates correct charge-light matching. 
		%The effective detector boundary is illustrated with dashed red lines. The nonfunctional volume is labeled as the gray area on the ``side view''. 
	}
	%\url{https://www.phy.bnl.gov/twister/bee/set/uboone/scan/2018-06/d7e6/event/0/} 
	%    See text for more discussions.
	
	\label{fig:tgm_example1}
\end{figure*}

Figure~\ref{fig:tgm_example1} shows a typical TGM from MicroBooNE data. The muon enters and exits through the TPC effective boundary due to the space charge effect. When a TGM is tagged,
the activities associated with the TGM in the same TPC cluster bundle are removed.

\subsection{Stopped muons (STM)}~\label{sec:STM}
%\begin{figure}[t!hb]
%	\centering
%	\includegraphics[width=0.5\figwidth]{STM_Diagram.jpg}
%	\caption{Illustration of the STM tagging algorithm.}
%	\label{fig:stm_diagram}
%\end{figure}
After rejecting the through-going muons, the largest remaining background comes from STMs, which enter the fiducial volume from outside and stop inside. All stopped $\mu^+$s, with a lifetime of about 2.2~$\mu$ sec, decays to a positron. Only about 25\% of stopped $\mu^-$s decay to an electron, with the rest captured by argon nuclei, reducing the total lifetime (capture and decay) to 0.57~$\mu$s. The event topology of a STM therefore contains either only one track from the muon, or sometimes an additional short track from the Michel electron (energy up to $\approx$50 MeV) attached to the end of the muon track. Figure~\ref{fig:perform2} in Sec.~\ref{sec:track_fitting} shows an example STM event from MicroBooNE data, and the best-fit \dqdx~along its track trajectory. Since the tracks from a neutrino interaction travel outward, the main discrimination of STMs relies on the determination of the track direction, which is through the identification of the entering point and then searching for a rise in \dqdx~consistent with the Bragg peak at the end of the track trajectory. 
%In addition, event topology can help to differentiate neutrino interaction candidates from STMs. 
%Figure~\ref{fig:stm_diagram} illustrates how the STM tagging algorithm is executed. 

%\subsubsection{Entering point(s) identification}\label{sec:STM_exiting_point}

The method used to identify the entering point of an STM is similar to that for a TGM as described in Sec.~\ref{sec:TGM}. The difference is that the first stage of the STM tagging requires exactly one extreme point of the corresponding TPC cluster to be outside the fiducial volume.

\begin{figure*}[!htb]
	\centering
	\includegraphics[width=\figwidth]{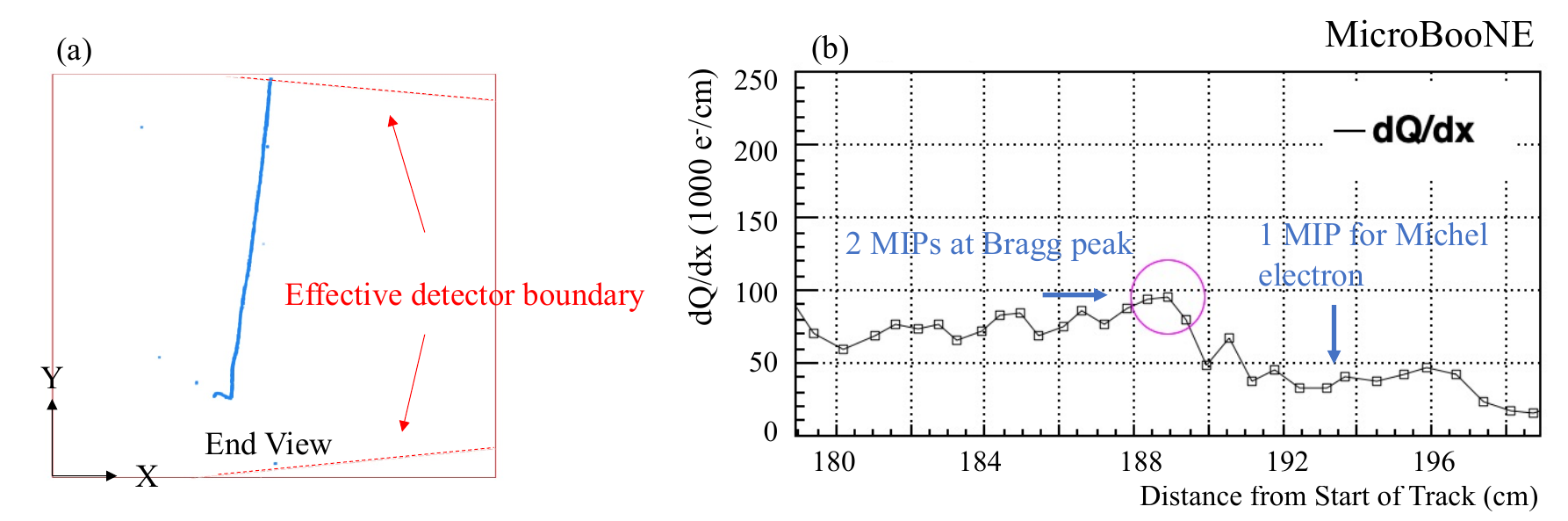}% Here is how to import EPS art
	\caption{\label{fig:STM_michel} A STM candidate with a short stopped muon track following by a Michel electron. A purple circle indicates the Bragg peak.
		%\url{https://www.phy.bnl.gov/twister/bee/set/uboone/scan/2018-06/7f1d/event/0/}
	}
\end{figure*}

The track trajectory and \dqdx~fits for the candidate STM cluster are then performed as described in Sec.~\ref{sec:track_fitting}. In order to correctly determine the stopping point of the STM, a search for a large angle change (i.e.~a kink) along the trajectory is carried out to identify a possible Michel electron track. If a kink is found, trajectory points from the entering point to the kink are labeled as belonging to the STM, while the rest of the trajectory points are labeled as belonging to its associated Michel electron. The track trajectory and \dqdx~fits are then repeated to further improve the accuracy of both. Figure~\ref{fig:STM_michel} shows an example of an STM with a Michel electron attached to the end.

With the entering and stopping points, the track trajectory, and the \dqdx~determined, the final stage of STM tagging is based on a comparison of the measured \dqdx~with the predicted mean \dqdx~along its trajectory. For the prediction (e.g. the analytical model shown in Fig.~\ref{fig:pid_example}b), the mean \dedx~of an STM is calculated from the PSTAR database~\cite{pstar}, and is checked for consistency with the GEANT4~\cite{Geant4} simulation. The modified-box model~\cite{Acciarri:2013met}, which takes into account the recombination effect of ionization electrons, is used to convert the \dedx~to \dqdx. The parameters of the model are taken from Ref.~\cite{Adams:2019ssg} and the calibration of the electronic response is taken from Ref.~\cite{Adams:2018gbi}.
A residual discrepancy between the predicted \dqdx~and data is observed and likely results from an imperfect recombination model. 
In order to mitigate the impact from uncertain overall normalization of the reconstructed \dqdx, Kolmogorov--Smirnov (KS) tests are utilized to determine the rise in a \dqdx~distribution.
%An additional 15\% overall suppression needs to be applied to the predicted \dqdx~in order to be consistent with the data. This normalization difference is partially explained by the usage of the mean instead of the most probable value of the \dqdx~distribution in the prediction, with the residual difference likely coming from the imperfect recombination model.

%Figure~\ref{fig:STM_pred_example} shows the comparison of the measured \dqdx~with the predicted \dqdx~near the stopping point of an STM. A good agreement is observed. A key of the STM identification is to find a Michel electron at the end of the track. KS tests are performed in order to find and confirm if the residual track is consistent with a Michel electron. More details can be found in Appendix.~\ref{app:stm}.
For each STM candidate track, two KS tests are performed for the residual 35~cm of the track trajectory, determined by counting backward from the stopping point to check the consistency between the measured \dqdx~distribution and the references. The first KS test, KS$_1$, is between the measured \dqdx~and the predicted \dqdx~of an STM. The second KS test, KS$_2$, is between the measured \dqdx~and a MIP hypothesis using a flat \dqdx~prediction (45k e$^-$/cm). The two KS scores are then used to build an empirical discriminator, $\alpha$: $\alpha = $KS$_1 - $KS$_2 + \left(|R_1-1| - |R_2-1| \right)/5$, where $R_1$ and $R_2$ are the two ratios between the prediction and the measurement of the integrated \dqdx. The candidate track is identified as an STM if $\alpha<0$. In addition, if there is a residual Michel electron track identified after the main STM track, the residual \dqdx~distribution and its track's (lack of) straightness are required to be consistent with the Michel electron hypothesis. 

%A search is performed along the track towards the starting point from a candidate stopping point, to find the space point with the maximal \dqdx~within a predefined range. The default range is 40~cm, but if a kink is found then the range is limited to 5~cm. 
%Two KS tests are performed; one between the measured \dqdx~and the predicted \dqdx, and another between the measured \dqdx~and a MIP hypothesis using a flat \dqdx~prediction (50k e$^-$/cm). Two KS distances are used to build a metric, $KS_{dis1} - KS_{dis2} + \frac{\left(|R_1-1| - |R_2-1| \right)}{5}$, where $R_1$ and $R_2$ are two ratios between the prediction and the measurement of the integrated \dqdx. An event is identified as an STM if a negative value is obtained. 
%In addition to this metric, other variables such as the residual track length, the direct distance between the starting and the end points of the residual track, and the average \dqdx~on this residual track are used to check if the residual track is consistent with those of a Michel electron. 

%Based on the variables listed above, a set of empirical cuts are developed. If the measured \dqdx~has a rise (independent of its normalization), an STM would generally be tagged. However, if the residual track is not consistent with a Michel electron, no STM would be tagged. If there are additional {\it clusters} associated with the main {\it cluster} (see discussions in Appendix.~\ref{app:stm_protection}), a more stringent test enforcing the consistency in normalization is performed instead. 

Several further checks are performed to increase the accuracy of the STM tagging: 1) Check for potential energetic delta rays on the trajectory path, which could impact the STM trajectory determination; 2) Take into account the cases where \dqdx~does not rise to its highest possible values when the muon decays in flight; and 3) Protect against a neutrino interaction being misidentified as an STM, similar to the cases in the TGM tagging. Figure~\ref{fig:pid_example} showed the measured \dqdx~distribution from a sample of $\approx$2000 STMs identified from MicroBooNE data. In Appendix~\ref{app:stm}, we show several representative STM examples with difficult topologies or unusual \dqdx~distributions.

\subsection{Light-mismatched (LMM) events}~\label{sec:incorrect_match}
The third largest in-beam background, next to the TGM and STM backgrounds, comes from light-mismatched (LMM) events, where the observed light pattern on the PMTs does not agree with the prediction from the matched TPC clusters. This could happen because the charge-light matching procedure as described in Sec.~\ref{sec:flash_tpc_match} is designed to be more inclusive when matching clusters, with an expectation that later reexamination is necessary to improve the matching accuracy.

The majority of the LMM events contain only small clusters that give very low-intensity predicted light. The typical energies of these clusters are a few MeV, and it is challenging to correctly match such low energy dot-like activities to their predicted light patterns. The length of the cluster and the intensity of the predicted and measured light are used to tag and remove those low-energy events. 

To tag and remove the LMM events with higher energy, a Kolmogorov--Smirnov (KS) test is performed between the observed and the predicted light pattern without any normalization constraint. LMM events are usually caused by the inefficiency of the PMT system to detect cathode-side events, the light production outside of the TPC active volume, or the inaccuracy of the photon library for anode-side events.
If the KS test score is extremely low, the cluster is directly tagged as an LMM event and rejected. 
If the KS test score indicates a modest inconsistency, a further check is performed to see if the LMM candidate can match a different light flash from the cosmic discriminator, and if it is consistent with either a through-going muon or a stopped muon. This check relies on the precise knowledge of the effective boundary that is distorted by the space charge effect, and so the Z-dependent effective boundary as shown in Fig.~\ref{fig:SCB_YX_2d} is used. Firstly, any such candidate LMM cluster is paired with the other flashes in the PMT readout window. 
%Flashes that are nearby in time and position, and not much brighter than predicted, are considered for possible replacement matches to the candidate neutrino cluster.  
Under the new pair of flash-cluster hypotheses, the LMM cluster is placed at a different drift location given the new flash time. Several scenarios follow:
\begin{itemize}
	\item If a new flash is found to be more consistent with the cluster prediction, and the cluster has two end points on the effective detector boundary, this cluster is then re-tagged as a TGM and rejected.
	\item If a new flash is found to be more consistent with the cluster prediction, and the cluster has only the entering point on the effective detector boundary, this cluster is tagged as a possible STM and then vetted by the STM tagger (Sec.~\ref{sec:STM}) to confirm and reject.
	\item If no new flashes are more consistent with the cluster prediction, but the cluster is moved along the drift direction such that both end points exactly touch the effective boundary, this cluster is also re-tagged as a TGM and rejected. The associated flash is assumed to be lost in the light detection or flash reconstruction (Sec.~\ref{sec:light_reco}). The boundary contact tolerance is made more stringent in order for the TGM to be determined purely by the geometric information.
	%\item If none of the previous scenarios hold, it is likely that either the clustering or the light detection has a mistake, and this cluster is tagged as an LM event and rejected.   
\end{itemize}

The TGM, STM, and LMM background taggers are applied, in this order, after the charge-light matching step to remove most of the in-beam backgrounds. The remaining events are the neutrino candidates. In the next section, we evaluate the performance of this generic neutrino detection procedure.

\section{Performance of the generic neutrino detection}\label{sec:performance}
%Beside various background taggers described in Sec.~\ref{sec:rejection_in_beam}, another cut used to select neutrino interactions is the length of the {\it cluster} larger than 15~cm. To calculate the {\it cluster}length, two extreme points (Sec.~\ref{sec:track_seed_steiner} and e.g. Figure~\ref{fig:steiner_example5}) are firstly identified. The track trajectory fit is performed (discussed in Sec.~\ref{sec:track_fitting}). The {\it cluster} length is essentially the integrated length of the track trajectory. 
In this section, we evaluate the performance of the generic neutrino detection, i.e., cosmic-ray background rejection, in terms of the neutrino selection efficiency, purity, and cosmic-ray rejection power. 
A series of selections are applied in sequence to select neutrino interactions and reject cosmic-ray backgrounds from the original hardware triggers initiated by beam spills: 1) software trigger, 2) offline light filter, 3) charge-light matching, 4) TGM rejection, 5) STM rejection, and 6) LMM event rejection. The software trigger is a DAQ trigger serving as a first-stage data reduction that analyzes PMT light signals to record a readout if beam-coincidence {\it flash} PE and multiplicity conditions are met. The techniques implemented in the other selections were described in previous sections. 

Three samples are used for this evaluation:
\begin{itemize}
	\item {Beam-on data (\it BNB)}: a partial set of on-beam data recorded from February to April 2016, triggered by BNB neutrino spills (30k events after software trigger). 
	\item {Beam-off data (\it EXT)}: a partial set of off-beam data (pure cosmic background) taken during the periods when no beam was received, recorded from February to April 2016 (30k events after software trigger).
	\item {\it MC}:simulated neutrino interactions with BNB overlaid with an ``unbiased EXT" data sample. 
	This unbiased EXT data are taken without neutrino beam and triggered by a random external trigger. The simulated TPC and PMT waveform are then overlaid with the data waveform.  This eliminates one source of systematic uncertainty related to the simulation of the cosmic ray backgrounds. This sample is also called \emph{Overlay-MC}. 
	Each Overlay-MC event has one simulated neutrino interaction uniformly distributed in the liquid argon volume inside the TPC cryostat, of which 44\% is the TPC active volume (540k events in the entire TPC cryostat). A special MC ``dirt'' sample is also used where each event has one simulated neutrino interaction outside the cryostat liquid argon volume, as far as 20 meters into the experimental hall (90k events simulated).
	MicroBooNE cross section modeling tuned from GENIE~\cite{Genie2015} v3 is used in the simulation.
\end{itemize} 
While the calculation of efficiency relies on the MC sample, the calculation of purity requires
both MC signal and EXT background samples. The BNB data is compared with the prediction from 
MC and EXT to demonstrate the similarity between data and MC.

\begin{table*}[htbp]
	\begin{center}
		\caption{\label{tab:cosmicrejection}
			Summary of the cumulative neutrino selection efficiency, over all energies, for $\nu_{\mu}$ CC and $\nu_{\mu}$ NC events in the fiducial volume (94.2\% of the active volume), the
			cosmic-ray reduction factor, and the neutrino signal to the cosmic-ray background ratio for each cut. The relative cosmic-ray reduction to the previous cut is shown in the parentheses. The last column shows the generic neutrino signal to cosmic-ray background ratio. The errors are statistical only uncertainties. Neutrinos that originate outside the fiducial volume are not counted in this table. See Fig.~\ref{fig:bnb_nu_breakdown} for more details of the selected neutrino candidates.}
		\begin{tabular}{ccccc}
			\hline
			\hline
			Selection & $\nu_{\mu}$ CC efficiency & $\nu_{\mu}$ NC efficiency & Cosmic-ray reduction & $\nu$ : cosmic-ray \\\hline
			Hardware trigger     & 100\% & 100\% & 1 (1) & 1 : 20000\\
			%Software trigger     &  ~\ref{sec:intro}       & 1:800 & 0.04 (0.04)\\
			Light filter & (98.31$\pm$0.03)\%  & (85.4$\pm$0.1)\%  & (0.998$\pm$0.002)$\times$10$^{-2}$ (0.01) & 1 : 210\\
			Charge-light matching & (92.1$\pm$0.1)\%  & (53.6$\pm$0.2)\%  & (2.62$\pm$0.04)$\times$10$^{-4}$ (0.026) & 1 : 6.4 \\
			Through-going muon rejection & (88.9$\pm$0.1)\%  & (52.1$\pm$0.2)\%  & (4.4$\pm$0.2)$\times$10$^{-5}$ (0.17) & 1.1 : 1\\
			Stopped muon rejection & (82.9$\pm$0.1)\%  & (50.3$\pm$0.2)\%  & (1.4$\pm$0.1)$\times$10$^{-5}$ (0.32) & 2.8 : 1\\
			Light-mismatch rejection  & (80.4$\pm$0.1)\%  & (35.9$\pm$0.2)\%  & (6.9$\pm$0.6)$\times$10$^{-6}$ (0.50) & 5.2 : 1\\
			\hline
			\hline
		\end{tabular}
	\end{center}
\end{table*}

% cosmic rejection

The neutrino selection efficiency for $\nu_{\mu}$ charged-current (CC) and $\nu_{\mu}$ neutral-current (NC) events, the cosmic-ray reduction factor, and the ratio of the neutrino signal to the cosmic ray background for each cut are evaluated from the MC and EXT samples, and summarized in Table~\ref{tab:cosmicrejection}. Since the software trigger is functionally a subset of the offline light filter algorithms in this work, the two cuts are combined together in the table. The efficiency for each
channel is calculated for events generated in the fiducial volume only. The fiducial volume requirement itself has an efficiency of 94.2\%, and is counted separately. The overall selection efficiency of the neutrino interactions in the fiducial volume, integrated over the entire energy range of the BNB spectrum, is 80.4\% for $\nu_{\mu}$ CC interactions and 35.9\% for $\nu_{\mu}$ NC interactions. The cosmic-ray reduction factor is calculated by counting individual cosmic-ray tracks
in the TPC readout window before and after each cut. An overall cosmic-ray rejection power of $6.9\times10^{-6}$ is achieved, resulting in a neutrino signal to cosmic-ray background ratio of 5.2 to 1. %In the final selected events, there are an additional 10\% neutrino interactions which originate outside the fiducial volume and are not taken into account in this ratio.

\begin{figure}[htpb]
	\centering
	\includegraphics[width=\columnwidth]{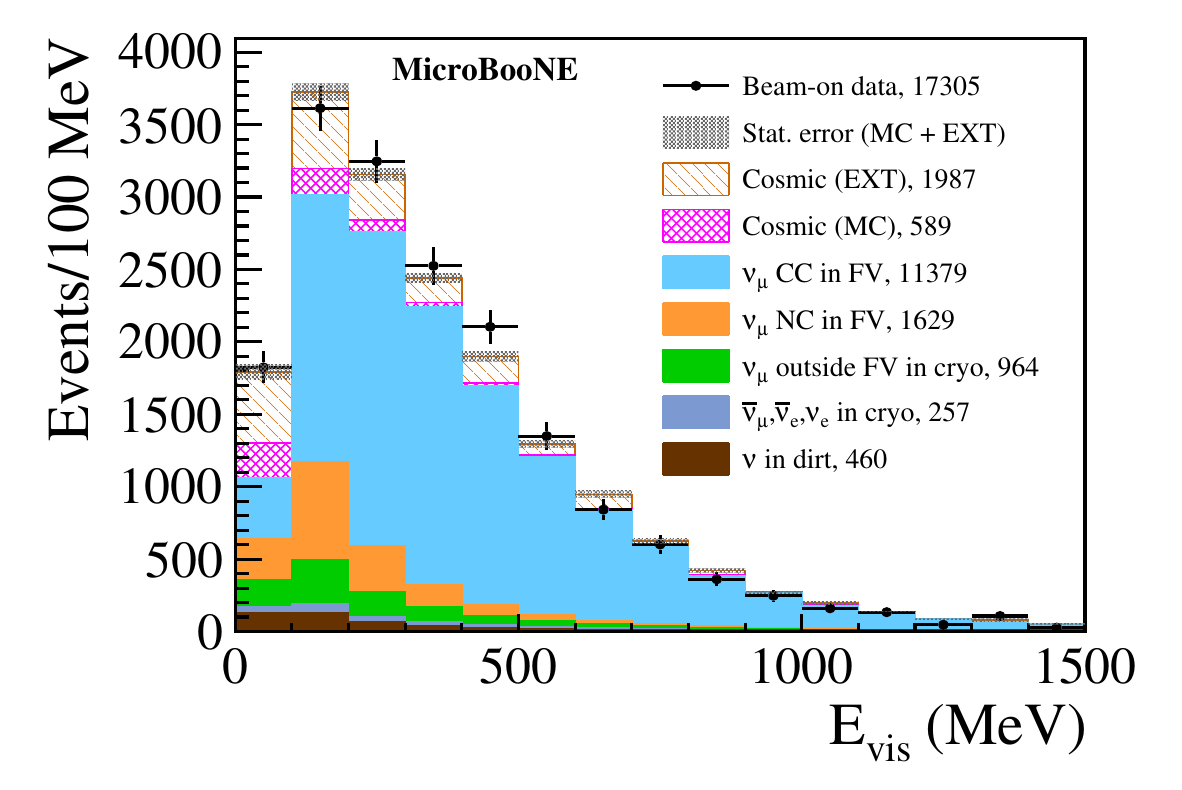} 
	\put(-192,147){(a)}\\
	\includegraphics[width=\columnwidth]{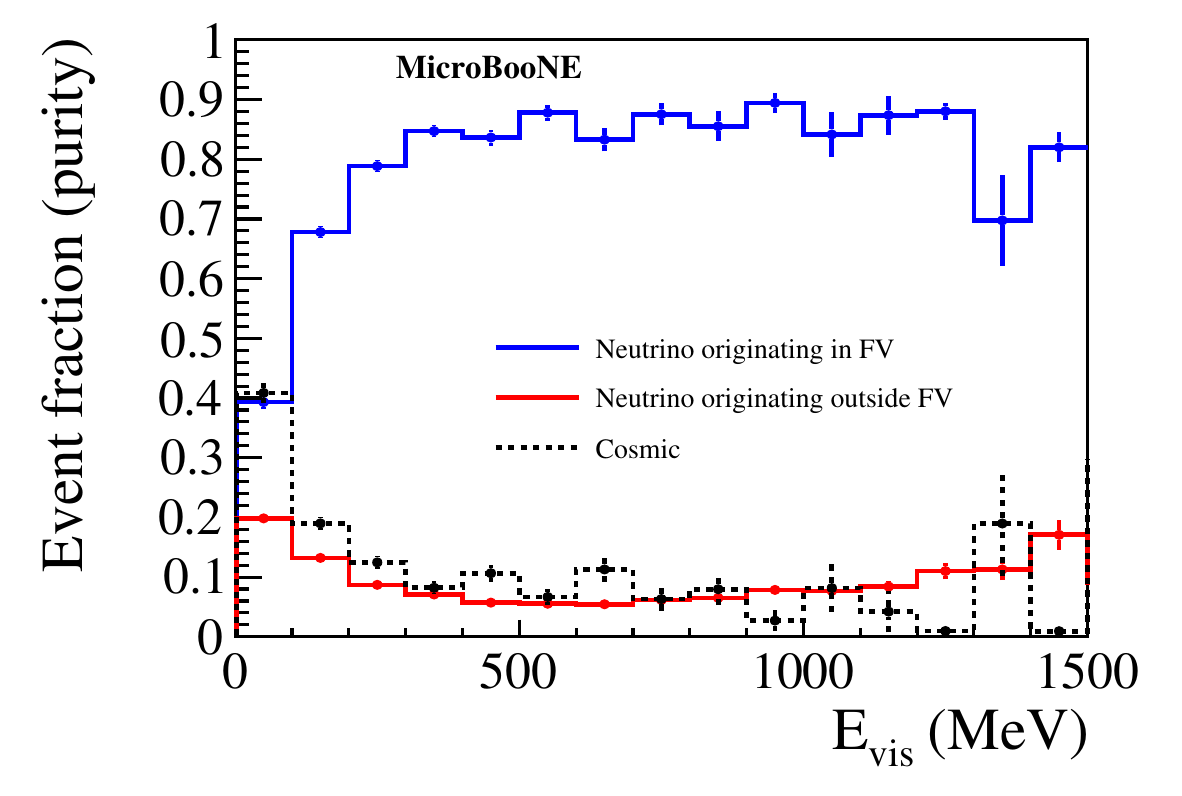} 
	\put(-192,147){(b)}\\
	\caption{\label{fig:bnb_nu_breakdown}(a) The stacked MC and EXT events are compared with the final selected events from the BNB sample. Selected events are further categorized based on MC truth. All numbers are scaled to $5 \times 10^{19}$ POT. (b) Event fraction for neutrino signal and cosmic-ray background events in the selected neutrino candidates. The dip (jump) in ``purity'' (``impurity'') curve around 1400 MeV is caused by the statistical fluctuation of cosmic-ray background events in that bin.}
\end{figure}

\begin{figure}[!htpb]
	\centering
	\includegraphics[width=\columnwidth]{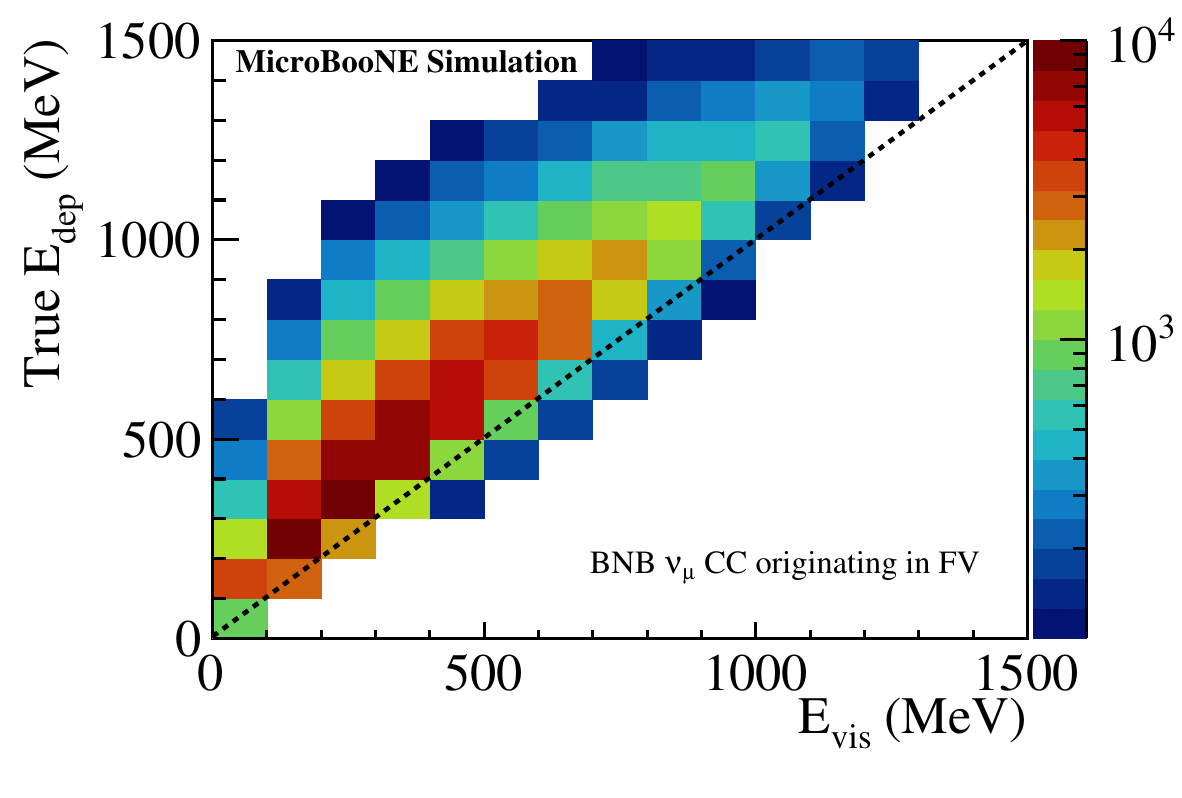} 
	\put(-197,149){(a)}\\
	\includegraphics[width=\columnwidth]{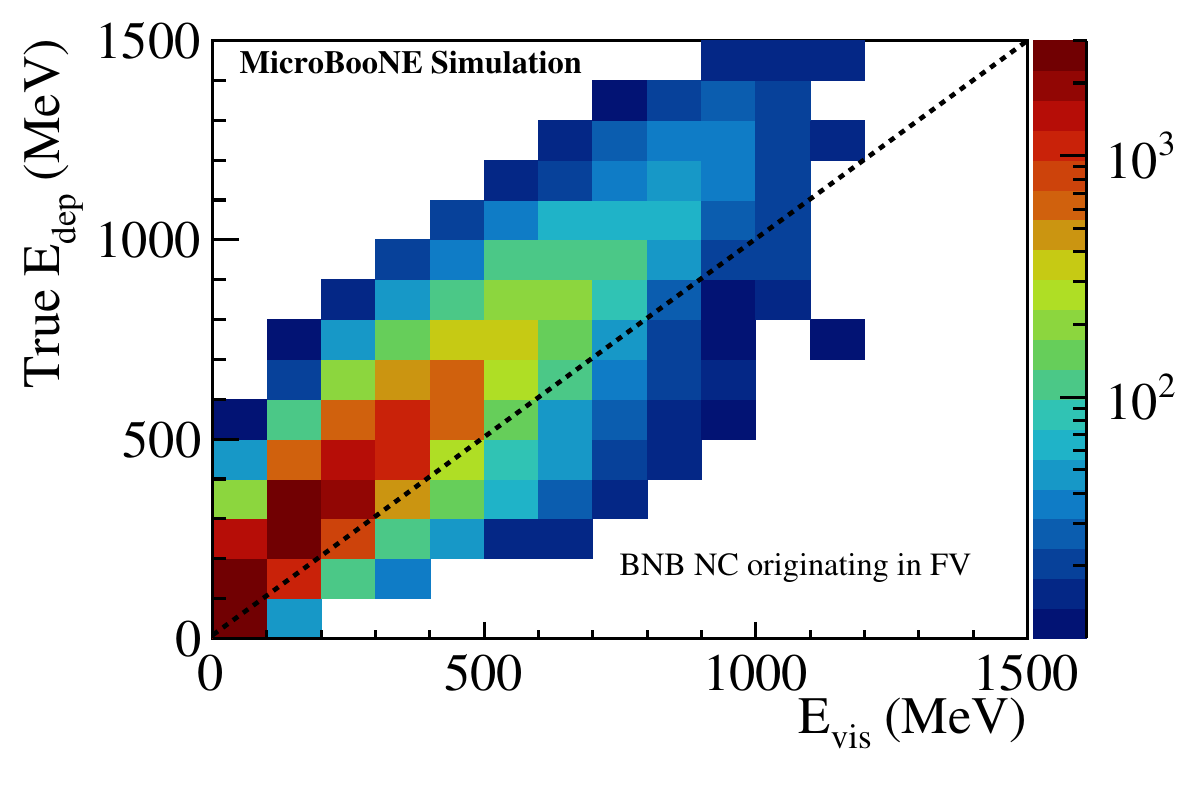} 
	\put(-197,149){(b)}\\
	\caption{\label{fig:bnb_nu_EdepEvis} Comparison of the true deposited energy ($E_{\textrm{dep}}$) and the reconstructed visible energy ($E_{\textrm{vis}}$) for (a) BNB $\nu_{\mu}$ CC interactions and (b) BNB NC interactions.  The deviation from the diagonal line shows the difference between the expected and observed charge scaling factor in calculating $E_{\textrm{vis}}$.
	}
\end{figure}

The final selected events showing all event categories are shown in Fig.~\ref{fig:bnb_nu_breakdown}(a). The selected MC and EXT events are stacked to compare with the BNB events as a function of the visible energy~\cite{Adams:2019ssg}, $E_{\mathrm{vis}}$, which is calculated from the total charge measured by the collection wire
plane with a universal scaling factor of $\sim4.3\times10^{-5}$ MeV/electron. 
This scaling factor takes into account the average expectation of the recombination and attenuation of the ionization electrons. While the reconstruction of visible energy is 
simple and sufficient to evaluate the performance of cosmic-ray background rejection, 
its performance (see. Fig.~\ref{fig:bnb_nu_EdepEvis}) is not sufficient to
reconstruct the neutrino energy. An improved reconstruction of neutrino energy will be used in future study.
All reported numbers are scaled to an integrated neutrino
beam intensity of $5 \times 10^{19}$ POT. The cosmic-ray background is estimated from the EXT beam-off data sample. An additional beam-on cosmic-ray background is estimated from the MC sample, which corresponds to a cosmic-ray cluster that is incorrectly matched to the neutrino-induced flash but passes the LMM cut. Neutrino events are categorized based on their interaction type: CC or NC, and their location: inside fiducial volume (FV), inside the liquid argon volume (cryo),
or outside the liquid argon volume (dirt). The error bars for the BNB and MC samples are statistical only. The event fraction, i.e.~purity, of the selected events is shown in Fig.~\ref{fig:bnb_nu_breakdown}(b) as a function of the visible energy. About 10\% of the selected events are from neutrinos originating outside the fiducial volume. They are not counted in the efficiency calculation in Table~\ref{tab:cosmicrejection}. For generic neutrino detection, if only cosmic-ray backgrounds are considered as impurity, an overall 85.1\% purity is achieved. The purity increases to 90.3\% for events with more than 200~MeV of visible energy.
%For fiducial neutrino detection, the overall purity is about 76\%, and it is more than 80\% for visible energy greater than 200~MeV. 

\begin{figure}[!htpb]
	\centering
	\includegraphics[width=\columnwidth]{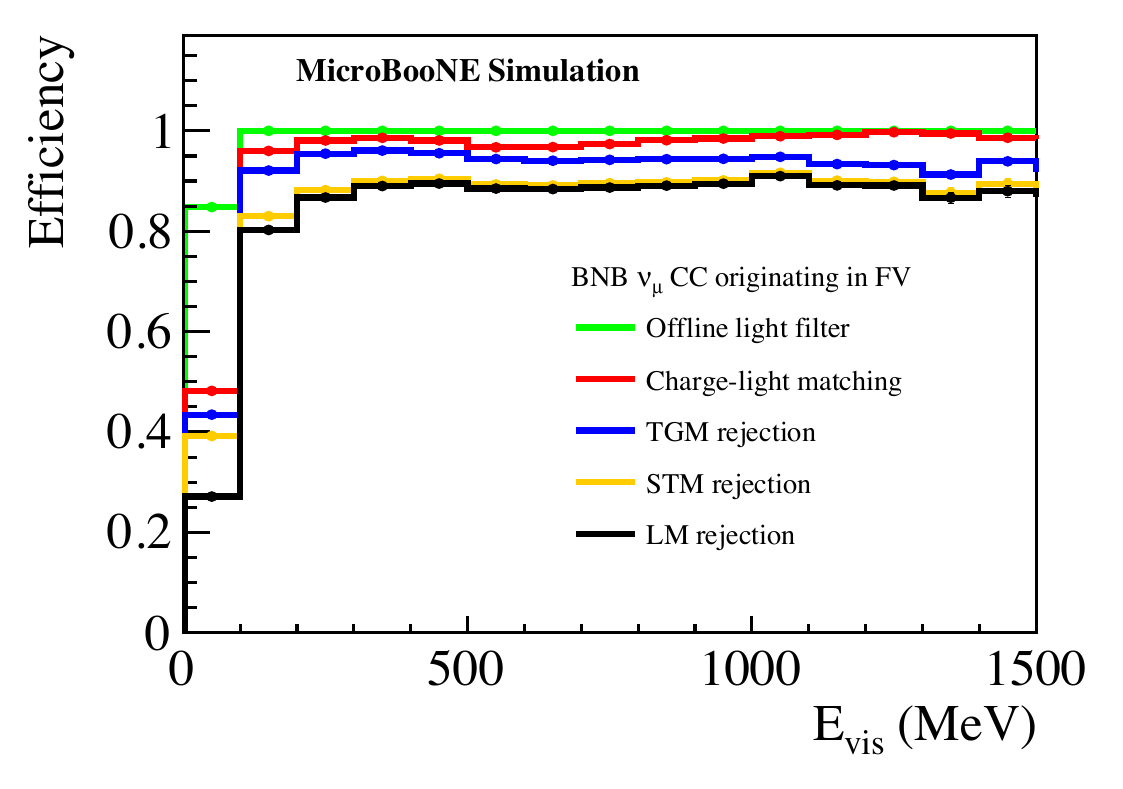} 
	\put(-197,154){(a)}\\
	\includegraphics[width=\columnwidth]{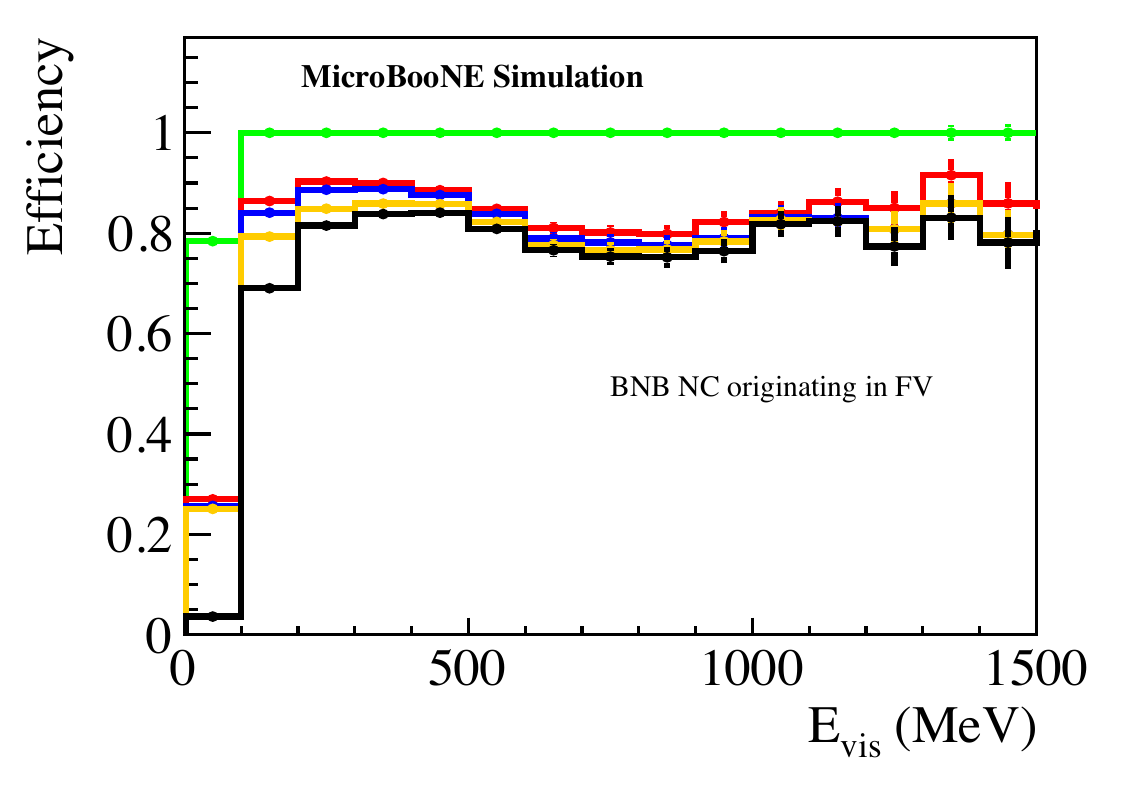}
	\put(-197,154){(b)}\\
	\caption{\label{fig:bnb_nu_efficiency} Progressive efficiencies of the neutrino interactions originating in the fiducial volume as a function of the visible energy. (a) Efficiency for BNB $\nu_{\mu}$ CC interactions. (b) Efficiency for BNB $\nu_{\mu}$ NC interactions. The efficiency drop in the $500\textrm{--}1000$ MeV region corresponds to improper charge-light matching for some low deposited energy($E_{\rm dep}$) NC events.
	}
\end{figure}

Figure~\ref{fig:bnb_nu_efficiency} shows the cumulative selection efficiencies after each cut as a function of the visible energy. The efficiency calculation is performed for $\nu_\mu$ CC and $\nu_\mu$ NC interactions in the fiducial volume separately. The overall efficiency for $\nu_\mu$ CC events is 80.4\%, where 88.4\% is achieved for visible energy greater than 200~MeV. The overall efficiency for $\nu_\mu$ NC events is 35.9\% because their typically low-energy depositions enhance the
contributions from the low-energy bins ($<$100 MeV). A comparison between the visible energy ($E_{\rm vis}$) and the deposited energy ($E_{\rm dep}$) using the MC sample is shown in Fig.~\ref{fig:bnb_nu_EdepEvis}. The efficiency drop in the $E_{\rm vis}$ region of 500-1000 MeV for NC events, as shown in Fig.~\ref{fig:bnb_nu_efficiency}(b), corresponds to an inefficient separation of cosmic activity and NC interaction final-state particles in the charge-light matching. Though the
selection of low $E_{\rm dep}$ NC events is most likely to fail in the matching stage, some of them could be clustered with cosmic activity and collectively matched to the in-beam PMT signals. This results in a much greater value of $E_{\rm vis}$ than $E_{\rm dep}$ as shown in Fig.~\ref{fig:bnb_nu_EdepEvis}(b). These events are the origin of the efficiency drop as mentioned above. Further improvements to the removal of residual cosmic activity and visible energy calibration are expected in the downstream pattern recognition and neutrino energy reconstruction.
%comprehensively taking into account the detector response, thresholding effect in the digital processing, particle interactions, and so on. 

%Figure~\ref{fig:bnb_nue_breakdown} show the selected events broken into different categories given the MC truth information in the overlay samples. Beside the pie chart showing the composition of the total selected events, the stacked plots of the EXT and overlay events are compared with the BNB events as a function of the calibrated visible energy~\cite{Adams:2019ssg}, which is deduced from the total charge from the collection wire plane. The final number of events has been scaled to an integrated neutrino beam intensity of 5e19 proton on target (POT). Cosmic events from beam-off data is estimated from the EXT sample, and cosmic events from $\nu$ in TPC cryo corresponds to the incorrect match in the overlay sample which has a beam neutrino interaction simulated in each event. Both CC and NC neutrino interaction events whose vertices are inside the fiducial volume (FV) are labeled. The external $\nu_\mu$ labels neutrino interaction events whose vertices are outside the fiducial volume but still inside the cryostat. Again, the simulation and the data shows a good agreement.

The generic neutrino detection procedures described in this article mark the beginning of a high-performance selection of individual neutrino interaction channels, which requires additional particle-level pattern recognition and reconstruction techniques. Several algorithms have been developed in MicroBooNE and applied in previous publications, such as Pandora~\cite{Acciarri:2017hat}, Deep Learning~\cite{Acciarri:2016ryt,Adams:2018bvi}, Multiple Coulomb Scattering~\cite{Abratenko:2017nki}, and
electromagnetic shower reconstruction~\cite{Adams:2019law}. Additional pattern recognition tools are in development, including those within the Wire-Cell reconstruction. Nevertheless, it is interesting to compare the performance of the cosmic rejection and generic neutrino selection in this work with those from previous results in Refs.~\cite{Adams:2018fud, Adams:2018sgn, Adams:2018lzd, Adams:2019iqc}. The cosmic rejection in this work is enhanced by a factor of 8 compared to the cosmic rejection
power (without kinematical requirements) published in Ref.~\cite{Adams:2018lzd}, which 
aims at selecting exclusive charged current quasi elastic neutrino interaction. In a separate comparison, the number of selected inclusive $\nu_{\mu}$ CC events for $5 \times 10^{19}$ POT is about 4300 in Ref.~\cite{Adams:2019iqc}, with an overall cosmic contamination of 35.5\% in the final selection. In comparison, the number of selected $\nu_{\mu}$ CC events is expected to be about 11300 with this generic neutrino detection procedure, with an overall cosmic contamination of 14.9\% in all neutrino candidates. The increase in number of
events comes from both the enhancement (a factor of 1.41) in the selection efficiency and the enlargement (a factor of 1.86) of the fiducial volume in this analysis.

%1) 5\%? numuCC efficiency and 21.2\% cosmic contamination for $\nu_\mu$-Ar multiplicity measurement in Ref.~\cite{Adams:2018fud}; 
%2) [irrelevant] 16\% efficiency and 56\% purity in the cross section measurement of $\nu_\mu$-Ar CC $\pi^\circ$ production~\cite{Adams:2018sgn}; 
%3) 26.9\% efficiency and 78.4\% purity, cosmic rejection without kinematical requirement is smaller by a factor of 8.2 for $\nu_\mu$-Ar CC $1p0\pi$ channel~\cite{Adams:2018lzd}; 
%4) 57\% efficiency and 50\% purity for cross section measurement of inclusive CC $\nu_\mu$ interactions~\cite{Adams:2019iqc}. 4.3k selected for 5e19 POT with 35.5\% cosmic-ray background.

\begin{figure}[!htpb]
	\centering
	\includegraphics[width=\columnwidth]{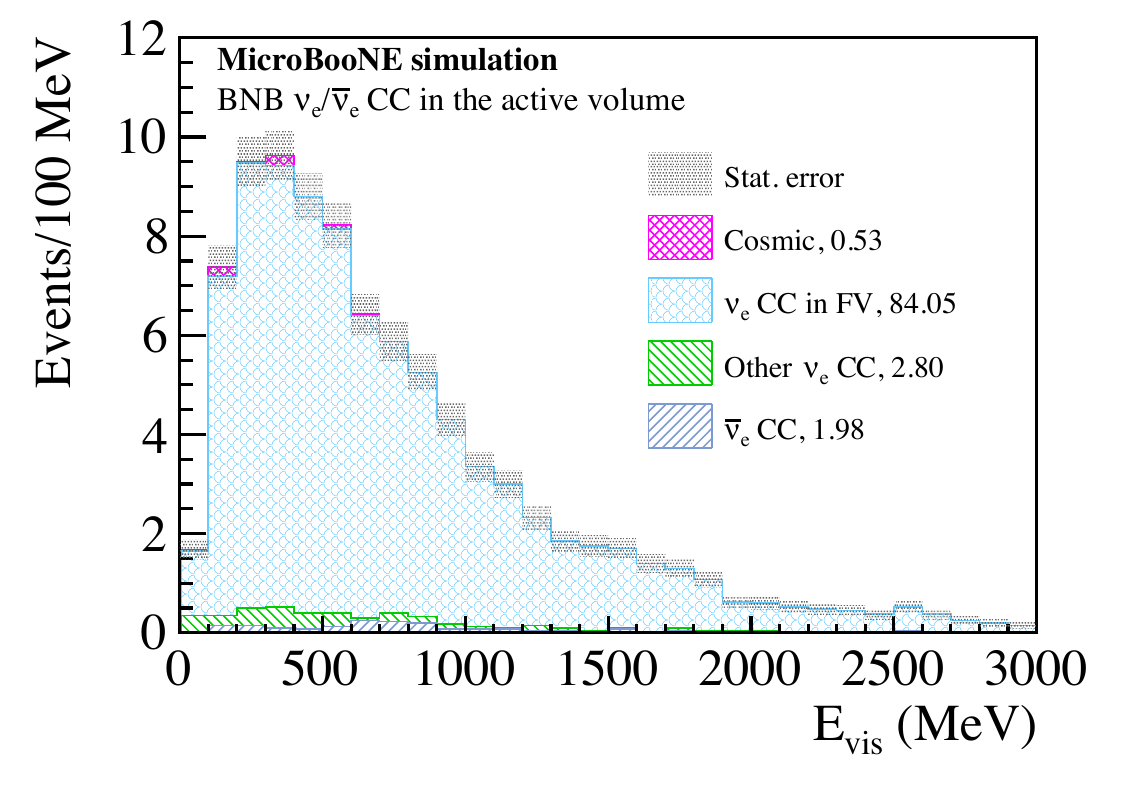} 
	\put(-197,40){(a)}\\
	\includegraphics[width=\columnwidth]{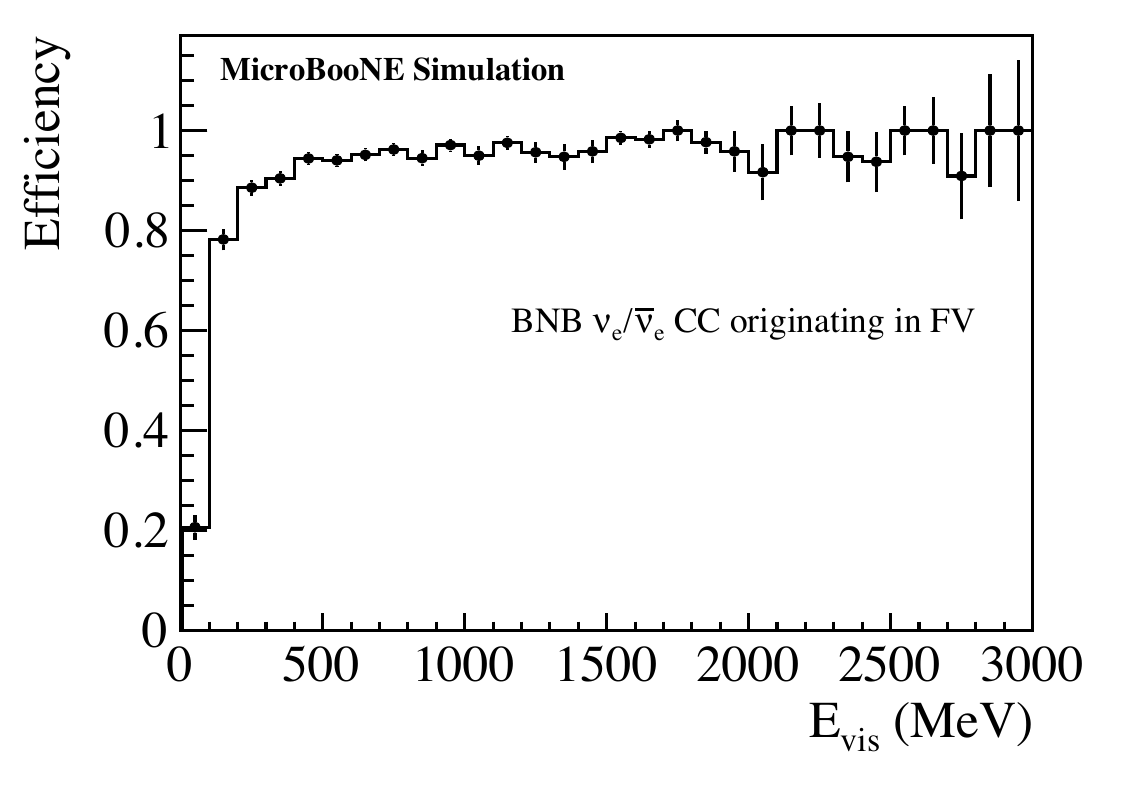} 
	\put(-197,40){(b)}\\	
	\caption{\label{fig:bnb_nue_efficiency}
		(a) Composition of the selected events from the simulated $\nu_e$ CC sample after scaling to $5 \times 10^{19}$ POT. Only $\nu_e$ CC interactions are simulated in 
		the TPC active volume. No other neutrino interaction backgrounds are 
		included. (b) Efficiency for $\nu_e$ CC interactions originating in the fiducial 
		volume as a function of the reconstructed visible energy. The low efficiency
		at the small $E_{vis}$ is the result of signal efficiency of the light 
		system. 
	}
\end{figure}

Finally, to evaluate the performance of the generic neutrino selection as a
pre-selection for the $\nu_e$ CC selection, a special simulated sample with only $\nu_e$ CC interactions from the BNB intrinsic $\nu_e$ flux, overlaid with EXT data, is used to estimate $\nu_e$ CC efficiency under this selection procedure. The expected number of $\nu_e$ CC events in the TPC active (fiducial) volume for $5 \times 10^{19}$ POT is 100 (95) in total. Figure~\ref{fig:bnb_nue_efficiency} shows the composition of the selected events and the selection efficiency as a function of the visible energy. The overall $\nu_e$ CC event selection efficiency is 87.6\%. This high efficiency for $\nu_e$ CC events is particularly important for future MicroBooNE analyses investigating the nature of the low-energy excess of $\nu_e$-like events observed in the MiniBooNE experiment~\cite{Aguilar-Arevalo:2012fmn}. The remaining challenge of improving the $\nu_e$ CC selection purity is an active research area. Recent progress built upon this work will be reported in future study.

\section{Summary and outlook}\label{sec:summary}
This article describes various new techniques developed in the Wire-Cell event reconstruction paradigm to achieve a high-performance generic neutrino detection in the MicroBooNE detector. In particular, about 99.98\% of the cosmic-ray backgrounds are rejected after software triggering, leading to a cosmic-ray impurity of 9.7\% (14.9\%) for reconstructed visible energy, E$_{\rm vis}$, greater than 200 (0) MeV. Compared to the result in Ref.~\cite{Adams:2019iqc}, the cosmic contamination is reduced by a factor of 2.4, while attaining a higher neutrino detection efficiency, e.g. for inclusive $\nu_\mu$ CC in the active volume, by a factor of 2.7 in this work.

This is the first analysis to achieve cosmic ray background rejection in excess of 99\% in a near-surface LArTPC, while keeping the neutrino detection efficiency high. The improved performance presented in this article provides a solid foundation for upcoming physics analyses in MicroBooNE and marks a major milestone in demonstrating the full capability of LArTPCs in neutrino physics.
Further development of particle-level pattern recognition and reconstruction techniques toward selections of individual neutrino interaction channels are in progress using Wire-Cell, and will be reported in future publications. Looking forward, the new analysis techniques summarized in this work utilizing the Wire-Cell reconstruction algorithms such as 3D image reconstruction, many-to-many charge-light matching, and track trajectory and $dQ/dx$ fitting can be naturally adopted into and expected to have a significant performance impact on the upcoming SBN~\cite{Antonello:2015lea} and DUNE~\cite{dune-tdr-1} experiments.

\begin{acknowledgments}
This document was prepared by the MicroBooNE collaboration using the
resources of the Fermi National Accelerator Laboratory (Fermilab), a
U.S. Department of Energy, Office of Science, HEP User Facility.
Fermilab is managed by Fermi Research Alliance, LLC (FRA), acting
under Contract No. DE-AC02-07CH11359.  MicroBooNE is supported by the
following: the U.S. Department of Energy, Office of Science, Offices
of High Energy Physics and Nuclear Physics; the U.S. National Science
Foundation; the Swiss National Science Foundation; the Science and
Technology Facilities Council (STFC), part of the United Kingdom 
Research and Innovation; and The Royal Society (United Kingdom).  
Additional support for the laser
calibration system and cosmic ray tagger was provided by the Albert
Einstein Center for Fundamental Physics, Bern, Switzerland.
\end{acknowledgments}

\appendix
\section{TRAJECTORY SEED FINDING}~\label{sec:trajectory_appendix}
\renewcommand{\theequation}{\thesection\arabic{equation}}
%\subsection*{Preparatory Work}~\label{sec:prep}
As described in Sec.~\ref{sec:track}, both the 2D images from wire plane measurements after signal processing (Sec.~\ref{sec:tpc_sp}) and the Wire-Cell 3D imaging results (Sec.~\ref{sec:wire-cell-imaging}) are used in the track trajectory and \dqdx~fitting. Graph theory plays an important role in constructing an initial seed for the 3D trajectory, which is essential for associating the nearby 2D pixels for the trajectory fit. The quality of the initial seed finding impacts the quality of the final fit. 
\begin{figure}[t!hb]
	\centering
	\includegraphics[width=0.292\figwidth]{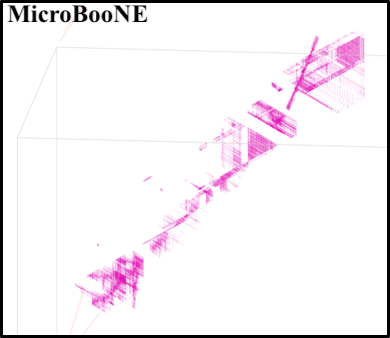}
	\includegraphics[width=0.19\figwidth]{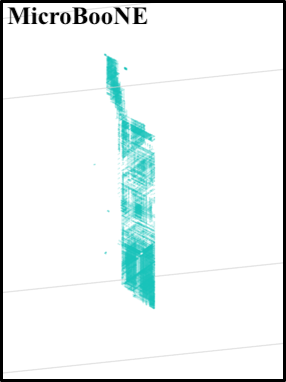}
	\put(-230,6){(a)}
	\put(-87,6){(b)}	
	\caption{(a) An example of an isochronous track with gaps. Some of the gaps are results of nonfunctional channels, while others are results of the coherent noise removal during the excess noise filtering step~\cite{Acciarri:2017sde}. (b) An example of an isochronous track that is also compact with respect to the vertical W collection wires.}
	\label{fig:seed_problem}
\end{figure}
Figure~\ref{fig:seed_problem} shows two examples illustrating the importance and challenges of constructing the 3D trajectory seed. When a track is traveling close to parallel to the wire planes (also referred to as an isochronous topology), the associations among wires from different planes at the same time slice are no longer obvious. This ambiguity typically leads to mistakes in forming associations, which further propagate to the trajectory and \dqdx~fits. In addition, gaps can occur in the reconstructed 3D images (Fig.~\ref{fig:seed_problem}(a)), which have two typical origins. First, gaps are the results of the $\approx$10\% nonfunctional channels~\cite{Acciarri:2017sde}, which are spread across different views. Second, gaps are created when small coherent signals (e.g.,~when a track travels parallel to the wire plane) are accidentally removed by the coherent noise removal~\cite{Acciarri:2017sde} or by the TPC signal processing procedure~\cite{Adams:2018dra,Adams:2018gbi}. For induction wire planes, the current TPC signal processing procedure shows inefficiency for extended signals in time, i.e. tracks parallel to the drift direction, which is referred to as the {\it prolonged} track topology. For this topology, the raw signal is typically small, leading to difficulties in constructing the signal region of interest. The algorithm that constructs the track seeds takes into account these imperfections by using advanced graph theory operations. The central idea is to find the shortest path between points of interest on a Steiner-tree-inspired graph, the construction of which takes into account the additional charge information. In the following, we describe the details of the related algorithms.  

\begin{figure}[t!hb]
	\centering
	\includegraphics[width=0.26\figwidth]{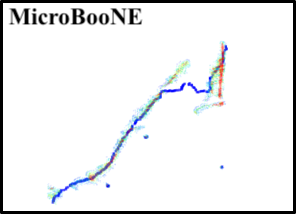}
	\includegraphics[width=0.22\figwidth]{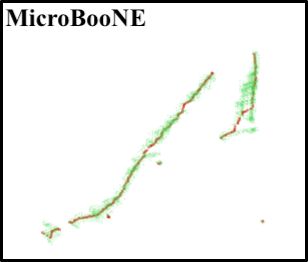}
	\put(-230,6){(a)}
	\put(-100,6){(b)}
	\caption{(a) Track seed before applying the overclustering protection. As a result of the overclustering, track seed (blue points) is wrongfully constructed going through two separated clusters. Red and blue colors represent high and low charge, respectively. (b) Track seed after applying the overclustering protection. Two clusters (red points) are now properly separated. Space points are displayed as green points without charge information. }
	\label{fig:overclustering}
\end{figure}

%Then the TPC-charge/PMT-light matched bundles described in Sec.~\ref{sec:flash_tpc_match} are examined further in order to avoid overclustering, which means separated {\it clusters} are merged into a single {\it cluster}. This typically is the result of incorrect gap mitigation. An algorithm is applied to the {\it blobs} from the matched {\it clusters}, by establishing edges between clearly connected {\it blobs} and setting a new {\it cluster} having the most overlap with the original main {\it cluster} using the edge information. 

%\subsubsection{Examination of Matched Bundles to Protect Against Overclustering}~\label{sec:prep_overclustering}
{\it Overclustering protection:} The result of the charge-light matching described in Sec.~\ref{sec:flash_tpc_match} is a matched bundle which consists of 1) a PMT flash, 2) the main TPC cluster, and 3) secondary clusters. As described in Sec.~\ref{sec:clustering}, the separation of different clusters is largely based on connectivity and proximity, with dedicated algorithms to mitigate gaps. The main cluster is defined to be the cluster which provides the largest contribution to the observed PMT flash. Therefore the track trajectory and \dqdx~fits are performed only on the main cluster. 

Since the clustering algorithm mainly focuses on the separation of different interactions, there are a small number of events in which the main cluster has been overclustered, that is, the grouping of separated clusters into a single cluster. One common cause of the overclustering is from incorrect gap mitigation. A re-examination of the matched bundle is performed to protect against overclustering. First, all blobs from the matched clusters are collected as the input to a new clustering algorithm. Each blob is treated as a vertex in a graph. Second, edges are established between blobs that are evidently connected. For example, if two blobs in adjacent time slices are overlapping in the transverse direction (parallel to the wire planes), they are defined to be connected. Third, additional edges are established to mitigate gaps. This algorithm improves upon the original clustering algorithm (Sec.~\ref{sec:clustering}) with slightly different criteria.  Finally, the new cluster with the most overlap with the original main cluster is set as the new main cluster for the matched bundle. Figure~\ref{fig:overclustering} shows the improvements after implementing the overclustering protection algorithm. 

\begin{figure*}[thb]
	\centering
	\includegraphics[width=\figwidth]{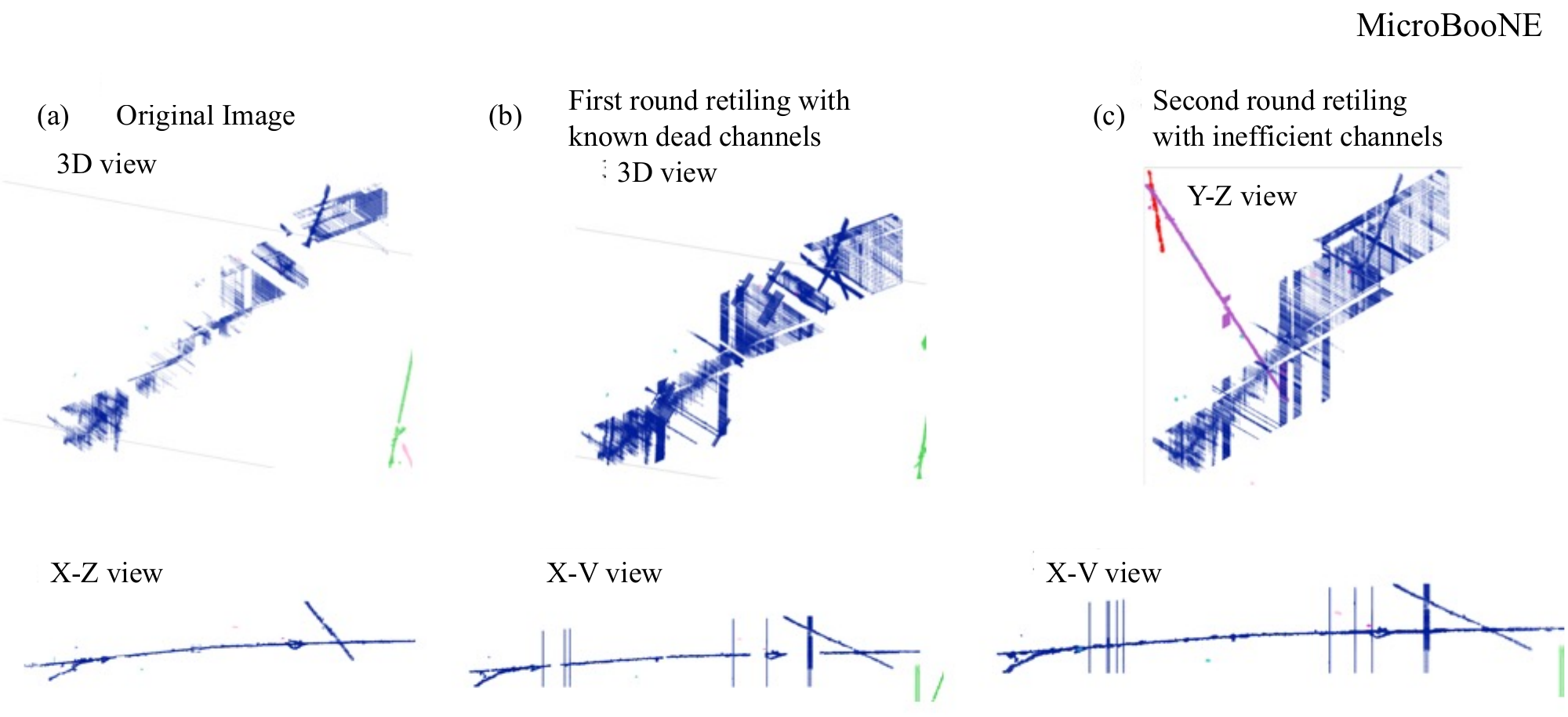}
	\caption{Illustration of the retiling step. (a) Original images are shown for 3D (top) and X-Z projection (bottom) views. (b) The same images for 3D (top) and X-V projection (bottom) after the first round of retiling when the known nonfunctional channels are filled. (c) Same images for Y-Z projection (top) and X-V projection (bottom) after the second round of retiling, which fills the inefficient channels.}
	\label{fig:retiling}
\end{figure*}

{\it Overcoming gaps:} As described previously, despite the LArTPC being a fully active detector, gaps can occur in a charged particle track due to various hardware or software issues. The gaps present a serious challenge to the trajectory fitting, and a retiling algorithm is introduced to overcome this problem. Figure~\ref{fig:retiling} shows the performance of two rounds of the retiling step. During the first round, the known nonfunctional channels are assumed to be live during the tiling step of the 3D image reconstruction (Sec.~\ref{sec:wire-cell-imaging}). Since the 3D image reconstruction is limited within the current cluster instead of the entire event, this procedure does not create many spurious blobs. The middle figure of Fig.~\ref{fig:retiling} shows the reconstructed 3D image after filling the known nonfunctional channels. Improvement in terms of removing gaps is obvious. The remaining gaps are the result of the inefficient channels, where signals are lost as the result of either the coherent noise removal or signal processing. A second round of retiling is performed to deal with inefficient channels. First, the highest and lowest space points in the vertical direction are found in the current cluster. On the associated graph, a Dijkstra's shortest path algorithm~\cite{graph_dijkstra_shortest_path} is used to find the shortest path between these two 3D points. The path can go through gaps in the image, although it may not be located at the correct place on the 3D image. Despite this issue, the shortest path is projected to each of the three 2D time-versus-wire views. 2D pixels close to the projected path are treated to be independent of their original states. This effectively fills the inefficient channels and a new round of tiling is performed. The right column of Fig.~\ref{fig:retiling} shows the reconstructed 3D image after the second round of retiling. Gaps from the inefficient channels are successfully filled.

%\subsection*{Construction of the 3D Point Graph}~\label{sec:graph}
\begin{figure}[t!hb]
	\centering
	\includegraphics[width=0.305\figwidth]{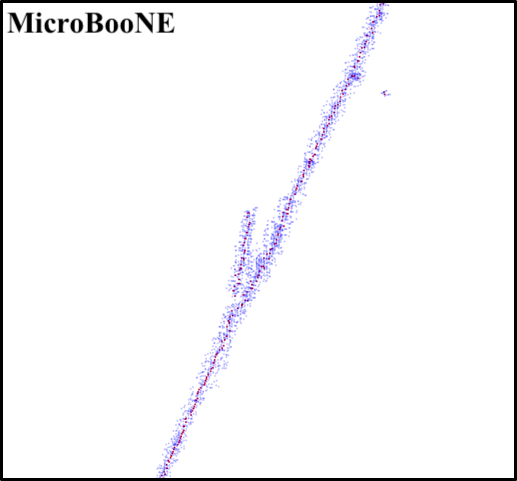}
	\includegraphics[width=0.19\figwidth]{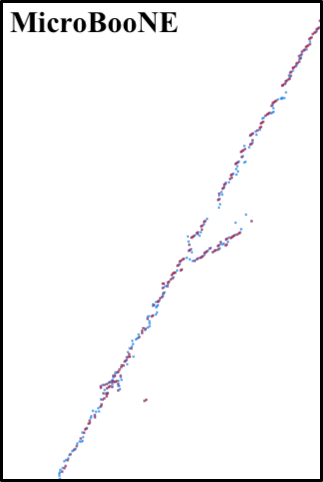}
	\put(-240,6){(a)}
	\put(-87,6){(b)}
	\caption{In both views Steiner terminals are shown in red.  All 3D space points are shown in blue in (a), while only those selected by the Steiner tree are shown in (b). }
	\label{fig:steiner_illu}
\end{figure}

{\it Steiner-tree-inspired graph construction:} Naively, the 3D trajectory seed is obtained by finding the shortest path on the constructed graph after two rounds of retiling in the previous step. However, the resulting seed sometimes significantly deviates from the true trajectory. The situation is improved by implementing a Steiner-tree-inspired graph, which forces the seed to go through important points in the graph. In mathematics, given an undirected graph with non-negative edge weights and a subset of selected vertices 
(terminals), the Steiner-tree problem is to find the tree with minimum total weights (the minimal spanning tree) that contains all selected terminals. Mapping to our problem, the Steiner-tree terminals are selected to be the 3D points (also vertices in the graph) that are associated with a large charges in the three views. The Steiner tree would then be guaranteed to go through these high-charge points, which are more likely to be close to the true track trajectory. Figure~\ref{fig:steiner_illu}(a) shows that the selected Steiner terminals are along the middle of the available 3D space points. Figure~\ref{fig:steiner_illu}(b) shows both the selected terminals and non-terminals in a Steiner tree.

\begin{figure}[t!hb]
	\centering
	\includegraphics[width=0.4\figwidth]{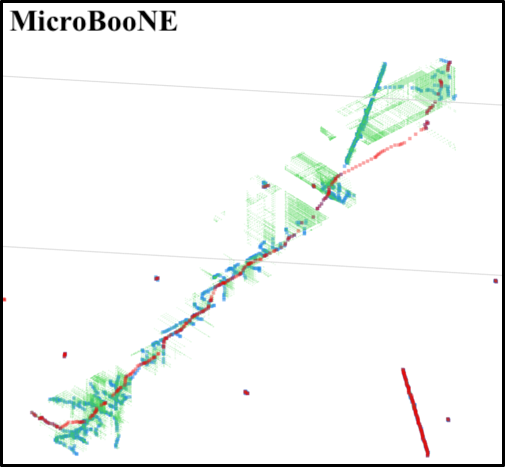}
	\caption{An example of the shortest path on the Steiner-tree-inspired graph (red points). 
		Green points represent original space points. Blue points are the 
		selected Steiner terminals. }
	\label{fig:steiner_example1}
\end{figure}

% actual implementation ...
Mathematically, the Steiner-tree problem is an NP-complete (non-deterministic polynomial-time complete) problem. Therefore, the actual implementation is through an approximated solution, because of the cost of the computation. The Steiner tree greedy algorithm in the practical approximation algorithm~\cite{Steiner} is used. In this algorithm, the Voronoi regions around the selected terminals are constructed (In mathematics, a Voronoi diagram is a partition of a plane into regions close to each of a given set of objects.) The shortest path between any two adjacent terminals with their Voronoi regions connected is constructed. The Steiner tree then becomes the minimal spanning tree of the newly constructed graph, which we call the Steiner-tree-inspired graph. Figure~\ref{fig:steiner_example1} shows an example of the shortest path on a Steiner-tree-inspired graph, which is used in this work as the initial trajectory seed. In the following, we describe in more detail how the Steiner-tree-inspired graph is constructed.

% path creation etc ...
A first-stage graph for a cluster is constructed as follows: First, the  cluster goes through two rounds of retiling to mitigate gaps. Second, each of the three 2D projection views of a blob 
is examined. The two views with the longest and shortest channel extensions are found. 3D points are created at the wire crossings from these two views with certain spacing, which reduces the amount of computer memory usage. Additional space points where the charge on wires is larger than a certain threshold (4000 electrons as a default) are added. The creation of these 3D points also considers the third view, which guarantees that all points with high charges are properly included. Third, a graph is created with these 3D points as vertices. Edges between vertices in the same blob are established when their distance is smaller than a predefined value. The weight of the edge is assigned as the distance between the two points. Edges connecting points from different blobs within two adjacent times are created under the same predefined distance threshold. Finally, the connected components algorithm~\cite{graph_connected_component} is used to find the disconnected subgraphs. Additional edges are established between these disconnected subgraphs according to the distance and directional information. First, between any two subgraphs, the closest pair of points is found. The direction is then calculated by performing a Hough transformation inside a subgraph with the selected point as the origin. If the directions of both subgraphs are aligned, an edge is created.

\begin{figure}[t!hb]
	\centering
	\includegraphics[width=0.29\figwidth]{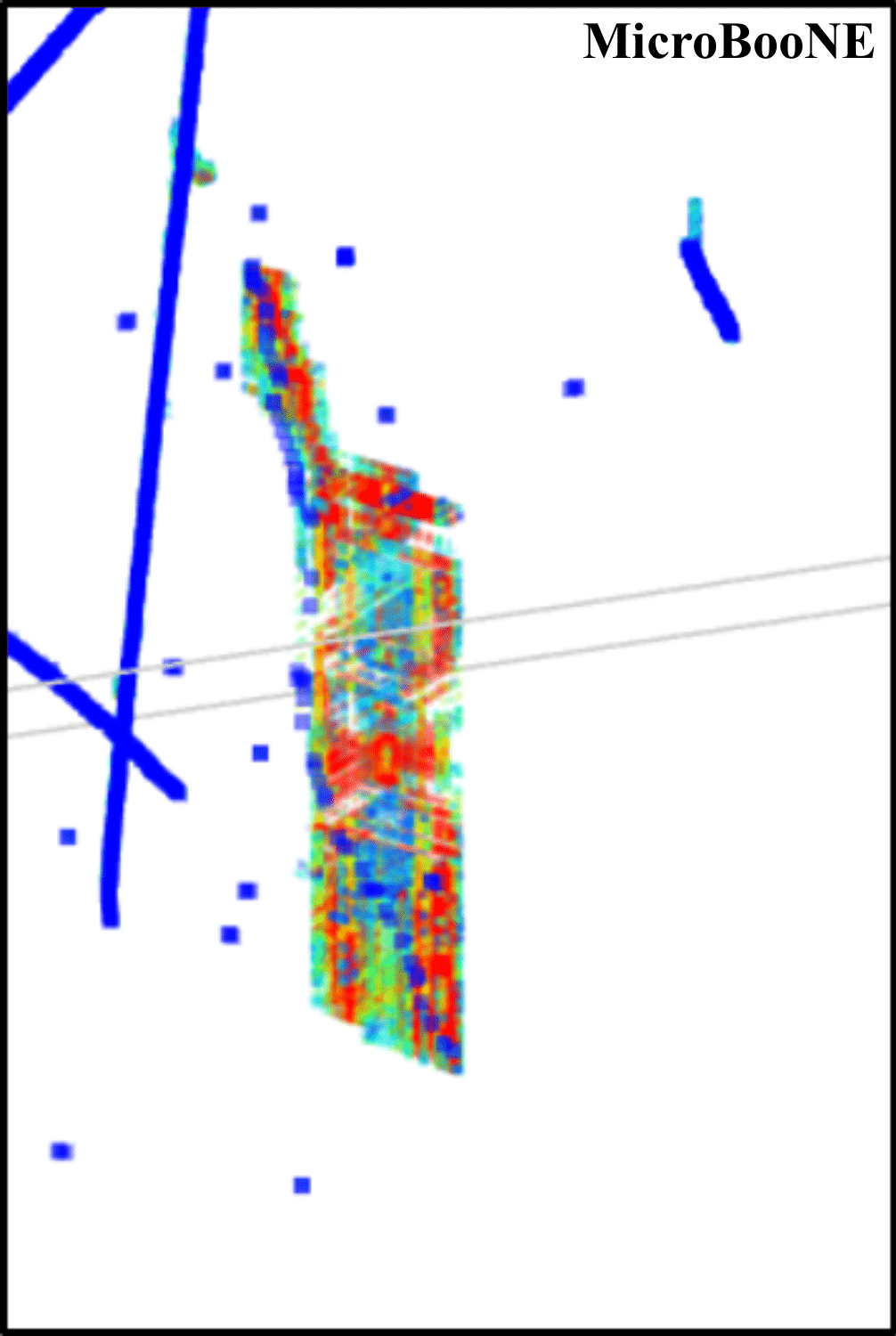}
	\includegraphics[width=0.2\figwidth]{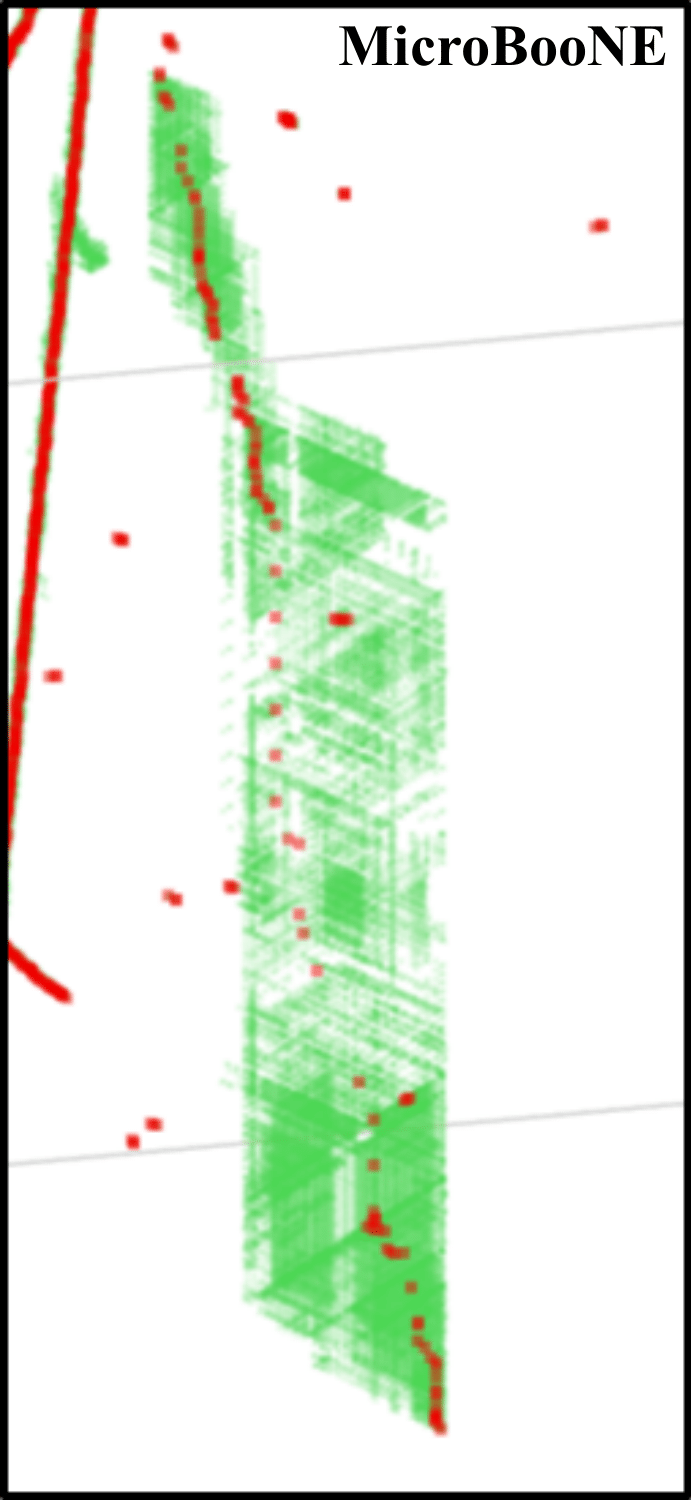}
	\put(-235,6){(a)}
	\put(-90,6){(b)}
	\caption{Impact of the same-blob Steiner edges. (a) The shortest path (blue points) 
		without adding the same-blob Steiner edges is along the boundary of the 3D image. This
		is partially caused by the grid structure of space points. Space points without charge information are displayed as green.
		(b) The shortest path after adding the same-blob Steiner edges (red points). 
		Space points are displayed as green without the charge information.}
	\label{fig:steiner_example2}
\end{figure}

\begin{figure}[t!hb]
	\centering
	\includegraphics[width=0.265\figwidth]{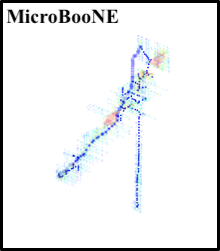}
	\includegraphics[width=0.233\figwidth]{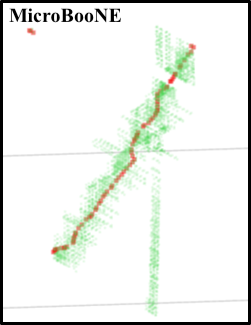}
	\put(-235,6){(a)}
	\put(-100,6){(b)}
	\caption{Improvement in searching for end points. (a) Extreme points identification 
		based on the principle component analysis. Space points are displayed in color with charge 
		information. The shortest path is shown as blue points. 
		(b) Current extreme points identification.  Space points are displayed as green points
		without charge information. The shortest path is shown as red points. }
	\label{fig:steiner_example5}
\end{figure}

\begin{figure}[t!hb]
	\centering
	\includegraphics[width=0.2\figwidth]{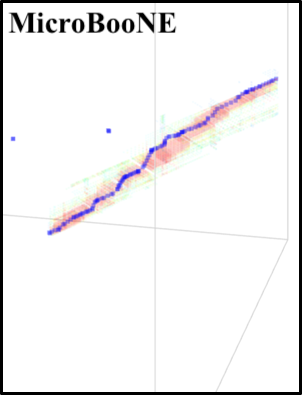}
	\includegraphics[width=0.28\figwidth]{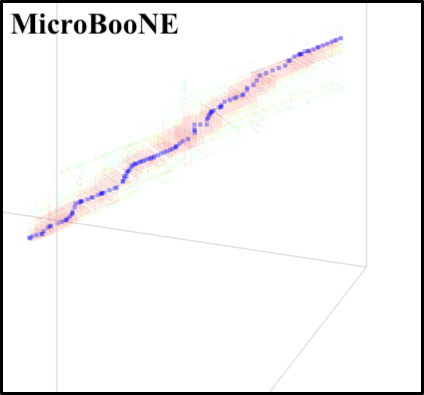}
	\put(-232,6){(a)}
	\put(-125,6){(b)}
	\caption{Impact of adding charge information to the weight calculation. Blue
		points represent the shortest path. Space points are displayed in colors with 
		their charge information. 
		(a) Weights are calculated only according to distances. 
		(b) Weights are calculated including the charge information. }
	\label{fig:steiner_example4}
\end{figure}

The Steiner-tree-inspired graph is then constructed. First, Steiner-tree terminals are found inside the first-stage graph. For each vertex (3D point), the three corresponding 2D pixels (one on each view) are found. The charge of each vertex is calculated to be the average charge of the three 2D pixels. Any vertex with its average charge higher or equal than all its neighbors on the first-stage graph is defined as a Steiner-tree terminal. A predefined threshold is applied to remove points with very low charge. Second, for each terminal, every other point inside the same blob is connected to it with an edge, which avoids the creation of grid points. Figure~\ref{fig:steiner_example2} shows the improvement in building the shortest path with these same-blob Steiner-tree edges. Third, two extreme space points of the first-stage graph, which are defined to cover the most live channels and time slices, are found. Figure~\ref{fig:steiner_example5} shows the comparison of the path construction using the old and new extreme-points searching algorithm. The original algorithm begins by finding the two points that have the largest separation along the main axis of the cluster. The shortest path between two extreme points is found using the Dijkstra shortest path algorithm. This path is essential for excluding spurious Steiner-tree terminals with isochronous track topologies. Terminals are excluded if these two conditions are met: (1) their 3D distance to the aforementioned shortest path is larger than a specified distance (6 cm as the default), and (2) their 2D distance to the shortest path in two projection views is smaller than a chosen distance (1.8 cm as the default). Finally, the Steiner-tree-inspired graph is constructed. For edges constructed on the Steiner-tree-inspired graph, a slightly different weight ($w$) is calculated, using the charge information as follows:
\begin{equation}
	w = \Delta r \left(0.8 + 0.2 \times \left( \frac{Q_0}{Q_s + Q_0} + \frac{Q_0}{Q_0 + Q_t} \right) \right), 
\end{equation}
with $Q_0=10^4$ electrons, and $Q_s$ and $Q_t$ being the average charge of the starting and ending vertices, respectively. $\Delta r$ represents the original distance between the two vertices. This choice leads to a slightly smaller weight for edges connecting two high-charge points. Figure~\ref{fig:steiner_example4} shows the impact of adding the charge information to the weight calculation. The shortest path is found on the Steiner-tree-inspired graph. The initial trajectory seed for the track trajectory fitting is chosen from the the shortest path such that the distance between two adjacent points is not too small nor too large (1~cm for the coarse-spacing fit and 0.6~cm for the fine-spacing fit, as described in Sec.~\ref{sec:track_fitting}). This operation leads to a more uniform set of seed points to produce the desired spacial granularity.

{\it 2D pixel association:} During the track trajectory fit, the associations between the 3D points and 2D pixels need to be formed, so that only a limited number of 2D pixels participate in determining each track trajectory point. This association is aided by the initial trajectory seed. The 3D points (vertices) on the first-stage graph close to the initial trajectory seed are found, and their parent blobs are saved. These 3D blobs are projected to the three 2D views to find the close-by 2D pixels (within 90\% of the projected 2D distance) to associate. This procedure is repeated on the Steiner-tree-inspired graph, which helps to bridge the gaps in the original 3D image. Since there are no blobs associated with the vertices on the Steiner-tree-inspired graph, the 2D pixels that are close to a projected 3D point are directly saved to form the association. If no 2D points are found to be associated with a particular 3D point in all three views, a virtual association from the projection of the 3D point is created as a regularization in the fit. 

The associations that have been formed are further examined. Only 2D pixels that are not associated with known nonfunctional channels and those with reconstructed charge higher than a threshold (2000 electrons as the default) are used during the examination. For any given view, the average location of the eligible 2D pixels is checked against the initial 2D projection of the 3D point. If the distance is larger than 75\% of the position spread and the number of eligible 2D pixels is small compared to the possible number of 2D pixels, the established association is replaced by a virtual association to avoid the bias in the trajectory fit near nonfunctional channels. If a 2D pixel is associated among multiple 3D candidate points, its charge is equally distributed amongst the 3D points. 

\begin{figure*}[!htb]
	\centering
	\includegraphics[width=\figwidth]{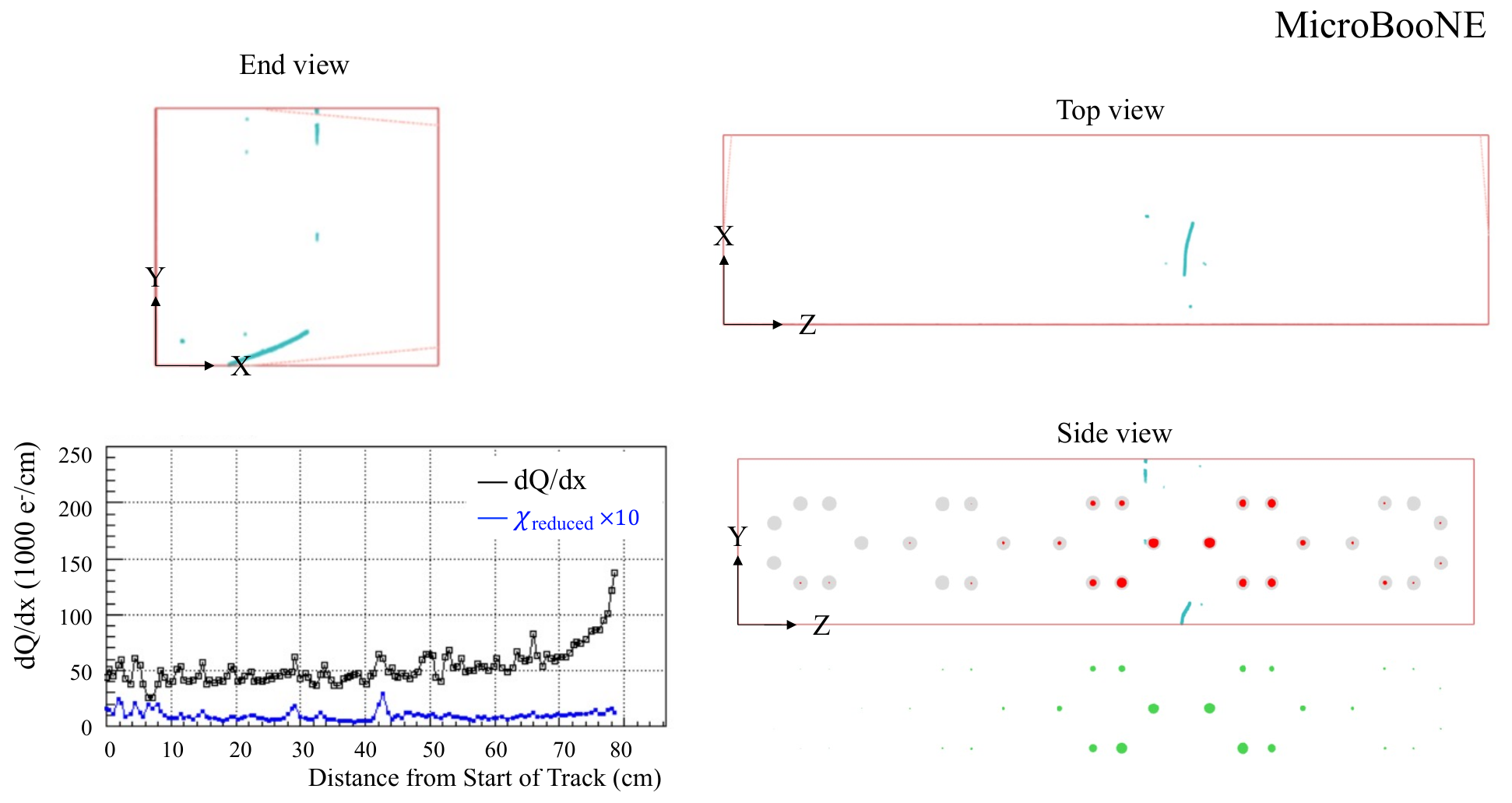}% Here is how to import EPS art
	\caption{\label{fig:STM_weird_angle} A STM candidate with a large angle with respect to the 
		TPC anode plane. See texts for more discussions. %\red{cz: remove $\chi_{\rm red}$ from all similar figures since we don't describe it in the text.} 
		%\url{https://www.phy.bnl.gov/twister/bee/set/uboone/scan/2018-06/b582/event/40/}
	}
\end{figure*}

\begin{figure*}[!htbp]
	\centering
	\includegraphics[width=\figwidth]{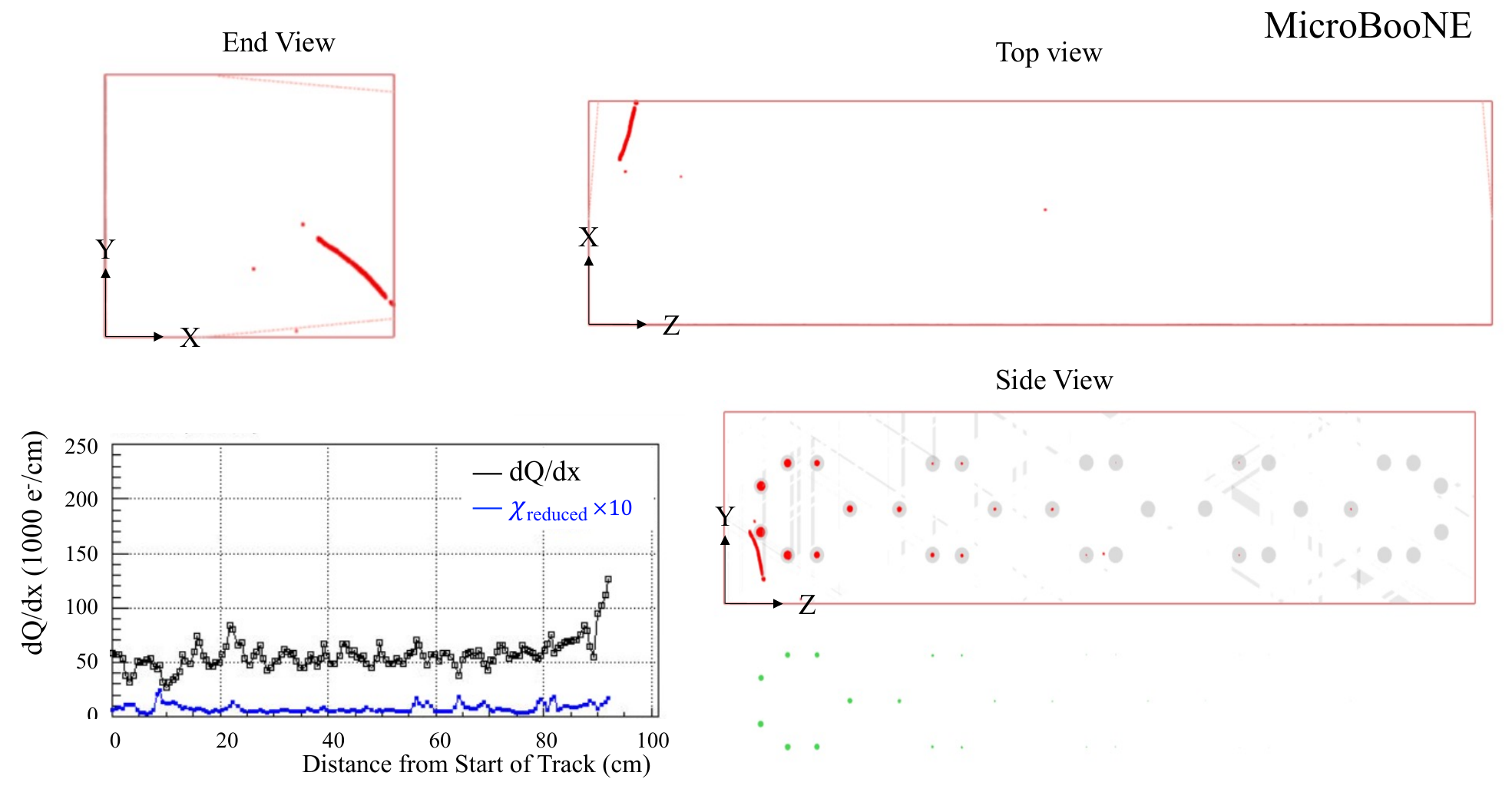}% Here is how to import EPS art
	\caption{\label{fig:STM_weird_angle2} A STM candidate with a large angle with respect to 
		the TPC anode plane. See texts for more discussions. 
		%\url{https://www.phy.bnl.gov/twister/bee/set/uboone/scan/2018-06/b582/event/49/}
	}
\end{figure*}

% short track and michel ...
\begin{figure*}[!htbp]
	\centering
	\includegraphics[width=\figwidth]{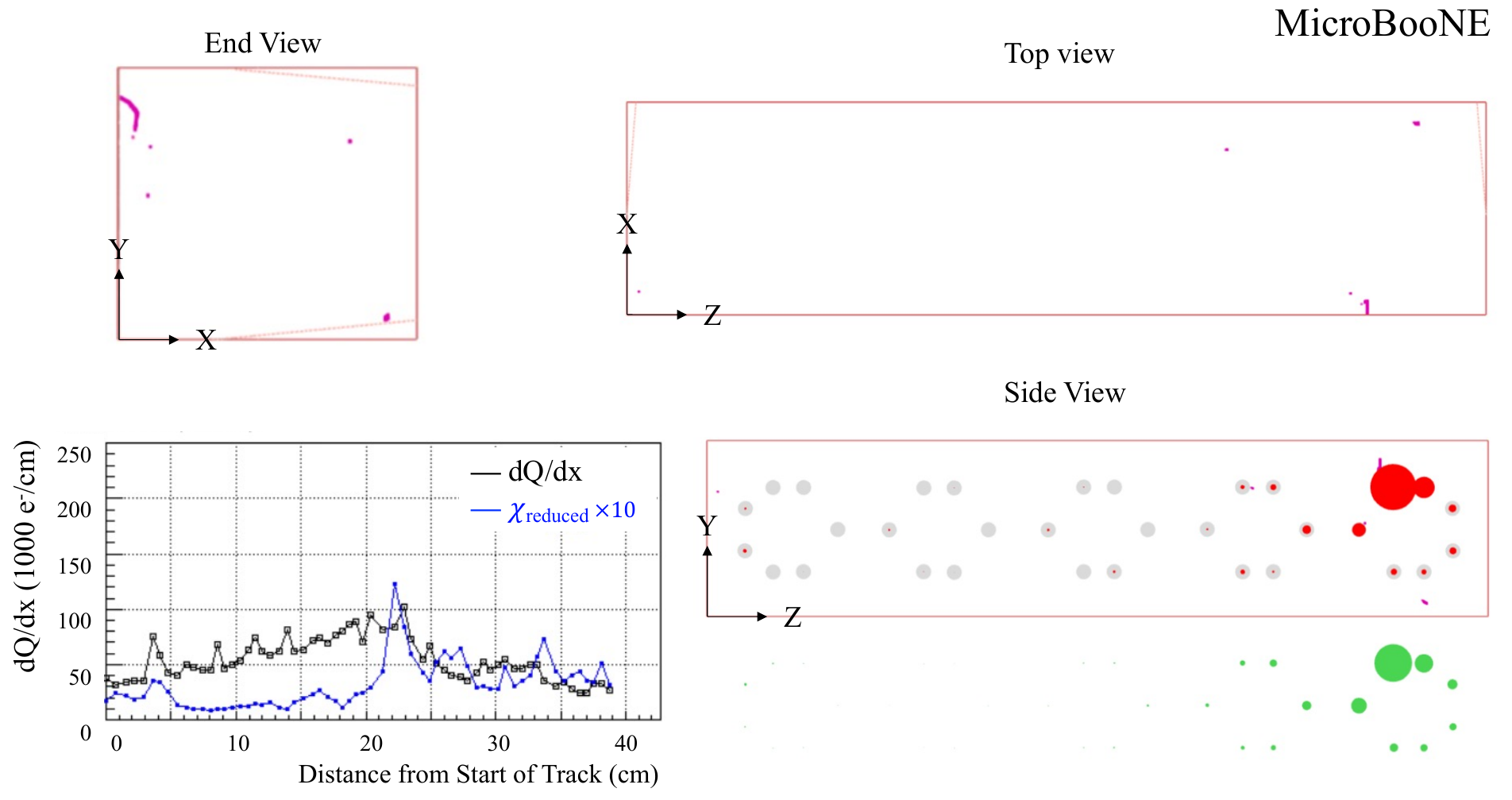}% Here is how to import EPS art
	\caption{\label{fig:STM_short_michel} A STM candidate with a short stopped muon track 
		following by a Michel electron.
		%\url{https://www.phy.bnl.gov/twister/bee/set/uboone/scan/2018-06/7f1d/event/0/}
	}
\end{figure*}

% low rise stm case?
\begin{figure*}[!htbp]
	\centering
	\includegraphics[width=\figwidth]{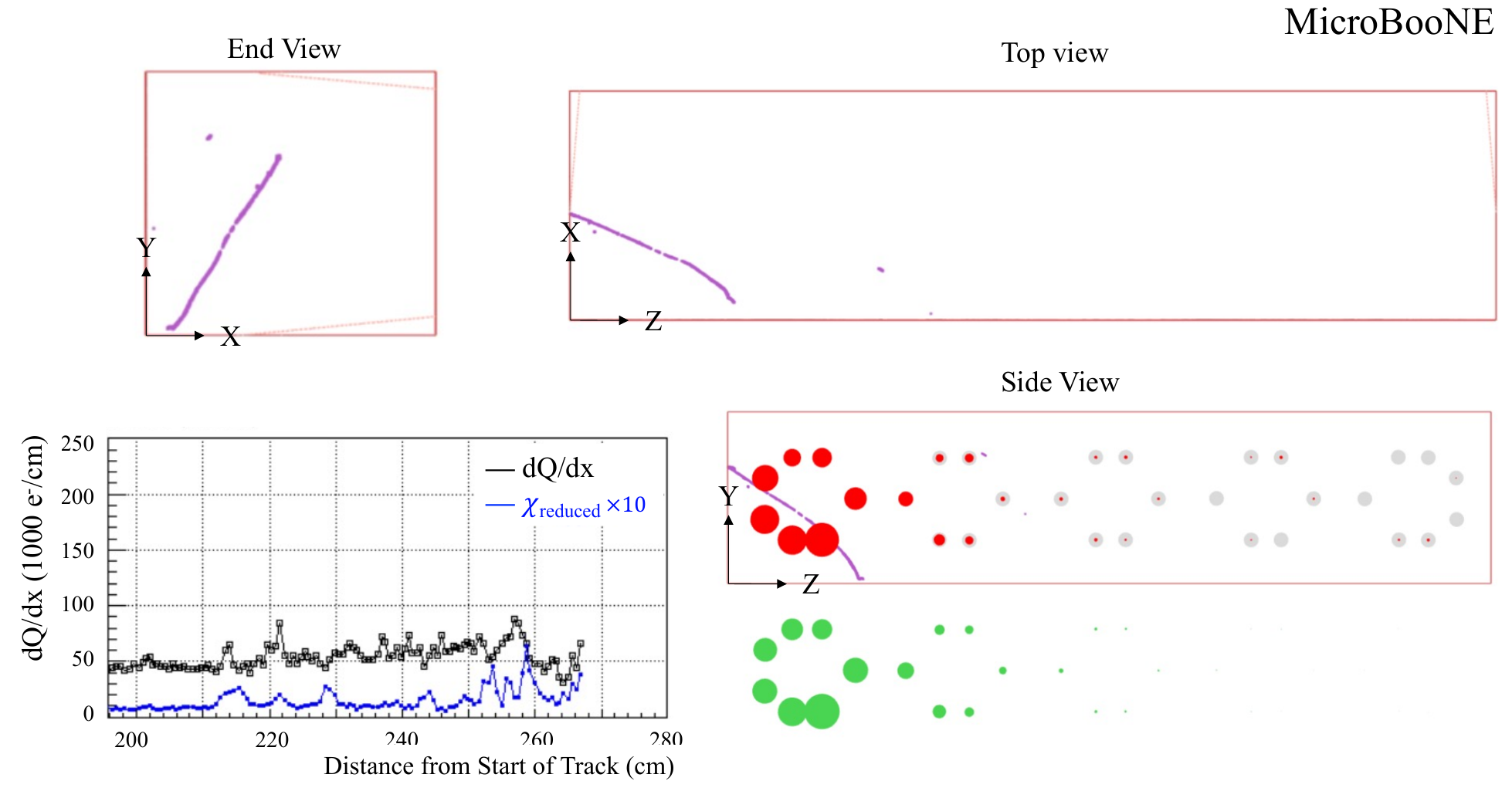}% Here is how to import EPS art
	\caption{\label{fig:STM_small_rise} A STM candidate with a small rise in \dqdx~near the end.
		%\url{https://www.phy.bnl.gov/twister/bee/set/uboone/scan/2018-06/7f1d/event/9/}
	}
\end{figure*}

\begin{figure*}[!htbp]
	\centering
	\includegraphics[width=\figwidth]{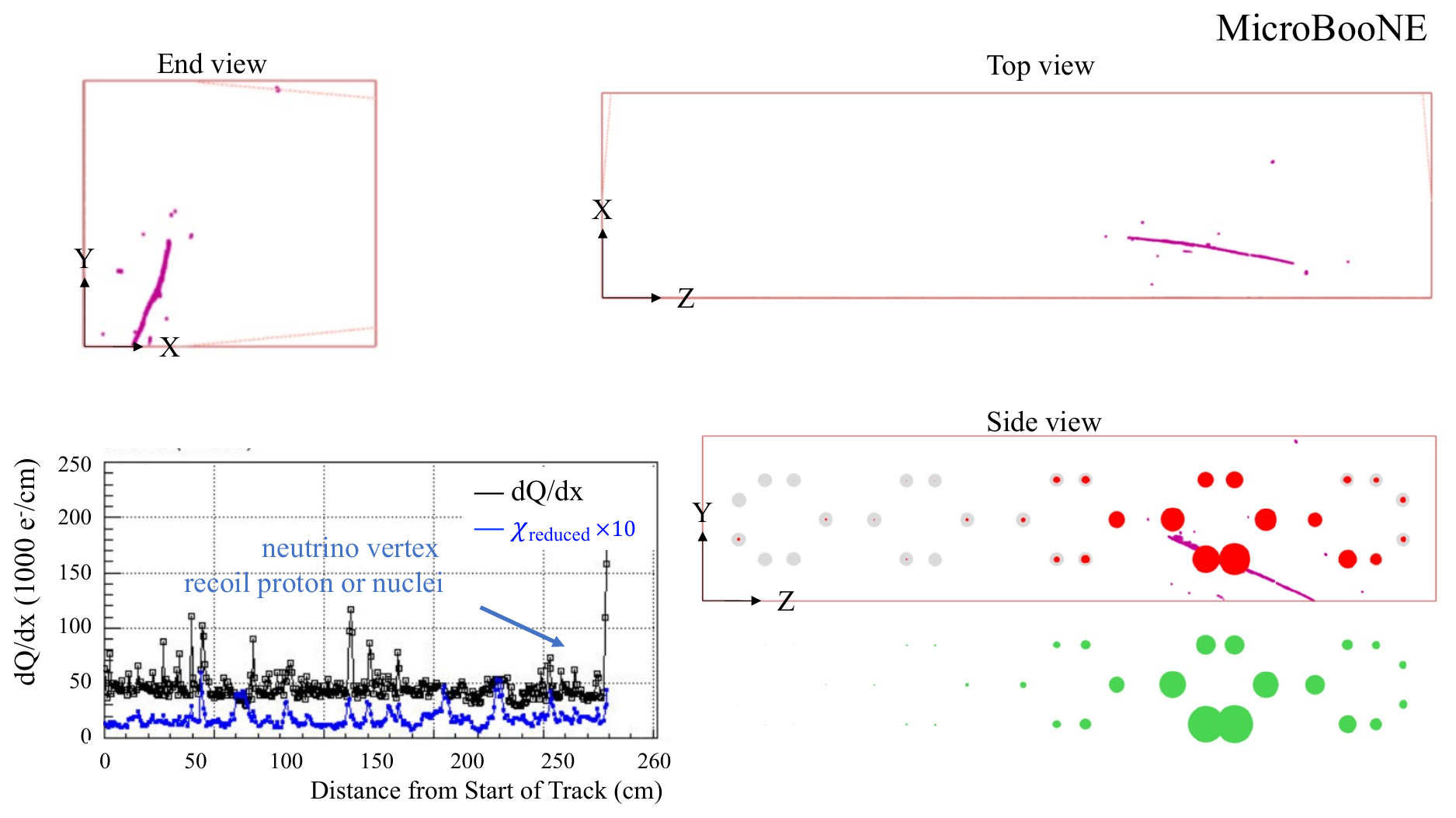}% Here is how to import EPS art
	\caption{\label{fig:STM_end_proton} A neutrino interaction candidate with a very large \dqdx~
		near the neutrino vertex. The sharp rise in \dqdx~(black line) near the end of track trajectory 
		is presumably the result of a recoil proton or nucleus (see the plot of \dqdx~vs. Distance from start of track).}
	%\url{https://www.phy.bnl.gov/twister/bee/set/uboone/scan/2018-06/b582/event/51/}
	
\end{figure*}

\begin{figure*}[!htbp]
	\centering
	\includegraphics[width=\figwidth]{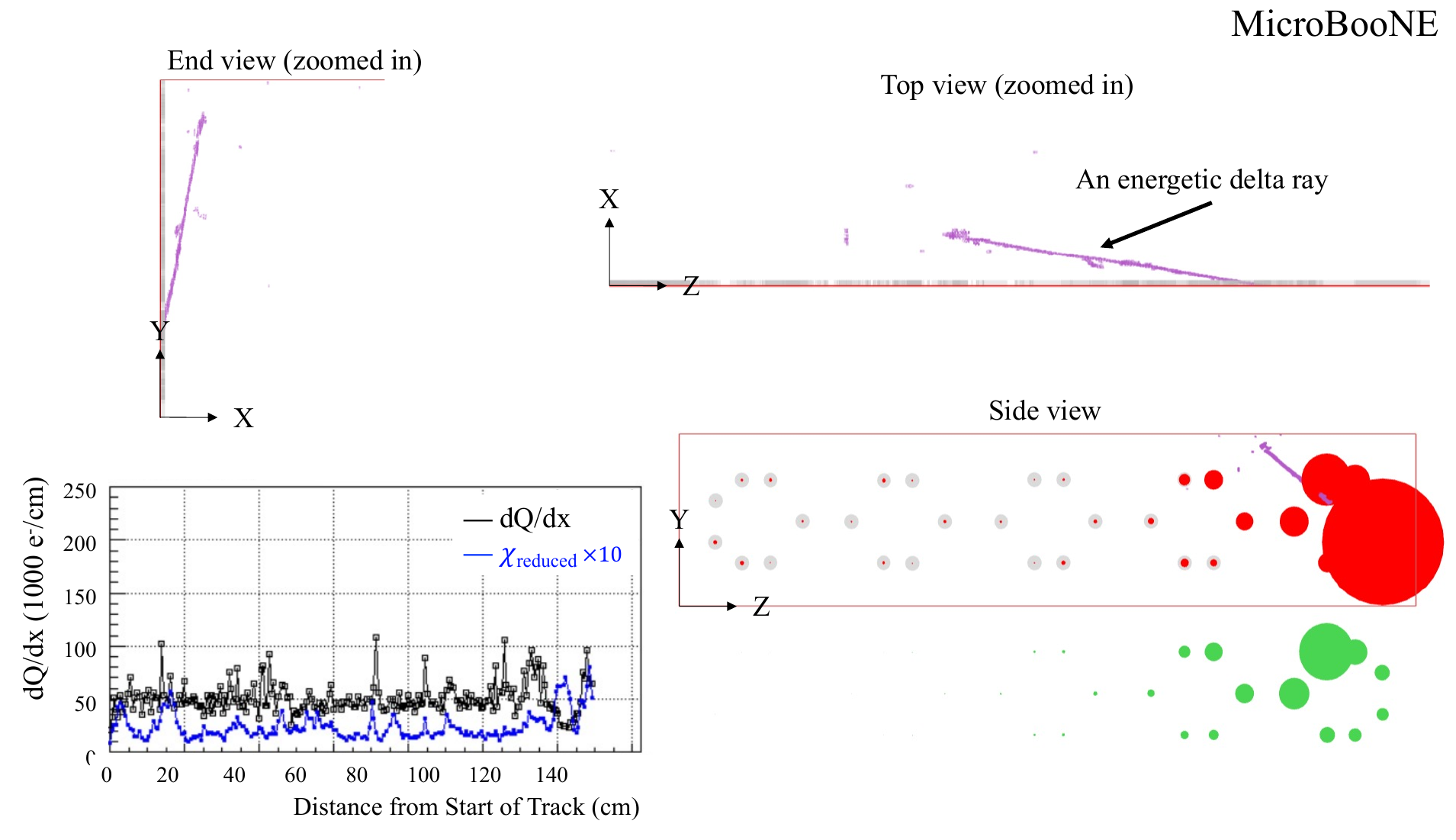}% Here is how to import EPS art
	\caption{\label{fig:STM_delta_ray} A neutrino interaction candidate with an energetic
		delta ray. The track direction is obvious after taking into account the direction of the 
		energetic delta ray. There is a rise in \dqdx~near the stopping point tagged by the STM 
		evaluation algorithm described in Sec.~\ref{sec:STM}. This rise is partially 
		because of the dip in \dqdx~in the region just before the stopping point. 
		%\url{https://www.phy.bnl.gov/twister/bee/set/uboone/scan/2018-06/b582/event/0/}
	}
\end{figure*}

\section{STOPPED MUON EXAMPLES}~\label{app:stm}
In this appendix, we show several representative STM examples with difficult topologies or unusual \dqdx~distributions, and some example neutrino interactions that could be mis-identified as an STM.

Figure~\ref{fig:STM_weird_angle} shows a tagged up-going STM that enters from the bottom of the detector. The black curve is the fitted \dqdx, and the blue curve is the reduced chi-squared ($\chi^2/ndf$) value comparing the predicted and measured charge from each 2D pixel. This track is clearly not a cosmic-ray muon which would enter the detector from the top or side. However, a clear rise in \dqdx~near the stopping point is seen. This track originates from a $\nu_\mu$CC interaction outside the TPC active volume. Only the muon enters the active TPC and is seen by the detector. 

A similar example is shown in Fig.~\ref{fig:STM_weird_angle2}. This track enters the detector from the cathode plane and travels toward the top of the detector. The angle of the track is not consistent with that of a cosmic-ray muon. However, a clear rise in \dqdx~near the stopping point is seen. This track, which also should originate from a $\nu_\mu$CC interaction outside the TPC active volume, is tagged as a STM background. Similarly, a $\nu_\mu$CC interaction is tagged as a TGM if its neutrino interaction vertex is outside the active volume and only the muon goes through the detector.

Figure~\ref{fig:STM_short_michel} shows an example of a STM with a Michel electron attached to the end. This track enters from the anode plane. The stopped muon is quite short, with a length of about 23~cm. The Michel electron is traveling vertically downward, leading to a very compact view in the collection W plane. The decay of the STM to a Michel electron (at $\approx$23~cm) is clearly seen in the \dqdx~distribution. The rise of \dqdx~before 23~cm is properly tagged by the STM tagger, and the residual \dqdx~is consistent with that of the Michel electron topology.

Figure~\ref{fig:STM_small_rise} shows another example of a STM with a Michel electron attached to the end. The rise of \dqdx~before 256~cm is smaller compared to that in Fig.~\ref{fig:STM_weird_angle}, but the residual track is consistent with a Michel electron. The algorithm discussed in Sec.~\ref{sec:STM} successfully tags this event as a STM background by considering the possibility of a muon decaying in flight.

Similarly to the TGM tagger described in Sec.~\ref{sec:TGM}, simplified pattern recognition algorithms are applied in the STM tagger to protect against neutrino interactions that may mimic STMs. Most neutrino interactions result in multiple tracks and are effectively removed by detecting a large angle deflection. Figure~\ref{fig:STM_end_proton} shows an example of a single-track-like neutrino interaction candidate. At the identified end point (kink) of the track, there is a very sharp rise in \dqdx. Such a sharp rise is not consistent with the expectation for an STM. Instead, this should be the vertex of the neutrio interaction, and the high \dqdx~comes from a very short recoil proton or nucleus. A dedicated algorithm identifies these neutrino interaction candidates based on the shape of the \dqdx~of the main track and the length and \dqdx~of the residual track. 
%\begin{itemize}
%	\item If the track length is longer than 4~cm beyond the kink point, the algorithm ends without tagging this event as a neutrino candidate. 
%	\item A number of tools are used to tag this kind of topology, including all of the following: the maximal bin of \dqdx, the maximal value of \dqdx, the KS distance between the measured and predicted STM \dqdx~distribution after excluding the last 3~cm as well as the ratio of the predicted vs. measured integrated \dqdx, the KS distance between the measured and predicted flat \dqdx~distribution after excluding the last 3~cm as well as the ratio of predicted vs. measured integrated \dqdx, and the KS distance between the measured and predicted STM \dqdx~with the full range as well as the ratio of predicted vs. measured integrated \dqdx. 
%	\item An empirical set of cuts are developed. Essentially, we require i) the rise of \dqdx~to be fast and high, ii) the \dqdx~before the rise is more consistent with the flat \dqdx~hypothesis (50k e$^-$/cm), and iii) the overall \dqdx~distribution is not consistent with that of an STM.
%\end{itemize}

Figure~\ref{fig:STM_delta_ray} shows another neutrino interaction example. There is a rise in \dqdx~near the stopping point tagged by the STM evaluation algorithm described in Sec.~\ref{sec:STM}, partially because of the dip in \dqdx~in the region just before the stopping point. However, a long delta ray is identified as a separated track candidate along the main track. The direction of the delta ray with respect to the main track is used to determine the direction of the track. Since the track direction is not consistent with that of an incoming STM, this event is successfully tagged as a neutrino interaction.

\bibliography{wire-cell-ns}% Produces the bibliography via BibTeX.

\begin{thebibliography}{62}%
\makeatletter
\providecommand \@ifxundefined [1]{%
 \@ifx{#1\undefined}
}%
\providecommand \@ifnum [1]{%
 \ifnum #1\expandafter \@firstoftwo
 \else \expandafter \@secondoftwo
 \fi
}%
\providecommand \@ifx [1]{%
 \ifx #1\expandafter \@firstoftwo
 \else \expandafter \@secondoftwo
 \fi
}%
\providecommand \natexlab [1]{#1}%
\providecommand \enquote  [1]{``#1''}%
\providecommand \bibnamefont  [1]{#1}%
\providecommand \bibfnamefont [1]{#1}%
\providecommand \citenamefont [1]{#1}%
\providecommand \href@noop [0]{\@secondoftwo}%
\providecommand \href [0]{\begingroup \@sanitize@url \@href}%
\providecommand \@href[1]{\@@startlink{#1}\@@href}%
\providecommand \@@href[1]{\endgroup#1\@@endlink}%
\providecommand \@sanitize@url [0]{\catcode `\\12\catcode `\$12\catcode
  `\&12\catcode `\#12\catcode `\^12\catcode `\_12\catcode `\%12\relax}%
\providecommand \@@startlink[1]{}%
\providecommand \@@endlink[0]{}%
\providecommand \url  [0]{\begingroup\@sanitize@url \@url }%
\providecommand \@url [1]{\endgroup\@href {#1}{\urlprefix }}%
\providecommand \urlprefix  [0]{URL }%
\providecommand \Eprint [0]{\href }%
\providecommand \doibase [0]{http://dx.doi.org/}%
\providecommand \selectlanguage [0]{\@gobble}%
\providecommand \bibinfo  [0]{\@secondoftwo}%
\providecommand \bibfield  [0]{\@secondoftwo}%
\providecommand \translation [1]{[#1]}%
\providecommand \BibitemOpen [0]{}%
\providecommand \bibitemStop [0]{}%
\providecommand \bibitemNoStop [0]{.\EOS\space}%
\providecommand \EOS [0]{\spacefactor3000\relax}%
\providecommand \BibitemShut  [1]{\csname bibitem#1\endcsname}%
\let\auto@bib@innerbib\@empty
%</preamble>
\bibitem [{\citenamefont {Rubbia}(1977)}]{rubbia77}%
  \BibitemOpen
  \bibfield  {author} {\bibinfo {author} {\bibfnamefont {C.}~\bibnamefont
  {Rubbia}},\ }\bibfield  {title} {\enquote {\bibinfo {title} {{The Liquid
  Argon Time Projection Chamber: A New Concept for Neutrino Detectors}},}\
  }\href@noop {} {\bibfield  {journal} {\bibinfo  {journal} {CERN-EP-INT-77-08,
  CERN-EP-77-08}\ } (\bibinfo {year} {1977})}\BibitemShut {NoStop}%
\bibitem [{\citenamefont {Chen}\ \emph {et~al.}(1976)\citenamefont {Chen},
  \citenamefont {Condon}, \citenamefont {Barish},\ and\ \citenamefont
  {Sciulli}}]{Chen:1976pp}%
  \BibitemOpen
  \bibfield  {author} {\bibinfo {author} {\bibfnamefont {H.H.}\ \bibnamefont
  {Chen}}, \bibinfo {author} {\bibfnamefont {P.E.}\ \bibnamefont {Condon}},
  \bibinfo {author} {\bibfnamefont {B.C.}\ \bibnamefont {Barish}}, \ and\
  \bibinfo {author} {\bibfnamefont {F.J.}\ \bibnamefont {Sciulli}},\ }\bibfield
   {title} {\enquote {\bibinfo {title} {{A Neutrino detector sensitive to rare
  processes. I. A Study of neutrino electron reactions}},}\ }\href@noop {}
  {\bibfield  {journal} {\bibinfo  {journal} {FERMILAB-PROPOSAL-0496}\ }
  (\bibinfo {year} {1976})}\BibitemShut {NoStop}%
\bibitem [{\citenamefont {Willis}\ and\ \citenamefont
  {Radeka}(1974)}]{willis74}%
  \BibitemOpen
  \bibfield  {author} {\bibinfo {author} {\bibfnamefont {W.J.}\ \bibnamefont
  {Willis}}\ and\ \bibinfo {author} {\bibfnamefont {V.}~\bibnamefont
  {Radeka}},\ }\bibfield  {title} {\enquote {\bibinfo {title} {{Liquid Argon
  Ionization Chambers as Total Absorption Detectors}},}\ }\href {\doibase
  10.1016/0029-554X(74)90039-1} {\bibfield  {journal} {\bibinfo  {journal}
  {Nucl. Instrum. Meth.}\ }\textbf {\bibinfo {volume} {120}},\ \bibinfo {pages}
  {221--236} (\bibinfo {year} {1974})}\BibitemShut {NoStop}%
\bibitem [{\citenamefont {Nygren}(1974)}]{Nygren:1976fe}%
  \BibitemOpen
  \bibfield  {author} {\bibinfo {author} {\bibfnamefont {D.R.}\ \bibnamefont
  {Nygren}},\ }\bibfield  {title} {\enquote {\bibinfo {title} {{The Time
  Projection Chamber: A New 4 pi Detector for Charged Particles}},}\
  }\href@noop {} {\bibfield  {journal} {\bibinfo  {journal} {eConf}\ }\textbf
  {\bibinfo {volume} {C740805}},\ \bibinfo {pages} {58} (\bibinfo {year}
  {1974})}\BibitemShut {NoStop}%
\bibitem [{\citenamefont {Amerio}\ \emph {et~al.}(2004)\citenamefont {Amerio}
  \emph {et~al.}}]{Amerio:2004ze}%
  \BibitemOpen
  \bibfield  {author} {\bibinfo {author} {\bibfnamefont {S.}~\bibnamefont
  {Amerio}} \emph {et~al.} (\bibinfo {collaboration} {ICARUS Collaboration}),\
  }\bibfield  {title} {\enquote {\bibinfo {title} {{Design, construction and
  tests of the ICARUS T600 detector}},}\ }\href {\doibase
  10.1016/j.nima.2004.02.044} {\bibfield  {journal} {\bibinfo  {journal} {Nucl.
  Instrum. Meth. A}\ }\textbf {\bibinfo {volume} {527}},\ \bibinfo {pages}
  {329--410} (\bibinfo {year} {2004})}\BibitemShut {NoStop}%
\bibitem [{\citenamefont {Anderson}\ \emph {et~al.}(2012)\citenamefont
  {Anderson} \emph {et~al.}}]{ArgoNeuT2012}%
  \BibitemOpen
  \bibfield  {author} {\bibinfo {author} {\bibfnamefont {C.}~\bibnamefont
  {Anderson}} \emph {et~al.} (\bibinfo {collaboration} {ArgoNeuT
  Collaboration}),\ }\bibfield  {title} {\enquote {\bibinfo {title} {{The
  ArgoNeuT Detector in the NuMI Low-Energy beam line at Fermilab}},}\ }\href
  {\doibase 10.1088/1748-0221/7/10/P10019} {\bibfield  {journal} {\bibinfo
  {journal} {JINST}\ }\textbf {\bibinfo {volume} {7}},\ \bibinfo {pages}
  {P10019} (\bibinfo {year} {2012})}\BibitemShut {NoStop}%
\bibitem [{\citenamefont {Acciarri}\ \emph
  {et~al.}(2017{\natexlab{a}})\citenamefont {Acciarri} \emph
  {et~al.}}]{Acciarri:2016smi}%
  \BibitemOpen
  \bibfield  {author} {\bibinfo {author} {\bibfnamefont {R.}~\bibnamefont
  {Acciarri}} \emph {et~al.} (\bibinfo {collaboration} {MicroBooNE
  Collaboration}),\ }\bibfield  {title} {\enquote {\bibinfo {title} {{Design
  and Construction of the MicroBooNE Detector}},}\ }\href {\doibase
  10.1088/1748-0221/12/02/P02017} {\bibfield  {journal} {\bibinfo  {journal}
  {JINST}\ }\textbf {\bibinfo {volume} {12}},\ \bibinfo {pages} {P02017}
  (\bibinfo {year} {2017}{\natexlab{a}})}\BibitemShut {NoStop}%
\bibitem [{\citenamefont {Badhrees}\ \emph {et~al.}(2012)\citenamefont
  {Badhrees} \emph {et~al.}}]{Badhrees:2012zz}%
  \BibitemOpen
  \bibfield  {author} {\bibinfo {author} {\bibfnamefont {I.}~\bibnamefont
  {Badhrees}} \emph {et~al.},\ }\bibfield  {title} {\enquote {\bibinfo {title}
  {{Argontube: An R\&D liquid Argon Time Projection Chamber}},}\ }\href
  {\doibase 10.1088/1748-0221/7/02/C02011} {\bibfield  {journal} {\bibinfo
  {journal} {JINST}\ }\textbf {\bibinfo {volume} {7}},\ \bibinfo {pages}
  {C02011} (\bibinfo {year} {2012})}\BibitemShut {NoStop}%
\bibitem [{\citenamefont {Bhandari}\ \emph {et~al.}(2019)\citenamefont
  {Bhandari} \emph {et~al.}}]{Bhandari:2019rat}%
  \BibitemOpen
  \bibfield  {author} {\bibinfo {author} {\bibfnamefont {B.}~\bibnamefont
  {Bhandari}} \emph {et~al.} (\bibinfo {collaboration} {CAPTAIN
  Collaboration}),\ }\bibfield  {title} {\enquote {\bibinfo {title} {{First
  Measurement of the Total Neutron Cross Section on Argon Between 100 and 800
  MeV}},}\ }\href {\doibase 10.1103/PhysRevLett.123.042502} {\bibfield
  {journal} {\bibinfo  {journal} {Phys. Rev. Lett.}\ }\textbf {\bibinfo
  {volume} {123}},\ \bibinfo {pages} {042502} (\bibinfo {year}
  {2019})}\BibitemShut {NoStop}%
\bibitem [{\citenamefont {Hahn}\ \emph {et~al.}(2016)\citenamefont {Hahn} \emph
  {et~al.}}]{Hahn:2016tia}%
  \BibitemOpen
  \bibfield  {author} {\bibinfo {author} {\bibfnamefont {Alan}\ \bibnamefont
  {Hahn}} \emph {et~al.} (\bibinfo {collaboration} {LBNE Collaboration}),\
  }\bibfield  {title} {\enquote {\bibinfo {title} {{The LBNE 35 Ton Prototype
  Cryostat}},}\ }in\ \href {\doibase 10.1109/NSSMIC.2014.7431158} {\emph
  {\bibinfo {booktitle} {{2014 IEEE Nuclear Science Symposium and Medical
  Imaging Conference and 21st Symposium on Room-Temperature Semiconductor X-ray
  and Gamma-ray Detectors}}}}\ (\bibinfo {year} {2016})\ p.\ \bibinfo {pages}
  {7431158}\BibitemShut {NoStop}%
\bibitem [{\citenamefont {Paley}\ \emph {et~al.}(2014)\citenamefont {Paley}
  \emph {et~al.}}]{Cavanna:2014iqa}%
  \BibitemOpen
  \bibfield  {author} {\bibinfo {author} {\bibfnamefont {J.}~\bibnamefont
  {Paley}} \emph {et~al.} (\bibinfo {collaboration} {LArIAT Collaboration}),\
  }\href@noop {} {\enquote {\bibinfo {title} {{LArIAT: Liquid Argon In A
  Testbeam}},}\ } (\bibinfo {year} {2014}),\ \Eprint
  {http://arxiv.org/abs/1406.5560} {arXiv:1406.5560 [physics.ins-det]}
  \BibitemShut {NoStop}%
\bibitem [{\citenamefont {Abi}\ \emph {et~al.}(2020{\natexlab{a}})\citenamefont
  {Abi} \emph {et~al.}}]{Abi:2020mwi}%
  \BibitemOpen
  \bibfield  {author} {\bibinfo {author} {\bibfnamefont {B.}~\bibnamefont
  {Abi}} \emph {et~al.} (\bibinfo {collaboration} {DUNE Collaboration}),\
  }\bibfield  {title} {\enquote {\bibinfo {title} {{First results on
  ProtoDUNE-SP liquid argon time projection chamber performance from a beam
  test at the CERN Neutrino Platform}},}\ }\href {\doibase
  10.1088/1748-0221/15/12/P12004} {\bibfield  {journal} {\bibinfo  {journal}
  {JINST}\ }\textbf {\bibinfo {volume} {15}},\ \bibinfo {pages} {P12004}
  (\bibinfo {year} {2020}{\natexlab{a}})},\ \Eprint
  {http://arxiv.org/abs/2007.06722} {arXiv:2007.06722 [physics.ins-det]}
  \BibitemShut {NoStop}%
\bibitem [{\citenamefont {Cavanna}\ \emph {et~al.}(2018)\citenamefont
  {Cavanna}, \citenamefont {Ereditato},\ and\ \citenamefont
  {Fleming}}]{Cavanna:2018yfk}%
  \BibitemOpen
  \bibfield  {author} {\bibinfo {author} {\bibfnamefont {F.}~\bibnamefont
  {Cavanna}}, \bibinfo {author} {\bibfnamefont {A.}~\bibnamefont {Ereditato}},
  \ and\ \bibinfo {author} {\bibfnamefont {B.~T.}\ \bibnamefont {Fleming}},\
  }\bibfield  {title} {\enquote {\bibinfo {title} {{Advances in liquid argon
  detectors}},}\ }\href {\doibase 10.1016/j.nima.2018.07.010} {\bibfield
  {journal} {\bibinfo  {journal} {Nucl. Instrum. Meth.}\ }\textbf {\bibinfo
  {volume} {A907}},\ \bibinfo {pages} {1--8} (\bibinfo {year}
  {2018})}\BibitemShut {NoStop}%
%%CITATION = NUIMA,A907,1;%%
\bibitem [{\citenamefont {Aguilar-Arevalo}\ \emph {et~al.}(2012)\citenamefont
  {Aguilar-Arevalo} \emph {et~al.}}]{Aguilar-Arevalo:2012fmn}%
  \BibitemOpen
  \bibfield  {author} {\bibinfo {author} {\bibfnamefont {A.A.}\ \bibnamefont
  {Aguilar-Arevalo}} \emph {et~al.} (\bibinfo {collaboration} {MiniBooNE}),\
  }\href@noop {} {\enquote {\bibinfo {title} {{A Combined $\nu_\mu \rightarrow
  \nu_e$ and $\bar \nu_\mu \rightarrow \bar \nu_e$ Oscillation Analysis of the
  MiniBooNE Excesses}},}\ } (\bibinfo {year} {2012}),\ \Eprint
  {http://arxiv.org/abs/1207.4809} {arXiv:1207.4809 [hep-ex]} \BibitemShut
  {NoStop}%
\bibitem [{\citenamefont {Abratenko}\ \emph
  {et~al.}(2019{\natexlab{a}})\citenamefont {Abratenko} \emph
  {et~al.}}]{Abratenko:2019jqo}%
  \BibitemOpen
  \bibfield  {author} {\bibinfo {author} {\bibfnamefont {P.}~\bibnamefont
  {Abratenko}} \emph {et~al.} (\bibinfo {collaboration} {MicroBooNE
  Collaboration}),\ }\bibfield  {title} {\enquote {\bibinfo {title} {{First
  Measurement of Inclusive Muon Neutrino Charged Current Differential Cross
  Sections on Argon at $E_\nu\sim$0.8 GeV with the MicroBooNE Detector}},}\
  }\href {\doibase 10.1103/PhysRevLett.123.131801} {\bibfield  {journal}
  {\bibinfo  {journal} {Phys. Rev. Lett.}\ }\textbf {\bibinfo {volume} {123}},\
  \bibinfo {pages} {131801} (\bibinfo {year} {2019}{\natexlab{a}})}\BibitemShut
  {NoStop}%
\bibitem [{\citenamefont {Adams}\ \emph
  {et~al.}(2019{\natexlab{a}})\citenamefont {Adams} \emph
  {et~al.}}]{Adams:2018sgn}%
  \BibitemOpen
  \bibfield  {author} {\bibinfo {author} {\bibfnamefont {C.}~\bibnamefont
  {Adams}} \emph {et~al.} (\bibinfo {collaboration} {MicroBooNE
  Collaboration}),\ }\bibfield  {title} {\enquote {\bibinfo {title} {{First
  measurement of $\nu_{\mu}$ charged-current $\pi^{0}$ production on argon with
  the MicroBooNE detector}},}\ }\href {\doibase 10.1103/PhysRevD.99.091102}
  {\bibfield  {journal} {\bibinfo  {journal} {Phys. Rev.}\ }\textbf {\bibinfo
  {volume} {D99}},\ \bibinfo {pages} {091102} (\bibinfo {year}
  {2019}{\natexlab{a}})}\BibitemShut {NoStop}%
%%CITATION = ARXIV:1811.02700;%%
\bibitem [{\citenamefont {Abratenko}\ \emph
  {et~al.}(2020{\natexlab{a}})\citenamefont {Abratenko} \emph
  {et~al.}}]{Abratenko:2020acr}%
  \BibitemOpen
  \bibfield  {author} {\bibinfo {author} {\bibfnamefont {P.}~\bibnamefont
  {Abratenko}} \emph {et~al.} (\bibinfo {collaboration} {MicroBooNE
  Collaboration}),\ }\bibfield  {title} {\enquote {\bibinfo {title} {{First
  Measurement of Differential Charged Current Quasielastic-like $\nu_\mu$-Argon
  Scattering Cross Sections with the MicroBooNE Detector}},}\ }\href {\doibase
  10.1103/PhysRevLett.125.201803} {\bibfield  {journal} {\bibinfo  {journal}
  {Phys. Rev. Lett.}\ }\textbf {\bibinfo {volume} {125}},\ \bibinfo {pages}
  {201803} (\bibinfo {year} {2020}{\natexlab{a}})},\ \Eprint
  {http://arxiv.org/abs/2006.00108} {arXiv:2006.00108 [hep-ex]} \BibitemShut
  {NoStop}%
\bibitem [{\citenamefont {Abratenko}\ \emph
  {et~al.}(2020{\natexlab{b}})\citenamefont {Abratenko} \emph
  {et~al.}}]{Abratenko:2020sga}%
  \BibitemOpen
  \bibfield  {author} {\bibinfo {author} {\bibfnamefont {P.}~\bibnamefont
  {Abratenko}} \emph {et~al.} (\bibinfo {collaboration} {MicroBooNE
  Collaboration}),\ }\bibfield  {title} {\enquote {\bibinfo {title}
  {{Measurement of Differential Cross Sections for $\nu_\mu$-Ar Charged-Current
  Interactions with Protons and no Pions in the Final State with the MicroBooNE
  Detector}},}\ }\href@noop {} {\  (\bibinfo {year} {2020}{\natexlab{b}})},\
  \Eprint {http://arxiv.org/abs/2010.02390} {arXiv:2010.02390 [hep-ex]}
  \BibitemShut {NoStop}%
\bibitem [{\citenamefont {Antonello}\ \emph {et~al.}(2015)\citenamefont
  {Antonello} \emph {et~al.}}]{Antonello:2015lea}%
  \BibitemOpen
  \bibfield  {author} {\bibinfo {author} {\bibfnamefont {M.}~\bibnamefont
  {Antonello}} \emph {et~al.} (\bibinfo {collaboration} {MicroBooNE, LAr1-ND,
  and ICARUS-WA104 Collaboration}),\ }\href@noop {} {\enquote {\bibinfo {title}
  {{A Proposal for a Three Detector Short-Baseline Neutrino Oscillation Program
  in the Fermilab Booster Neutrino Beam}},}\ } (\bibinfo {year} {2015}),\
  \Eprint {http://arxiv.org/abs/1503.01520} {arXiv:1503.01520
  [physics.ins-det]} \BibitemShut {NoStop}%
\bibitem [{\citenamefont {Machado}\ \emph {et~al.}(2019)\citenamefont
  {Machado}, \citenamefont {Palamara},\ and\ \citenamefont
  {Schmitz}}]{Machado:2019oxb}%
  \BibitemOpen
  \bibfield  {author} {\bibinfo {author} {\bibfnamefont {Pedro~A.N.}\
  \bibnamefont {Machado}}, \bibinfo {author} {\bibfnamefont {Ornella}\
  \bibnamefont {Palamara}}, \ and\ \bibinfo {author} {\bibfnamefont {David~W.}\
  \bibnamefont {Schmitz}},\ }\bibfield  {title} {\enquote {\bibinfo {title}
  {{The Short-Baseline Neutrino Program at Fermilab}},}\ }\href@noop {}
  {\bibfield  {journal} {\bibinfo  {journal} {Ann. Rev. Nucl. Part. Sci.}\
  }\textbf {\bibinfo {volume} {69}} (\bibinfo {year} {2019})}\BibitemShut
  {NoStop}%
%%CITATION = ARXIV:1903.04608;%%
\bibitem [{\citenamefont {Abi}\ \emph {et~al.}(2020{\natexlab{b}})\citenamefont
  {Abi} \emph {et~al.}}]{dune-tdr-1}%
  \BibitemOpen
  \bibfield  {author} {\bibinfo {author} {\bibfnamefont {Babak}\ \bibnamefont
  {Abi}} \emph {et~al.} (\bibinfo {collaboration} {DUNE}),\ }\bibfield  {title}
  {\enquote {\bibinfo {title} {{Deep Underground Neutrino Experiment (DUNE),
  Far Detector Technical Design Report, Volume I Introduction to DUNE}},}\
  }\href {\doibase 10.1088/1748-0221/15/08/T08008} {\bibfield  {journal}
  {\bibinfo  {journal} {JINST}\ }\textbf {\bibinfo {volume} {15}},\ \bibinfo
  {pages} {T08008} (\bibinfo {year} {2020}{\natexlab{b}})},\ \Eprint
  {http://arxiv.org/abs/2002.02967} {arXiv:2002.02967 [physics.ins-det]}
  \BibitemShut {NoStop}%
\bibitem [{\citenamefont {Abi}\ \emph {et~al.}(2020{\natexlab{c}})\citenamefont
  {Abi} \emph {et~al.}}]{Abi:2020evt}%
  \BibitemOpen
  \bibfield  {author} {\bibinfo {author} {\bibfnamefont {Babak}\ \bibnamefont
  {Abi}} \emph {et~al.} (\bibinfo {collaboration} {DUNE Collaboration}),\
  }\bibfield  {title} {\enquote {\bibinfo {title} {{Deep Underground Neutrino
  Experiment (DUNE), Far Detector Technical Design Report, Volume II DUNE
  Physics}},}\ }\href@noop {} {\  (\bibinfo {year} {2020}{\natexlab{c}})},\
  \Eprint {http://arxiv.org/abs/2002.03005} {arXiv:2002.03005 [hep-ex]}
  \BibitemShut {NoStop}%
\bibitem [{\citenamefont {Qian}\ and\ \citenamefont
  {Vogel}(2015)}]{Qian:2015waa}%
  \BibitemOpen
  \bibfield  {author} {\bibinfo {author} {\bibfnamefont {X.}~\bibnamefont
  {Qian}}\ and\ \bibinfo {author} {\bibfnamefont {P.}~\bibnamefont {Vogel}},\
  }\bibfield  {title} {\enquote {\bibinfo {title} {{Neutrino Mass
  Hierarchy}},}\ }\href {\doibase 10.1016/j.ppnp.2015.05.002} {\bibfield
  {journal} {\bibinfo  {journal} {Prog. Part. Nucl. Phys.}\ }\textbf {\bibinfo
  {volume} {83}},\ \bibinfo {pages} {1--30} (\bibinfo {year}
  {2015})}\BibitemShut {NoStop}%
%%CITATION = ARXIV:1505.01891;%%
\bibitem [{\citenamefont {Qian}\ \emph {et~al.}(2018)\citenamefont {Qian},
  \citenamefont {Zhang}, \citenamefont {Viren},\ and\ \citenamefont
  {Diwan}}]{Qian:2018qbv}%
  \BibitemOpen
  \bibfield  {author} {\bibinfo {author} {\bibfnamefont {Xin}\ \bibnamefont
  {Qian}}, \bibinfo {author} {\bibfnamefont {Chao}\ \bibnamefont {Zhang}},
  \bibinfo {author} {\bibfnamefont {Brett}\ \bibnamefont {Viren}}, \ and\
  \bibinfo {author} {\bibfnamefont {Milind}\ \bibnamefont {Diwan}},\ }\bibfield
   {title} {\enquote {\bibinfo {title} {{Three-dimensional Imaging for Large
  LArTPCs}},}\ }\href {\doibase 10.1088/1748-0221/13/05/P05032} {\bibfield
  {journal} {\bibinfo  {journal} {JINST}\ }\textbf {\bibinfo {volume} {13}},\
  \bibinfo {pages} {P05032} (\bibinfo {year} {2018})}\BibitemShut {NoStop}%
\bibitem [{\citenamefont {Briese}\ \emph {et~al.}(2013)\citenamefont {Briese}
  \emph {et~al.}}]{Briese:2013wua}%
  \BibitemOpen
  \bibfield  {author} {\bibinfo {author} {\bibfnamefont {T.}~\bibnamefont
  {Briese}} \emph {et~al.},\ }\bibfield  {title} {\enquote {\bibinfo {title}
  {{Testing of Cryogenic Photomultiplier Tubes for the MicroBooNE
  Experiment}},}\ }\href {\doibase 10.1088/1748-0221/8/07/T07005} {\bibfield
  {journal} {\bibinfo  {journal} {JINST}\ }\textbf {\bibinfo {volume} {8}},\
  \bibinfo {pages} {T07005} (\bibinfo {year} {2013})}\BibitemShut {NoStop}%
%%CITATION = ARXIV:1304.0821;%%
\bibitem [{\citenamefont {Aguilar-Arevalo}\ \emph {et~al.}(2009)\citenamefont
  {Aguilar-Arevalo} \emph {et~al.}}]{AguilarArevalo:2008yp}%
  \BibitemOpen
  \bibfield  {author} {\bibinfo {author} {\bibfnamefont {A.~A.}\ \bibnamefont
  {Aguilar-Arevalo}} \emph {et~al.} (\bibinfo {collaboration} {MiniBooNE
  Collaboration}),\ }\bibfield  {title} {\enquote {\bibinfo {title} {{The
  Neutrino Flux prediction at MiniBooNE}},}\ }\href {\doibase
  10.1103/PhysRevD.79.072002} {\bibfield  {journal} {\bibinfo  {journal} {Phys.
  Rev. D.}\ }\textbf {\bibinfo {volume} {79}},\ \bibinfo {pages} {072002}
  (\bibinfo {year} {2009})}\BibitemShut {NoStop}%
%%CITATION = ARXIV:0806.1449;%%
\bibitem [{\citenamefont {Acciarri}\ \emph
  {et~al.}(2017{\natexlab{b}})\citenamefont {Acciarri} \emph
  {et~al.}}]{Acciarri:2017rnj}%
  \BibitemOpen
  \bibfield  {author} {\bibinfo {author} {\bibfnamefont {R.}~\bibnamefont
  {Acciarri}} \emph {et~al.} (\bibinfo {collaboration} {MicroBooNE
  Collaboration}),\ }\bibfield  {title} {\enquote {\bibinfo {title}
  {{Measurement of cosmic-ray reconstruction efficiencies in the MicroBooNE
  LArTPC using a small external cosmic-ray counter}},}\ }\href {\doibase
  10.1088/1748-0221/12/12/P12030} {\bibfield  {journal} {\bibinfo  {journal}
  {JINST}\ }\textbf {\bibinfo {volume} {12}},\ \bibinfo {pages} {P12030}
  (\bibinfo {year} {2017}{\natexlab{b}})}\BibitemShut {NoStop}%
%%CITATION = ARXIV:1707.09903;%%
\bibitem [{\citenamefont {Adams}\ \emph
  {et~al.}(2019{\natexlab{b}})\citenamefont {Adams} \emph
  {et~al.}}]{Adams:2018fud}%
  \BibitemOpen
  \bibfield  {author} {\bibinfo {author} {\bibfnamefont {C.}~\bibnamefont
  {Adams}} \emph {et~al.} (\bibinfo {collaboration} {MicroBooNE
  Collaboration}),\ }\bibfield  {title} {\enquote {\bibinfo {title}
  {{Comparison of $\nu_\mu$-Ar multiplicity distributions observed by
  MicroBooNE to GENIE model predictions}},}\ }\href {\doibase
  10.1140/epjc/s10052-019-6742-3} {\bibfield  {journal} {\bibinfo  {journal}
  {Eur. Phys. J.}\ }\textbf {\bibinfo {volume} {C79}},\ \bibinfo {pages} {248}
  (\bibinfo {year} {2019}{\natexlab{b}})}\BibitemShut {NoStop}%
%%CITATION = ARXIV:1805.06887;%%
\bibitem [{\citenamefont {Adams}\ \emph
  {et~al.}(2019{\natexlab{c}})\citenamefont {Adams} \emph
  {et~al.}}]{Adams:2018lzd}%
  \BibitemOpen
  \bibfield  {author} {\bibinfo {author} {\bibfnamefont {C.}~\bibnamefont
  {Adams}} \emph {et~al.} (\bibinfo {collaboration} {MicroBooNE
  Collaboration}),\ }\bibfield  {title} {\enquote {\bibinfo {title} {{Rejecting
  cosmic background for exclusive charged current quasi elastic neutrino
  interaction studies with Liquid Argon TPCs; a case study with the MicroBooNE
  detector}},}\ }\href {\doibase 10.1140/epjc/s10052-019-7184-7} {\bibfield
  {journal} {\bibinfo  {journal} {Eur. Phys. J.}\ }\textbf {\bibinfo {volume}
  {C79}},\ \bibinfo {pages} {673} (\bibinfo {year}
  {2019}{\natexlab{c}})}\BibitemShut {NoStop}%
%%CITATION = ARXIV:1812.05679;%%
\bibitem [{\citenamefont {Abratenko}\ \emph
  {et~al.}(2019{\natexlab{b}})\citenamefont {Abratenko} \emph
  {et~al.}}]{Adams:2019iqc}%
  \BibitemOpen
  \bibfield  {author} {\bibinfo {author} {\bibfnamefont {P.}~\bibnamefont
  {Abratenko}} \emph {et~al.} (\bibinfo {collaboration} {MicroBooNE
  Collaboration}),\ }\bibfield  {title} {\enquote {\bibinfo {title} {{First
  Measurement of Inclusive Muon Neutrino Charged Current Differential Cross
  Sections on Argon at $E_\nu\sim$0.8 GeV with the MicroBooNE Detector}},}\
  }\href {\doibase 10.1103/PhysRevLett.123.131801} {\bibfield  {journal}
  {\bibinfo  {journal} {Phys. Rev. Lett.}\ }\textbf {\bibinfo {volume} {123}},\
  \bibinfo {pages} {131801} (\bibinfo {year} {2019}{\natexlab{b}})}\BibitemShut
  {NoStop}%
%%CITATION = ARXIV:1905.09694;%%
\bibitem [{\citenamefont {Adamson}\ \emph
  {et~al.}(2016{\natexlab{a}})\citenamefont {Adamson} \emph
  {et~al.}}]{Adamson:2015dkw}%
  \BibitemOpen
  \bibfield  {author} {\bibinfo {author} {\bibfnamefont {P.}~\bibnamefont
  {Adamson}} \emph {et~al.},\ }\bibfield  {title} {\enquote {\bibinfo {title}
  {{The NuMI Neutrino Beam}},}\ }\href {\doibase 10.1016/j.nima.2015.08.063}
  {\bibfield  {journal} {\bibinfo  {journal} {Nucl. Instrum. Meth.}\ }\textbf
  {\bibinfo {volume} {A806}},\ \bibinfo {pages} {279--306} (\bibinfo {year}
  {2016}{\natexlab{a}})}\BibitemShut {NoStop}%
%%CITATION = ARXIV:1507.06690;%%
\bibitem [{\citenamefont {Tutto}(2017)}]{DelTutto:2017vtk}%
  \BibitemOpen
  \bibfield  {author} {\bibinfo {author} {\bibfnamefont {M.}~\bibnamefont
  {Tutto}},\ }\bibfield  {title} {\enquote {\bibinfo {title} {Venu: The virtual
  environment for neutrinos},}\ }\href@noop {} {\bibfield  {journal} {\bibinfo
  {journal} {ArXiv}\ }\textbf {\bibinfo {volume} {abs/1709.10120}} (\bibinfo
  {year} {2017})}\BibitemShut {NoStop}%
\bibitem [{\citenamefont {Li}\ \emph {et~al.}(2016)\citenamefont {Li} \emph
  {et~al.}}]{Li:2015rqa}%
  \BibitemOpen
  \bibfield  {author} {\bibinfo {author} {\bibfnamefont {Yichen}\ \bibnamefont
  {Li}} \emph {et~al.} (\bibinfo {collaboration} {MicroBooNE Collaboration}),\
  }\bibfield  {title} {\enquote {\bibinfo {title} {{Measurement of Longitudinal
  Electron Diffusion in Liquid Argon}},}\ }\href {\doibase
  10.1016/j.nima.2016.01.094} {\bibfield  {journal} {\bibinfo  {journal} {Nucl.
  Instrum. Meth.}\ }\textbf {\bibinfo {volume} {A816}},\ \bibinfo {pages}
  {160--170} (\bibinfo {year} {2016})}\BibitemShut {NoStop}%
%%CITATION = ARXIV:1508.07059;%%
\bibitem [{\citenamefont {Radeka}\ \emph {et~al.}(2011)\citenamefont {Radeka}
  \emph {et~al.}}]{Radeka:2011zz}%
  \BibitemOpen
  \bibfield  {author} {\bibinfo {author} {\bibfnamefont {Veljko}\ \bibnamefont
  {Radeka}} \emph {et~al.},\ }\bibfield  {title} {\enquote {\bibinfo {title}
  {{Cold electronics for 'Giant' Liquid Argon Time Projection Chambers}},}\
  }\bibfield  {booktitle} {\emph {\bibinfo {booktitle} {{Giant liquid argon
  charge imaging experiment. Proceedings, 1st International Workshop, GLA2010,
  Tsukuba, Japan, March 29-31, 2010}}},\ }\href {\doibase
  10.1088/1742-6596/308/1/012021} {\bibfield  {journal} {\bibinfo  {journal}
  {J. Phys. Conf. Ser.}\ }\textbf {\bibinfo {volume} {308}},\ \bibinfo {pages}
  {012021} (\bibinfo {year} {2011})}\BibitemShut {NoStop}%
%%CITATION = 00462,308,012021;%%
\bibitem [{\citenamefont {Acciarri}\ \emph
  {et~al.}(2017{\natexlab{c}})\citenamefont {Acciarri} \emph
  {et~al.}}]{Acciarri:2017sde}%
  \BibitemOpen
  \bibfield  {author} {\bibinfo {author} {\bibfnamefont {R.}~\bibnamefont
  {Acciarri}} \emph {et~al.} (\bibinfo {collaboration} {MicroBooNE
  Collaboration}),\ }\bibfield  {title} {\enquote {\bibinfo {title} {{Noise
  Characterization and Filtering in the MicroBooNE Liquid Argon TPC}},}\ }\href
  {\doibase 10.1088/1748-0221/12/08/P08003} {\bibfield  {journal} {\bibinfo
  {journal} {JINST}\ }\textbf {\bibinfo {volume} {12}},\ \bibinfo {pages}
  {P08003} (\bibinfo {year} {2017}{\natexlab{c}})}\BibitemShut {NoStop}%
%%CITATION = ARXIV:1705.07341;%%
\bibitem [{\citenamefont {Abratenko}\ \emph {et~al.}()\citenamefont {Abratenko}
  \emph {et~al.}}]{wire-cell-uboone}%
  \BibitemOpen
  \bibfield  {author} {\bibinfo {author} {\bibfnamefont {P.}~\bibnamefont
  {Abratenko}} \emph {et~al.} (\bibinfo {collaboration} {MicroBooNE
  Collaboration}),\ }\href@noop {} {\enquote {\bibinfo {title} {{Neutrino Event
  Selection in the MicroBooNE Liquid Argon Time Projection Chamber using
  Wire-Cell 3-D Imaging, Clustering and Charge-Light Matching}},}\ }\bibinfo
  {note} {Submitted to JINST}\BibitemShut {NoStop}%
\bibitem [{\citenamefont {Adams}\ \emph
  {et~al.}(2018{\natexlab{a}})\citenamefont {Adams} \emph
  {et~al.}}]{Adams:2018dra}%
  \BibitemOpen
  \bibfield  {author} {\bibinfo {author} {\bibfnamefont {C.}~\bibnamefont
  {Adams}} \emph {et~al.} (\bibinfo {collaboration} {MicroBooNE
  Collaboration}),\ }\bibfield  {title} {\enquote {\bibinfo {title}
  {{Ionization electron signal processing in single phase LArTPCs. Part I.
  Algorithm Description and quantitative evaluation with MicroBooNE
  simulation}},}\ }\href {\doibase 10.1088/1748-0221/13/07/P07006} {\bibfield
  {journal} {\bibinfo  {journal} {JINST}\ }\textbf {\bibinfo {volume} {13}},\
  \bibinfo {pages} {P07006--P07006} (\bibinfo {year}
  {2018}{\natexlab{a}})}\BibitemShut {NoStop}%
%%CITATION = ARXIV:1802.08709;%%
\bibitem [{\citenamefont {Adams}\ \emph
  {et~al.}(2018{\natexlab{b}})\citenamefont {Adams} \emph
  {et~al.}}]{Adams:2018gbi}%
  \BibitemOpen
  \bibfield  {author} {\bibinfo {author} {\bibfnamefont {C.}~\bibnamefont
  {Adams}} \emph {et~al.} (\bibinfo {collaboration} {MicroBooNE
  Collaboration}),\ }\bibfield  {title} {\enquote {\bibinfo {title}
  {{Ionization electron signal processing in single phase LArTPCs. Part II.
  Data/simulation comparison and performance in MicroBooNE}},}\ }\href
  {\doibase 10.1088/1748-0221/13/07/P07007} {\bibfield  {journal} {\bibinfo
  {journal} {JINST}\ }\textbf {\bibinfo {volume} {13}},\ \bibinfo {pages}
  {P07007} (\bibinfo {year} {2018}{\natexlab{b}})}\BibitemShut {NoStop}%
%%CITATION = ARXIV:1804.02583;%%
\bibitem [{\citenamefont {Baller}(2017)}]{Baller:2017ugz}%
  \BibitemOpen
  \bibfield  {author} {\bibinfo {author} {\bibfnamefont {Bruce}\ \bibnamefont
  {Baller}},\ }\bibfield  {title} {\enquote {\bibinfo {title} {{Liquid argon
  TPC signal formation, signal processing and reconstruction techniques}},}\
  }\href {\doibase 10.1088/1748-0221/12/07/P07010} {\bibfield  {journal}
  {\bibinfo  {journal} {JINST}\ }\textbf {\bibinfo {volume} {12}},\ \bibinfo
  {pages} {P07010} (\bibinfo {year} {2017})}\BibitemShut {NoStop}%
%%CITATION = ARXIV:1703.04024;%%
\bibitem [{\citenamefont {Cand\`{e}s}\ \emph {et~al.}(2006)\citenamefont
  {Cand\`{e}s}, \citenamefont {Romberg},\ and\ \citenamefont {Tao}}]{cs}%
  \BibitemOpen
  \bibfield  {author} {\bibinfo {author} {\bibfnamefont {E.~J.}\ \bibnamefont
  {Cand\`{e}s}}, \bibinfo {author} {\bibfnamefont {J.~K.}\ \bibnamefont
  {Romberg}}, \ and\ \bibinfo {author} {\bibfnamefont {T.}~\bibnamefont
  {Tao}},\ }\bibfield  {title} {\enquote {\bibinfo {title} {Stable signal
  recovery from incomplete and inaccurate measurements},}\ }\href {\doibase
  10.1002/cpa.20124} {\bibfield  {journal} {\bibinfo  {journal} {Communications
  on Pure and Applied Mathematics}\ }\textbf {\bibinfo {volume} {59}},\
  \bibinfo {pages} {1207--1223} (\bibinfo {year} {2006})}\BibitemShut {NoStop}%
\bibitem [{\citenamefont {Adamson}\ \emph
  {et~al.}(2016{\natexlab{b}})\citenamefont {Adamson} \emph
  {et~al.}}]{Adamson:2016xxw}%
  \BibitemOpen
  \bibfield  {author} {\bibinfo {author} {\bibfnamefont {P.}~\bibnamefont
  {Adamson}} \emph {et~al.} (\bibinfo {collaboration} {NOvA Collaboration}),\
  }\bibfield  {title} {\enquote {\bibinfo {title} {{First measurement of
  muon-neutrino disappearance in NOvA}},}\ }\href {\doibase
  10.1103/PhysRevD.93.051104} {\bibfield  {journal} {\bibinfo  {journal} {Phys.
  Rev.}\ }\textbf {\bibinfo {volume} {D93}},\ \bibinfo {pages} {051104}
  (\bibinfo {year} {2016}{\natexlab{b}})}\BibitemShut {NoStop}%
%%CITATION = ARXIV:1601.05037;%%
\bibitem [{\citenamefont {Aliaga}\ \emph {et~al.}(2014)\citenamefont {Aliaga}
  \emph {et~al.}}]{Aliaga:2013uqz}%
  \BibitemOpen
  \bibfield  {author} {\bibinfo {author} {\bibfnamefont {L.}~\bibnamefont
  {Aliaga}} \emph {et~al.} (\bibinfo {collaboration} {MINERvA Collaboration}),\
  }\bibfield  {title} {\enquote {\bibinfo {title} {{Design, Calibration, and
  Performance of the MINERvA Detector}},}\ }\href {\doibase
  10.1016/j.nima.2013.12.053} {\bibfield  {journal} {\bibinfo  {journal} {Nucl.
  Instrum. Meth.}\ }\textbf {\bibinfo {volume} {A743}},\ \bibinfo {pages}
  {130--159} (\bibinfo {year} {2014})}\BibitemShut {NoStop}%
%%CITATION = ARXIV:1305.5199;%%
\bibitem [{\citenamefont {Michael}\ \emph {et~al.}(2008)\citenamefont {Michael}
  \emph {et~al.}}]{Michael:2008bc}%
  \BibitemOpen
  \bibfield  {author} {\bibinfo {author} {\bibfnamefont {D.~G.}\ \bibnamefont
  {Michael}} \emph {et~al.} (\bibinfo {collaboration} {MINOS Collaboration}),\
  }\bibfield  {title} {\enquote {\bibinfo {title} {{The Magnetized steel and
  scintillator calorimeters of the MINOS experiment}},}\ }\href {\doibase
  10.1016/j.nima.2008.08.003} {\bibfield  {journal} {\bibinfo  {journal} {Nucl.
  Instrum. Meth.}\ }\textbf {\bibinfo {volume} {A596}},\ \bibinfo {pages}
  {190--228} (\bibinfo {year} {2008})}\BibitemShut {NoStop}%
%%CITATION = ARXIV:0805.3170;%%
\bibitem [{\citenamefont {Agostinelli}\ \emph {et~al.}(2003)\citenamefont
  {Agostinelli} \emph {et~al.}}]{Geant4}%
  \BibitemOpen
  \bibfield  {author} {\bibinfo {author} {\bibfnamefont {S.}~\bibnamefont
  {Agostinelli}} \emph {et~al.},\ }\bibfield  {title} {\enquote {\bibinfo
  {title} {{Geant4 -- a simulation toolkit}},}\ }\href {\doibase
  10.1016/S0168-9002(03)01368-8} {\bibfield  {journal} {\bibinfo  {journal}
  {Nucl. Instrum. Meth.}\ }\textbf {\bibinfo {volume} {A506}},\ \bibinfo
  {pages} {250--303} (\bibinfo {year} {2003})}\BibitemShut {NoStop}%
\bibitem [{\citenamefont {Antonello}\ \emph {et~al.}(2017)\citenamefont
  {Antonello} \emph {et~al.}}]{Antonello:2016niy}%
  \BibitemOpen
  \bibfield  {author} {\bibinfo {author} {\bibfnamefont {M.}~\bibnamefont
  {Antonello}} \emph {et~al.} (\bibinfo {collaboration} {ICARUS
  Collaboration}),\ }\bibfield  {title} {\enquote {\bibinfo {title} {{Muon
  momentum measurement in ICARUS-T600 LAr-TPC via multiple scattering in
  few-GeV range}},}\ }\href {\doibase 10.1088/1748-0221/12/04/P04010}
  {\bibfield  {journal} {\bibinfo  {journal} {JINST}\ }\textbf {\bibinfo
  {volume} {12}},\ \bibinfo {pages} {P04010} (\bibinfo {year}
  {2017})}\BibitemShut {NoStop}%
%%CITATION = ARXIV:1612.07715;%%
\bibitem [{\citenamefont {Abratenko}\ \emph {et~al.}(2017)\citenamefont
  {Abratenko} \emph {et~al.}}]{Abratenko:2017nki}%
  \BibitemOpen
  \bibfield  {author} {\bibinfo {author} {\bibfnamefont {P.}~\bibnamefont
  {Abratenko}} \emph {et~al.} (\bibinfo {collaboration} {MicroBooNE
  Collaboration}),\ }\bibfield  {title} {\enquote {\bibinfo {title}
  {{Determination of muon momentum in the MicroBooNE LArTPC using an improved
  model of multiple Coulomb scattering}},}\ }\href {\doibase
  10.1088/1748-0221/12/10/P10010} {\bibfield  {journal} {\bibinfo  {journal}
  {JINST}\ }\textbf {\bibinfo {volume} {12}},\ \bibinfo {pages} {P10010}
  (\bibinfo {year} {2017})}\BibitemShut {NoStop}%
%%CITATION = ARXIV:1703.06187;%%
\bibitem [{\citenamefont {Kalman}(1960)}]{kalman_filter}%
  \BibitemOpen
  \bibfield  {author} {\bibinfo {author} {\bibfnamefont {R.~E.}\ \bibnamefont
  {Kalman}},\ }\bibfield  {title} {\enquote {\bibinfo {title} {A new approach
  to linear filtering and prediction problems},}\ }\href@noop {} {\bibfield
  {journal} {\bibinfo  {journal} {Journal of Basic Engineering}\ }\textbf
  {\bibinfo {volume} {{\bf 82}}},\ \bibinfo {pages} {35} (\bibinfo {year}
  {1960})}\BibitemShut {NoStop}%
\bibitem [{\citenamefont {Antonello}\ \emph {et~al.}(2013)\citenamefont
  {Antonello} \emph {et~al.}}]{Antonello:2012hu}%
  \BibitemOpen
  \bibfield  {author} {\bibinfo {author} {\bibfnamefont {M.}~\bibnamefont
  {Antonello}} \emph {et~al.},\ }\bibfield  {title} {\enquote {\bibinfo {title}
  {{Precise 3D track reconstruction algorithm for the ICARUS T600 liquid argon
  time projection chamber detector}},}\ }\href {\doibase 10.1155/2013/260820}
  {\bibfield  {journal} {\bibinfo  {journal} {Adv. High Energy Phys.}\ }\textbf
  {\bibinfo {volume} {2013}},\ \bibinfo {pages} {260820} (\bibinfo {year}
  {2013})}\BibitemShut {NoStop}%
%%CITATION = ARXIV:1210.5089;%%
\bibitem [{BiC(2019)}]{BiCGSTAB}%
  \BibitemOpen
  \href@noop {} {\enquote {\bibinfo {title} {{Biconjugate gradient stablized
  method (BiCGSTAB)}},}\ }\bibinfo {howpublished}
  {\url{https://eigen.tuxfamily.org/dox/classEigen_1_1BiCGSTAB.html}} (\bibinfo
  {year} {2019})\BibitemShut {NoStop}%
\bibitem [{\citenamefont {Adams}\ \emph
  {et~al.}(2020{\natexlab{a}})\citenamefont {Adams} \emph
  {et~al.}}]{Adams:2019qrr}%
  \BibitemOpen
  \bibfield  {author} {\bibinfo {author} {\bibfnamefont {C.}~\bibnamefont
  {Adams}} \emph {et~al.} (\bibinfo {collaboration} {MicroBooNE
  Collaboration}),\ }\bibfield  {title} {\enquote {\bibinfo {title} {{A Method
  to Determine the Electric Field of Liquid Argon Time Projection Chambers
  Using a UV Laser System and its Application in MicroBooNE}},}\ }\href
  {\doibase 10.1088/1748-0221/15/07/P07010} {\bibfield  {journal} {\bibinfo
  {journal} {JINST}\ }\textbf {\bibinfo {volume} {15}},\ \bibinfo {pages}
  {P07010} (\bibinfo {year} {2020}{\natexlab{a}})}\BibitemShut {NoStop}%
%%CITATION = ARXIV:1910.01430;%%
\bibitem [{\citenamefont {Abratenko}\ \emph
  {et~al.}(2020{\natexlab{c}})\citenamefont {Abratenko} \emph
  {et~al.}}]{Abratenko:2020bbx}%
  \BibitemOpen
  \bibfield  {author} {\bibinfo {author} {\bibfnamefont {P.}~\bibnamefont
  {Abratenko}} \emph {et~al.} (\bibinfo {collaboration} {MicroBooNE}),\
  }\bibfield  {title} {\enquote {\bibinfo {title} {{Measurement of Space Charge
  Effects in the MicroBooNE LArTPC Using Cosmic Muons}},}\ }\href@noop {}
  {\bibfield  {journal} {\bibinfo  {journal} {JINST}\ }\textbf {\bibinfo
  {volume} {15}},\ \bibinfo {pages} {P12037} (\bibinfo {year}
  {2020}{\natexlab{c}})}\BibitemShut {NoStop}%
\bibitem [{pst(2019)}]{pstar}%
  \BibitemOpen
  \href@noop {} {\enquote {\bibinfo {title} {{PSTAR at NIST}},}\ }\bibinfo
  {howpublished}
  {\url{https://physics.nist.gov/PhysRefData/Star/Text/PSTAR.html}} (\bibinfo
  {year} {2019})\BibitemShut {NoStop}%
\bibitem [{\citenamefont {Acciarri}\ \emph {et~al.}(2013)\citenamefont
  {Acciarri} \emph {et~al.}}]{Acciarri:2013met}%
  \BibitemOpen
  \bibfield  {author} {\bibinfo {author} {\bibfnamefont {R.}~\bibnamefont
  {Acciarri}} \emph {et~al.} (\bibinfo {collaboration} {ArgoNeuT
  Collaboration}),\ }\bibfield  {title} {\enquote {\bibinfo {title} {{A Study
  of Electron Recombination Using Highly Ionizing Particles in the ArgoNeuT
  Liquid Argon TPC}},}\ }\href {\doibase 10.1088/1748-0221/8/08/P08005}
  {\bibfield  {journal} {\bibinfo  {journal} {JINST}\ }\textbf {\bibinfo
  {volume} {8}},\ \bibinfo {pages} {P08005} (\bibinfo {year}
  {2013})}\BibitemShut {NoStop}%
%%CITATION = ARXIV:1306.1712;%%
\bibitem [{\citenamefont {Adams}\ \emph
  {et~al.}(2020{\natexlab{b}})\citenamefont {Adams} \emph
  {et~al.}}]{Adams:2019ssg}%
  \BibitemOpen
  \bibfield  {author} {\bibinfo {author} {\bibfnamefont {C.}~\bibnamefont
  {Adams}} \emph {et~al.} (\bibinfo {collaboration} {MicroBooNE
  Collaboration}),\ }\bibfield  {title} {\enquote {\bibinfo {title}
  {{Calibration of the Charge and Energy Response of the MicroBooNE Liquid
  Argon Time Projection Chamber using Muons and Protons}},}\ }\href {\doibase
  10.1088/1748-0221/15/03/P03022} {\bibfield  {journal} {\bibinfo  {journal}
  {JINST}\ }\textbf {\bibinfo {volume} {15}},\ \bibinfo {pages} {P03022}
  (\bibinfo {year} {2020}{\natexlab{b}})}\BibitemShut {NoStop}%
%%CITATION = ARXIV:1907.11736;%%
\bibitem [{\citenamefont {Andreopoulos}\ \emph {et~al.}(2015)\citenamefont
  {Andreopoulos} \emph {et~al.}}]{Genie2015}%
  \BibitemOpen
  \bibfield  {author} {\bibinfo {author} {\bibfnamefont {C.}~\bibnamefont
  {Andreopoulos}} \emph {et~al.},\ }\href@noop {} {\enquote {\bibinfo {title}
  {{The GENIE Neutrino Monte Carlo Generator: Physics and User Manual}},}\ }
  (\bibinfo {year} {2015}),\ \Eprint {http://arxiv.org/abs/1510.05494}
  {arXiv:1510.05494 [hep-ph]} \BibitemShut {NoStop}%
\bibitem [{\citenamefont {Acciarri}\ \emph {et~al.}(2018)\citenamefont
  {Acciarri} \emph {et~al.}}]{Acciarri:2017hat}%
  \BibitemOpen
  \bibfield  {author} {\bibinfo {author} {\bibfnamefont {R.}~\bibnamefont
  {Acciarri}} \emph {et~al.} (\bibinfo {collaboration} {MicroBooNE
  Collaboration}),\ }\bibfield  {title} {\enquote {\bibinfo {title} {{The
  Pandora multi-algorithm approach to automated pattern recognition of
  cosmic-ray muon and neutrino events in the MicroBooNE detector}},}\ }\href
  {\doibase 10.1140/epjc/s10052-017-5481-6} {\bibfield  {journal} {\bibinfo
  {journal} {Eur. Phys. J.}\ }\textbf {\bibinfo {volume} {C78}},\ \bibinfo
  {pages} {82} (\bibinfo {year} {2018})}\BibitemShut {NoStop}%
%%CITATION = ARXIV:1708.03135;%%
\bibitem [{\citenamefont {Acciarri}\ \emph
  {et~al.}(2017{\natexlab{d}})\citenamefont {Acciarri} \emph
  {et~al.}}]{Acciarri:2016ryt}%
  \BibitemOpen
  \bibfield  {author} {\bibinfo {author} {\bibfnamefont {R.}~\bibnamefont
  {Acciarri}} \emph {et~al.} (\bibinfo {collaboration} {MicroBooNE
  Collaboration}),\ }\bibfield  {title} {\enquote {\bibinfo {title}
  {{Convolutional Neural Networks Applied to Neutrino Events in a Liquid Argon
  Time Projection Chamber}},}\ }\href {\doibase 10.1088/1748-0221/12/03/P03011}
  {\bibfield  {journal} {\bibinfo  {journal} {JINST}\ }\textbf {\bibinfo
  {volume} {12}},\ \bibinfo {pages} {P03011} (\bibinfo {year}
  {2017}{\natexlab{d}})}\BibitemShut {NoStop}%
%%CITATION = ARXIV:1611.05531;%%
\bibitem [{\citenamefont {Adams}\ \emph
  {et~al.}(2019{\natexlab{d}})\citenamefont {Adams} \emph
  {et~al.}}]{Adams:2018bvi}%
  \BibitemOpen
  \bibfield  {author} {\bibinfo {author} {\bibfnamefont {C.}~\bibnamefont
  {Adams}} \emph {et~al.} (\bibinfo {collaboration} {MicroBooNE
  Collaboration}),\ }\bibfield  {title} {\enquote {\bibinfo {title} {{Deep
  neural network for pixel-level electromagnetic particle identification in the
  MicroBooNE liquid argon time projection chamber}},}\ }\href {\doibase
  10.1103/PhysRevD.99.092001} {\bibfield  {journal} {\bibinfo  {journal} {Phys.
  Rev.}\ }\textbf {\bibinfo {volume} {D99}},\ \bibinfo {pages} {092001}
  (\bibinfo {year} {2019}{\natexlab{d}})}\BibitemShut {NoStop}%
%%CITATION = ARXIV:1808.07269;%%
\bibitem [{\citenamefont {Adams}\ \emph
  {et~al.}(2020{\natexlab{c}})\citenamefont {Adams} \emph
  {et~al.}}]{Adams:2019law}%
  \BibitemOpen
  \bibfield  {author} {\bibinfo {author} {\bibfnamefont {C.}~\bibnamefont
  {Adams}} \emph {et~al.} (\bibinfo {collaboration} {MicroBooNE
  Collaboration}),\ }\bibfield  {title} {\enquote {\bibinfo {title}
  {{Reconstruction and Measurement of $\mathcal{O}$(100) MeV Energy
  Electromagnetic Activity from $\pi^0 \rightarrow \gamma\gamma$ Decays in the
  MicroBooNE LArTPC}},}\ }\href {\doibase 10.1088/1748-0221/15/02/p02007}
  {\bibfield  {journal} {\bibinfo  {journal} {JINST}\ }\textbf {\bibinfo
  {volume} {15}},\ \bibinfo {pages} {P02007} (\bibinfo {year}
  {2020}{\natexlab{c}})}\BibitemShut {NoStop}%
%%CITATION = ARXIV:1910.02166;%%
\bibitem [{gra(2019{\natexlab{a}})}]{graph_dijkstra_shortest_path}%
  \BibitemOpen
  \href@noop {} {\enquote {\bibinfo {title} {{Dijkstra's shortest path
  algorithm in graph theory}},}\ }\bibinfo {howpublished}
  {\url{https://www.boost.org/doc/libs/1_41_0/libs/graph/doc/dijkstra_shortest_paths.html}}
  (\bibinfo {year} {2019}{\natexlab{a}})\BibitemShut {NoStop}%
\bibitem [{Ste(2019)}]{Steiner}%
  \BibitemOpen
  \href@noop {} {\enquote {\bibinfo {title} {{Steiner tree greedy
  algorithm}},}\ }\bibinfo {howpublished}
  {\url{http://paal.mimuw.edu.pl/docs/index.html}} (\bibinfo {year}
  {2019})\BibitemShut {NoStop}%
\bibitem [{gra(2019{\natexlab{b}})}]{graph_connected_component}%
  \BibitemOpen
  \href@noop {} {\enquote {\bibinfo {title} {{Connected component algorithm in
  graph theory}},}\ }\bibinfo {howpublished}
  {\url{https://www.boost.org/doc/libs/1_68_0/libs/graph/doc/connected_components.html}}
  (\bibinfo {year} {2019}{\natexlab{b}})\BibitemShut {NoStop}%
\end{thebibliography}%

\end{document}